\documentclass[12pt,MSc]{muthesis}
\usepackage{listings}
\usepackage{verbatim}
\usepackage{graphicx}
\usepackage{url} % typeset URL's reasonably

\usepackage{fancyhdr}
\begin{document}
% Uncomment the following lines to leave out list of figures, tables
% and copyright until final printing
%\figurespagefalse
%\tablespagefalse
%\copyrightfalse

\title{Stochastic Volatility in a Quantitative Model of Stock Market Returns}
\author{Gilles Daniel}
\principaladviser{Professor David S. Br\'ee}

\beforeabstract
\prefacesection{Abstract}
\nocite{Hull}
\nocite{Dragu}
\nocite{Fama}
\nocite{Poncet}
\nocite{MathWorld}
\nocite{Yahoo}
\nocite{arXiv}
\nocite{Micciche}
\nocite{Johansen}
\nocite{Johansen2}
\nocite{Mikosch}
\nocite{Gardiner}
\nocite{Bree}

Standard quantitative models of the stock market predict a log-normal distribution for stock returns (Bachelier 1900, Osborne 1959), but it is recognised (Fama 1965) that empirical data, in comparison with a Gaussian, exhibit leptokurtosis (it has more probability mass in its tails and centre) and fat tails (probabilities of extreme events are underestimated). Different attempts to explain this departure from normality have coexisted. In particular, since one of the strong assumptions of the Gaussian model concerns the volatility, considered finite and constant, the new models were built on a non finite (Mandelbrot 1963) or non constant (Cox, Ingersoll and Ross 1985) volatility. \\

We investigate in this thesis a very recent model (Dragulescu et al. 2002) based on a Brownian motion process for the returns, and a stochastic mean-reverting process for the volatility. In this model, the forward Kolmogorov equation that governs the time evolution of returns is solved analytically. We test this new theory against different stock indexes (Dow Jones Industrial Average, Standard and Poor's and Footsie), over different periods (from 20 to 105 years). Our aim is to compare this model with the classical Gaussian and with a simple Neural Network, used as a benchmark. \\

We perform the usual statistical tests on the kurtosis and tails of the expected distributions, paying particular attention to the outliers. As claimed by the authors, the new model outperforms the Gaussian for any time lag, but is artificially too complex for medium and low frequencies, where the Gaussian is preferable. Moreover this model is still rejected for high frequencies, at a 0.05 level of significance, due to the kurtosis, incorrectly handled.

\afterabstract

\prefacesection{Acknowledgements}
I would like to express my thanks and appreciation to my supervisor, Professor David S. Br\'ee, for his valuable advice and guidance, and my gratitude to Reader Nathan L. Joseph from the Accounting and Finance Department of the University of Manchester, for his continuous support and assistance. \\

I would also like to thank Robert Woolfson (Ph.D. Student) for his help and enlightening comments and Bin Yang (M.Sc. Student) for his patience and skills as a team worker. \\

I feel a deep sense of gratitude for my father and mother, Hubert and Gina, who formed part of my vision and taught me the good things that really matter in life. I am grateful for my brother, Vincent, for rendering me the sense and the value of brotherhood. \\

Finally, the chain of my gratitude would be definitely incomplete if I would forget to thank my fellows in Manchester, Laure, Bruno, Fabrice, Vivek, Matthias, Arnaud, ToLiS, Darko and Tom, and more generally my very close friends, Guillaume, Jer\^ome, Max and Hugues. \\

Without the help of all these people none of the current work would have been feasible.

\prefacesection{Dedication}
\vspace{3cm}
\begin{flushright}
\emph{A Paul,}

\emph{Qui vit d\'esormais \`a travers nous}
\end{flushright}

\afterpreface

% These include the actual text
\chapter{Introduction}

\section{A two century old paradigm}

Many attempts have been made, since the first agreement to trade on the NYSE in 1792, to model the stock market's behaviour. Understanding the patterns that govern this heart of the capitalism is more than challenging, it is a crusade. But so far, can anybody claim to have found out the rules enabling them to predict tomorrow's move for instance? And do these rules even exist? Actually, the \emph{efficient markets theory} states that market prices reflect the knowledge and expectations of all investors. As a consequence, this theory predicts, according to E. Fama \cite{Fama}, that the market would react quickly to such a discovery and these patterns would be modified instantly. Contrary to natural laws that govern physics, laws of the market adjust themselves to new discoveries. 

Anyhow, financial and scientific communities persist in building new models, not only because we are eager to understand, but mainly because predicting tomorrow's move is not the only way to make money. Where the study of Fundamentals or Technical Analysis are broadly used by traders on the market floor, Quantitative Finance tries to evaluate risk and hence price options using statistical models of the market. 

\section{Aims}

We will present in this thesis the theory underlying most of these models, the Theory of Random Walks. Then we will discuss the different hypotheses that have been proposed, in the Bachelier-Osborne and Mandelbrot models, concerning a major parameter, the standard deviation. This parameter is crucial since it represents the volatility of the market, and is used for instance in Value-at-Risk. After having confronted these models with empirical data, we will see that the structure of the market implies neither a constant standard deviation, as thought initially, nor an infinite standard deviation: the Probability Density Function (PDF) of the log-returns exhibits fat tails and kurtosis (peakedness or flatness compared to a Gaussian), with finite standard deviation.

Then we will focus on one of these models, published very recently by A. A. Dragulescu and V. M. Yakovenko, from the University of Maryland. Their paper \cite{Dragu}, ``Probability distribution of returns with stochastic volatility'' introduces a new model for volatility of stock market indexes. The proposed probability density function of stock returns seems to fit empirical data much better than previous models. We will double-check their results and propose a methodology to test their model against empirical data.

But first, let's have a look at the different models of the market. Some are used by traders to try to predict tomorrow's move, others by derivative traders to price options or by risk managers to set the global policy of an Investment Bank, for instance. 

\section{Usual Stock Market Prediction Methods}

On market floors, two radically different types of traders usually coexist: fundamentalists and chartists. Fundamentalists believe that the stock price of a company reflects its intrinsic value. This intrinsic value depends on the present and forecast economic situation of the company (its ``fundamentals''), and is mainly influenced by any new piece of information about these fundamentals. The problem is to evaluate this intrinsic value.

On the other hand, chartists only analyse historical data (the ``charts'') of the stock, mainly the historical price, but other indicators as well, such as traded volumes, volatility, past resistance and support thresholds, etc. They try hard to find out hidden patterns that replicate over days, weeks, month or years, according to their speculative or investment needs. The assumption is that the market should have a short/long term memory, so we could use the past to predict the future. But here, the \emph{efficient market hypothesis} asserts that all information which can be learned from technical analysis of stock prices is already reflected in those prices. According to this hypothesis, past stock prices may be useful to estimate the parameters of the distribution of future returns, but they do not provide information which permits an investor to outperform the market.

\section{Quantitative Methods}

Broadly speaking, none of those stock traders (fundamentalists or chartists) daily use the quantitative models we are describing in this thesis. But these models are used on the floor by derivative traders and in risk divisions of investment banks to elaborate the global trading policy of the bank, the risk aversion, the over-night limits of individual traders, etc.

Now the reader won't be surprised to learn that quantitative models are neither fundamentalists nor chartists. They are much more deeply involved with maths. In fact, the theory underlying most of these models, called the ``Theory of Random Walks'',  claims that successive price changes ($P_\tau - P_0$) or price returns ($P_\tau / P_0$) are independent, identically distributed random variables. This i.i.d. hypothesis has been studied by E. Fama in 1965 \cite{Fama} and is still discussed today. We will discuss the independence of successive price changes in Chapter 2, but at the moment we will study different models that make this very strong assumption.  

Under this assumption of i.i.d. price returns, many models have been developed, but two of them are used commonly nowadays. The first and most common one, called the Bachelier-Osborne model and elaborated in 1959, states that price returns have a constant finite volatility over a given period of time (``time lag $\tau$''), e.g. one day, one week, one month, etc. This theory results in a log-normal distribution for price returns and a volatility proportional to the square root of the time lag, i.e. the weekly volatility will be about $\sqrt{5}$ times higher than the daily one. But it is now known that price returns do not follow a Gaussian distribution, since they exhibit kurtosis and fat tails: dramatic draw downs and spectacular jumps arise far more often than predicted by a Normal distribution. Hence, the idea of infinite volatility appeared. It was introduced by Mandelbrot in 1963 and leads to stable Pareto-Levy distributions that can exhibit fat tails. Unfortunately, the hypothesis of infinite volatility supposes that the variance increases indefinitely with sample size, which is not verified by empirical data. Variance first increases then reaches a bound \cite{HerdBehavior}.

\section{Conclusion}

For centuries, practitioner's have tried to model and predict the financial markets, using diverse techniques such as fundamental study or technical analysis. With the rapid growth of statistics and stochastic calculus fifty years ago, new quantitative methods were born that seem to be able to handle the complexity of the stock market.

The main model, Bachelier-Osborne, will be detailed in Chapter 2. We will see that it suffers mainly from two imperfections, high kurtosis and fat tails, that still remain to be explained. A. Dragulescu and V. Yakovenko published in March 2002 an improvement of this model, based on a stochastic finite volatility, that seems to fit the data perfectly. We will analyse this model in Chapter 3 and test it in Chapter 4.

But first, let us present the basis of most of stock returns models, the Theory of Random Walks.   

\chapter{The Theory of Random Walks}
\section{Introduction}

The Theory of Random Walks has been used for the last 35 years by the main statistical models of the stock markets. It was first introduced by Bachelier in his 1900 dissertation written in Paris, "Th\'eorie de la Sp\'eculation" (and in his subsequent work, esp. 1906, 1913), in which he anticipated much of what was to become standard fare in financial theory: the random walk of financial market prices, Brownian motion and martingales (all before both Einstein and Wiener!). His innovativeness, however, was not appreciated by his professors or contemporaries. His dissertation received poor marks from his teachers and, consequently blackballed, he quickly dropped into the shadows of the academic underground. After a series of minor posts, he ended up obscurely teaching in Besancon for much of the rest of his life. Virtually nothing else is known of this pioneer - his work being largely ignored until the 1960s when Osborne introduced his model based on Bachelier's work. 

A random walk is a random process consisting of a sequence of discrete steps of fixed length totally independent one from another. For instance, the random thermal perturbations in a liquid are responsible for a random walk phenomenon known as Brownian motion, and the collisions of molecules in a gas are a random walk responsible for diffusion.

Applied to our problem, this theory is founded on two strong hypothesis: price returns are independent (tomorrow's price return does not depend on today's or on any other price return) and identically distributed (they all follow the same distribution). This is called the i.i.d. hypothesis.

Throughout this thesis, we will use these notations
\begin{center}
\begin{tabular}{ccc}
price change & price return & log return \\
$c_t = S_t - S_{t-\tau}$ & $p_t = \frac{S_t - S_{t-\tau}}{S_{t-\tau}}$ & $r_t = log(\frac{S_t}{S_{t-\tau}})$ \\
\end{tabular}
\end{center}

\begin{tabular}{ll}
where & log is the natural logarithm \\
&$S_t$ is the close price of the security at time $t$ \\
&$S_{t-\tau}$ is the close price of the security at time $t-\tau$ \\
&$\tau$ is the time lag \\
\end{tabular}

For instance, if the close price of the studied index is 106 today and was 105 
yesterday, the daily (time lag  $\tau$= 1) values are

\begin{center}
\begin{tabular}{ccc}
price change & price return & log return \\
$106 - 105 = 1$ & $\frac{106}{105} = 1.009$ & $log(\frac{106}{105}) = 9.48*10^{-3}$ \\
\end{tabular}
\end{center}

By ``day'', we mean trading day, since all of our datasets are composed of trading days only: week-ends and bank holidays have been removed. By ``time lag'', we mean the number of days between two points used to compute a log return. If our initial dataset is composed of 1000 close prices, then for a time lag of five days, we will take one point every five to compute the log returns. As a consequence, our final dataset will will be composed of $[\frac{1000}{5}]=200$ log returns only. Nevertheless, we can begin by the first, the second, the third, the fourth or the fifth close price, so that finally, we can use five different datasets of 200 log returns. This will allow us to give, for any computation, the average value and an estimate of the variance of the result, which will give greater robustness to our statistical tests.

We will usually use the log return, mainly for two reasons:
\begin{enumerate}
\item financially, it corresponds to the continuously compounded return of the asset S;\footnote{If $\alpha$ is the continuously compounded return of an asset S, then the value $S_t$ of S at time \emph{t} is $S_t$ = $S_o$ $e^{\alpha t}$ for t $\in$ [0,T]} 
\item numerically, it has the advantage of guaranteeing the positivity of the price.
\end{enumerate}

Obviously, any hypothesis about the independence and identical distribution of price changes is directly applicable to price returns and log returns.

According to the theory of random walks, the price change series is a collection of random variables having the property that, given the present, the future is conditionally independent of the past. In other words,
$$
P(c_t=c | c_{t-1},c_{t-2}, ...) = P(c_t=c)
$$
or, formulated directly in terms of price $S_t$ at time t
$$
P(S_t=S | S_{t-1},S_{t-2}, ...) = P(S_t=S | S_{t-1})
$$

To sum up, in such a process without memory, the last realisation contains all of the information. This process is known as a Markov process.

Before we go more deeply in the mathematics of the theory of random walks, and study the main statistical model of stock market behaviour, the Bachelier-Osborne model, let us discuss the i.i.d. hypothesis itself.
 
\section{Independence of price returns}

As said before, all of the assumptions about price returns can be applied to price changes and log returns. Since we will not consider the mathematical aspect of the theory in this paragraph, we will prefer to use the price returns, simpler to tackle and often discussed in the financial press in terms of percentage of variation.

The hypothesis of \emph{independent} price returns is extremely important - and controversial - since it underlies all of the theory of random walks, and so all of the models developed around it. E. Fama discussed abundantly this hypothesis in his paper "The Behavior of Stock-Market Prices" \cite{Fama} and states that the independence of price returns is the result of a noisy price mechanism. By noise, one should understand the psychology of different traders and the uncertainty or disagreement about the intrinsic value of the security, which depends on new information arrived or about to arrive. If successive bits of new information arise independently across time and if noise or uncertainty concerning intrinsic value does not tend to follow any consistent pattern, then successive price returns in a common stock will be independent.

A third and crucial condition for independence of price returns is the existence of "superior traders", viz. traders who will detect abnormalities on stock prices - departure of the security price from its intrinsic value - and will correct them by buying (resp. selling) the security if it is underestimated (resp. overestimated). If there are enough such traders, then the price will tend to stabilise around its intrinsic value, reducing risks of bubbles or crashes.

In the light of the recent scandals about the conflicts of interests of financial analysts working for the largest Investment Banks that participated in the creation of the speculative bubble around the "new economy", it is legitimate to wonder if this last condition enunciated by Fama is still respected, and then if the hypothesis of independent price returns still holds. But this problem is out of the scope of this thesis, and from now on we will make the assumption that the hypothesis underlying the Random Walk are respected: price returns will be considered independent and identically distributed.

Let us have a look now at the classical statistical model of the stock market, the Geometric Brownian Motion. 

\section{Bachelier-Osborne Model}

The basic theory, known as the Bachelier-Osborne model, states that the stock index prices $S_t$ follows a Geometric Brownian Motion (GBM). The description "Brownian motion" comes from the fact that the same process describes the physical motion of a particle subject to random shocks, a phenomenon first noted by the British physicist Brown in 1828, observing irregular movement of pollen suspended in water. The first mathematically rigorous construction of Brownian motion was carried out by Wiener in 1923. This theory is based on Markov Processes, Wiener processes and It\^o processes, which are detailed in Appendix A. We  summarise here the basic idea of the GBM.

Let the price $S=\{S_t; t=0,1,...,T\}$ be a non negative stochastic process.\\
And the log return $r_t=log(\frac{S_t}{S_0})$

The idea, first introduced by Bachelier \cite{Bachelier}, even if he used a Brownian motion and not a geometric Brownian motion, is that for a given time lag \emph{t}, the log return $r_t$ is the sum of a large number of i.i.d. random variables $\Delta r_i$

$$
r_t = \sum_{\tau=1}^{n}\Delta r_{t,\tau}
$$   

Then if we assume that the distribution followed by the $\Delta r_{t,\tau}$ has finite moments, and specially finite variance, then the Central Limit Theorem states directly that $r_t$ must follow a Normal distribution. Osborne formalised this in 1959 using the following equation, called Geometric Brownian Motion, for the stock price $S_t$ 

\begin{equation} \label{eqn:gbm1}
dS_t = \mu S_t dt + \sigma S_t dW_t 
\end{equation}
where $\mu$ and $\sigma$ are two constants called the drift and the volatility, and W is a standard Wiener process.\footnote{See Appendix A. The increments of W, $dW_t$, are normally distributed with $E[dW_t]=0$ and $Var[dW_t]=dt$}

We will demonstrate that $Log S_t$ obeys a simple Brownian motion with an instantaneous expectation  $\mu-\frac{\sigma^2}{2}$ and an instantaneous volatility $\sigma^2$. 

We start from \ref{eqn:gbm1} and apply It\^o's lemma to the following function
$$
f
\left\{
\begin{array}{rcl}
\Re^2 & \stackrel{f}{\longrightarrow} & \Re \\
(S,t) & \longmapsto & Log (S(t)) = Log S_t
\end{array}
\right.
$$

We obtain

\begin{eqnarray}\label{eqn:gbm2}
dLog S_t & = & \frac{\delta Log S_t}{\delta t}dt + \frac{\delta Log S_t}{\delta s}dS + \frac{1}{2}\frac{\delta^2 Log S_t}{\delta s^2}(dS)^2 \nonumber \\
& = & 0 + \frac{dS_t}{S_t} - \frac{1}{2}\frac{(dS_t)^2}{S_t^2} \ \ \ \mbox{since} \ \frac{\delta Log X}{\delta X}=\frac{1}{X} \ \mbox{and} \ \frac{\delta^2 Log X}{\delta X^2}=-\frac{1}{X^2}, \ \forall X>0\nonumber \\
& = & \frac{dS_t}{S_t} - \frac{1}{2}\frac{(dS_t)^2}{S_t^2}
\end{eqnarray}

Besides 

\begin{eqnarray}\label{eqn:gbm3}
(\frac{dS_t}{S_t})^2 & = & \mu^2(dt)^2 + \sigma^2(dW_t)^2+2 \mu \sigma dtdW_t \nonumber \\
& = & \sigma^2(dW_t)^2 + O(dt)^{\frac{3}{2}} \ \ \ \mbox{since} \ (dW_t)^2=O(dt) \nonumber \\
& = & \sigma^2dt + O(dt)^{\frac{3}{2}} \nonumber \\
& = & \sigma^2dt + o(dt)
\end{eqnarray} 
\begin{tabular}{rl}
where & $ \displaystyle q(t)=O(t^n) \Leftrightarrow \lim_{t \rightarrow 0}\frac{t^n}{q(t)} = 1$ \\
and & $ \displaystyle q(t)=o(t^n) \Leftrightarrow \lim_{t \rightarrow 0}\frac{t^n}{q(t)} = 0$ \\
\end{tabular}

From \ref{eqn:gbm1}, \ref{eqn:gbm2} and \ref{eqn:gbm3}, we deduce 

\begin{eqnarray}\label{eqn:gbm4}
d LogS_t & = & \frac{dS_t}{S_t} - \frac{1}{2}\frac{(dS_t)^2}{S_t^2} \nonumber \\
& = & \mu dt + \sigma dW_t - \frac{1}{2}\sigma^2dt + o(dt) \nonumber \\
& = & (\mu - \frac{\sigma^2}{2})dt + \sigma dW_t + o(dt) 
\end{eqnarray}

From \ref{eqn:gbm4}, we see that $d LogS_t$ follows a simple Brownian motion. If $r_t=Log\frac{S_t}{S_0}$, then $r_t$ is governed by the following equation:
\begin{equation}\label{eqn:gbm}
dr_t = (\mu - \frac{\sigma^2}{2}) dt + \sigma dW_t
\end{equation}

Log returns $r_t$ follow a simple Brownian motion, and are then normally distributed. Indeed, \ref{eqn:gbm} admits the solution
\begin{equation}\label{eqn:gbm5}
r_t = (\mu - \frac{\sigma^2}{2}) t + \sigma W_t
\end{equation}

Or formulated differently
\begin{equation}\label{eqn:gbm6}
S_t = S_0 e^{(\mu - \frac{\sigma^2}{2}) t + \sigma W_t}
\end{equation}

This model is known as the \emph{Bachelier-Osborne model}, and predicts a log-normal distribution for the price $S_t$.

\section{Departure from normality}

According to the GBM model, stock prices should be log-normally distributed, i.e. stock log-returns $r_t$ should be normally distributed:

$$
r_t = (\mu - \frac{\sigma^2}{2}) t + \sigma W_t
$$

Nevertheless, it is now well known that log returns exhibit two specific kinds of departure from a Gaussian: kurtosis and fat tails. A high kurtosis means that the model distribution is more peaked than a Gaussian around the mean. Fat tails mean that crashes and huge increases appear far more often than predicted by the normal law. Let us consider the probability density function (PDF) of log returns for the different datasets we have:
\begin{itemize}
\item DJIA1982: Dow Jones Industrial Average from January 04, 1982 to December 31, 2001
\item DJIA1988: Dow Jones Industrial Average from January 04, 1988 to December 31, 2001
\item DJIA1930: Dow Jones Industrial Average from January 02, 1930 to December 31, 2001
\item DJIA1896: Dow Jones Industrial Average from May 26, 1896 to December 31, 2001
\item SP1965: Standard and Poor's 500 from January 04, 1965 to December 31, 2001
\item FTSE1984: FTSE100 from January 04, 1984 to December 31, 2001
\end{itemize}

We will perform a few tests to exhibit more precisely the kurtosis and fat tails. These tests will be reproduced later on Dragulescu's model.
Before we perform our tests, let us have a look at a few examples of PDFs: we plot the PDFs for different time lags ($\tau=1, 5, 20 \ \mbox{and} \ 250$ trading days) against a Normal distribution based on the sample mean and variance.

\begin{figure}
\begin{center}
\begin{tabular}{cc}
\includegraphics[width=5cm]{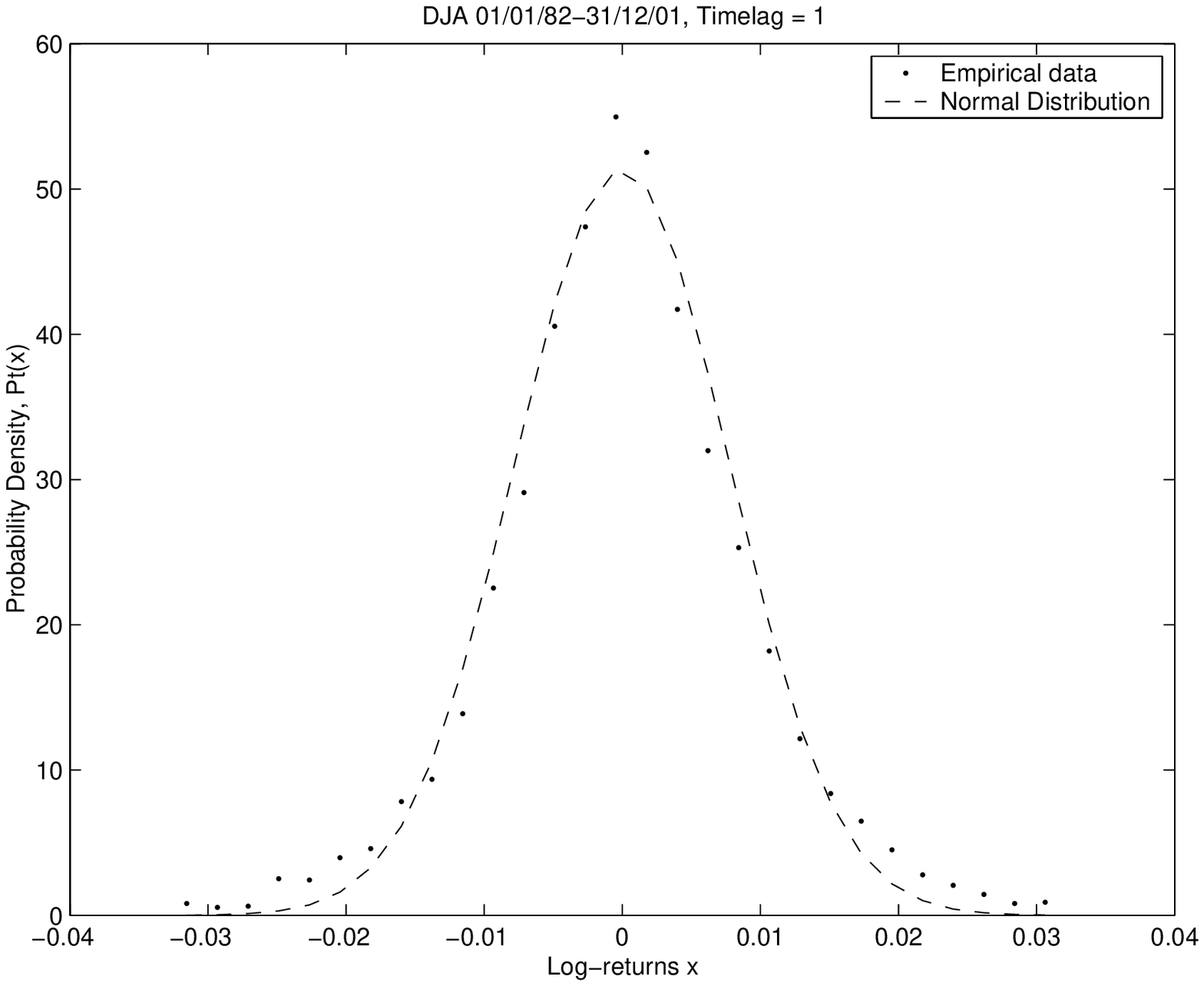} & \includegraphics[width=5cm]{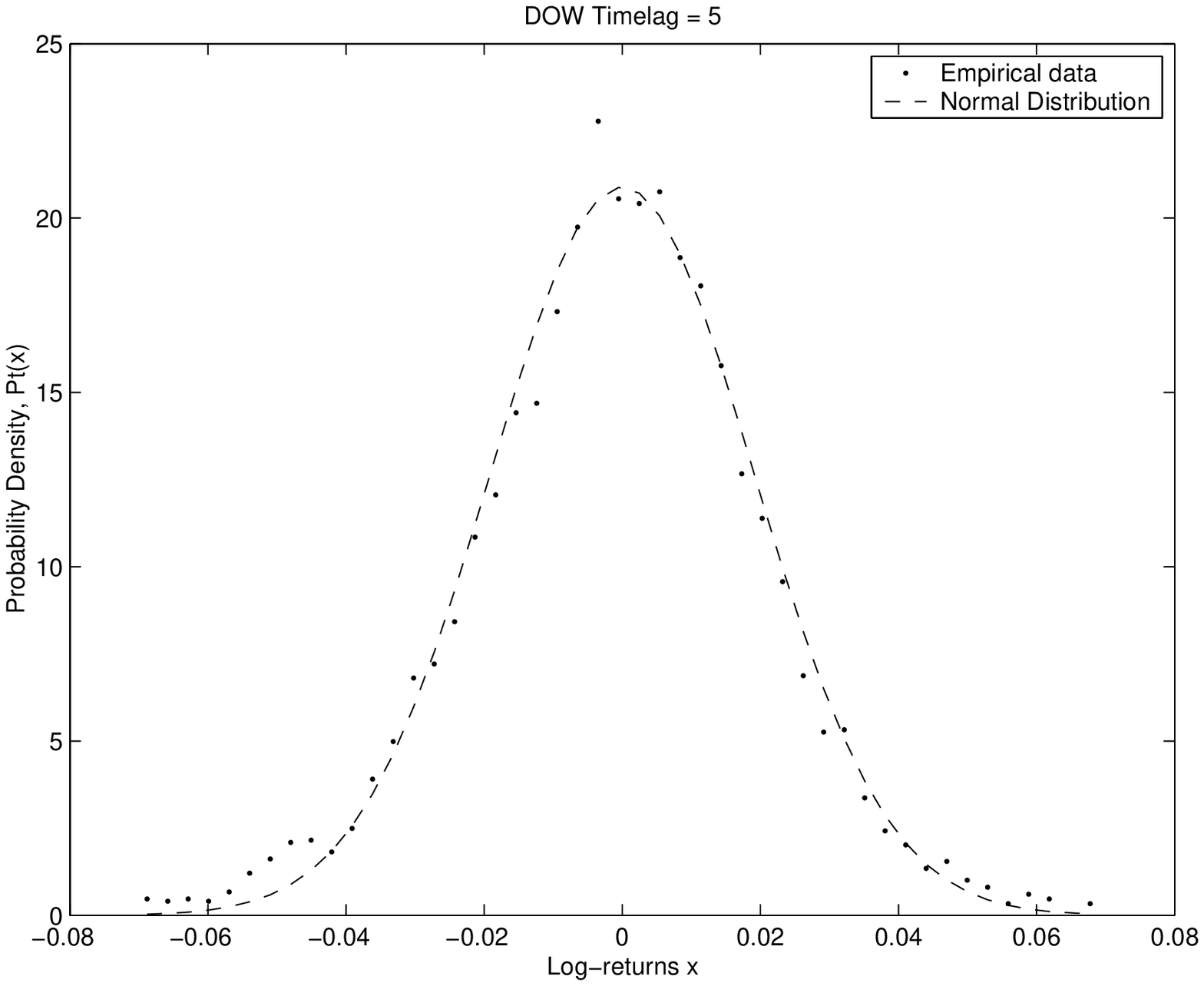} \\
\includegraphics[width=5cm]{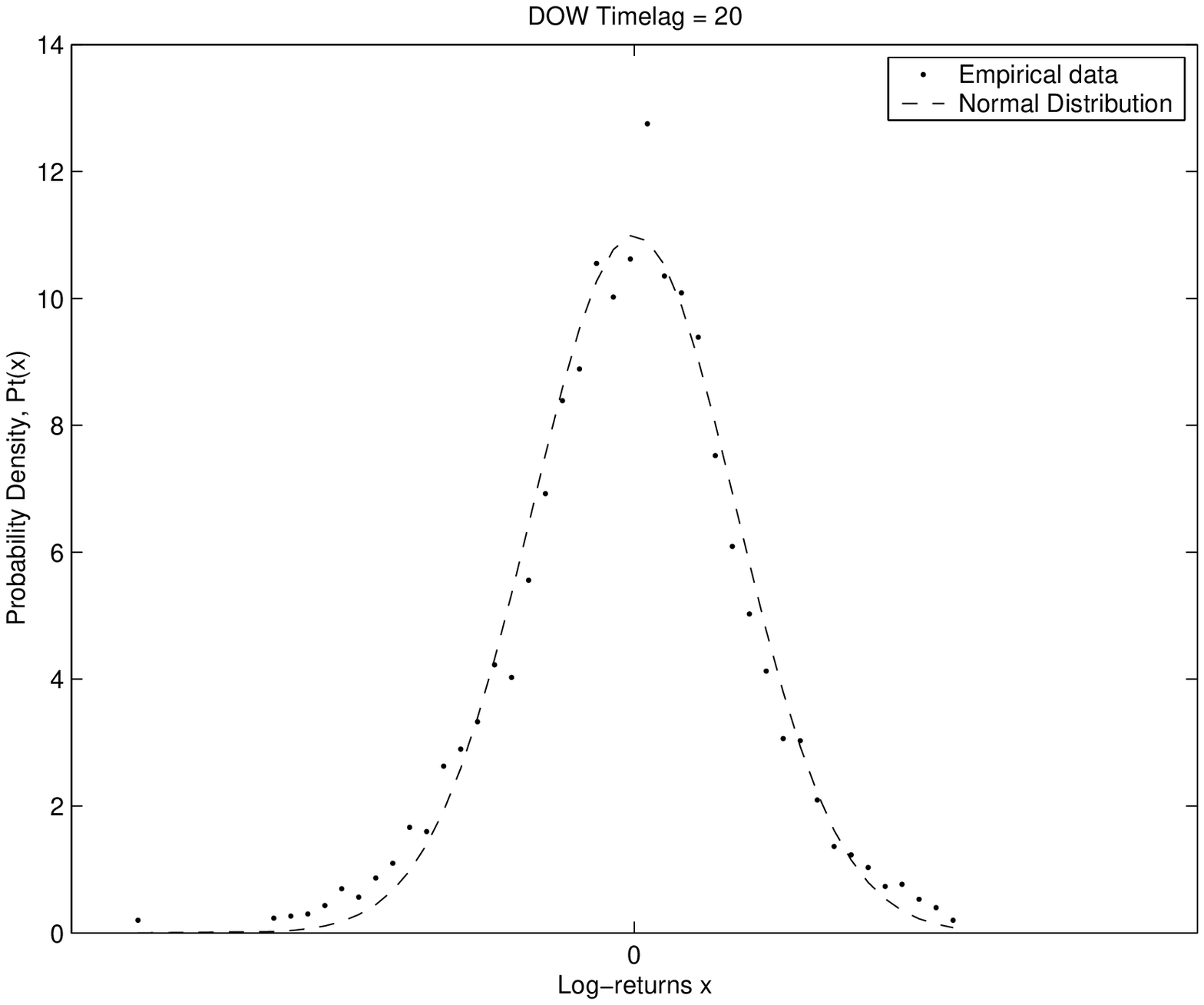} & \includegraphics[width=5cm]{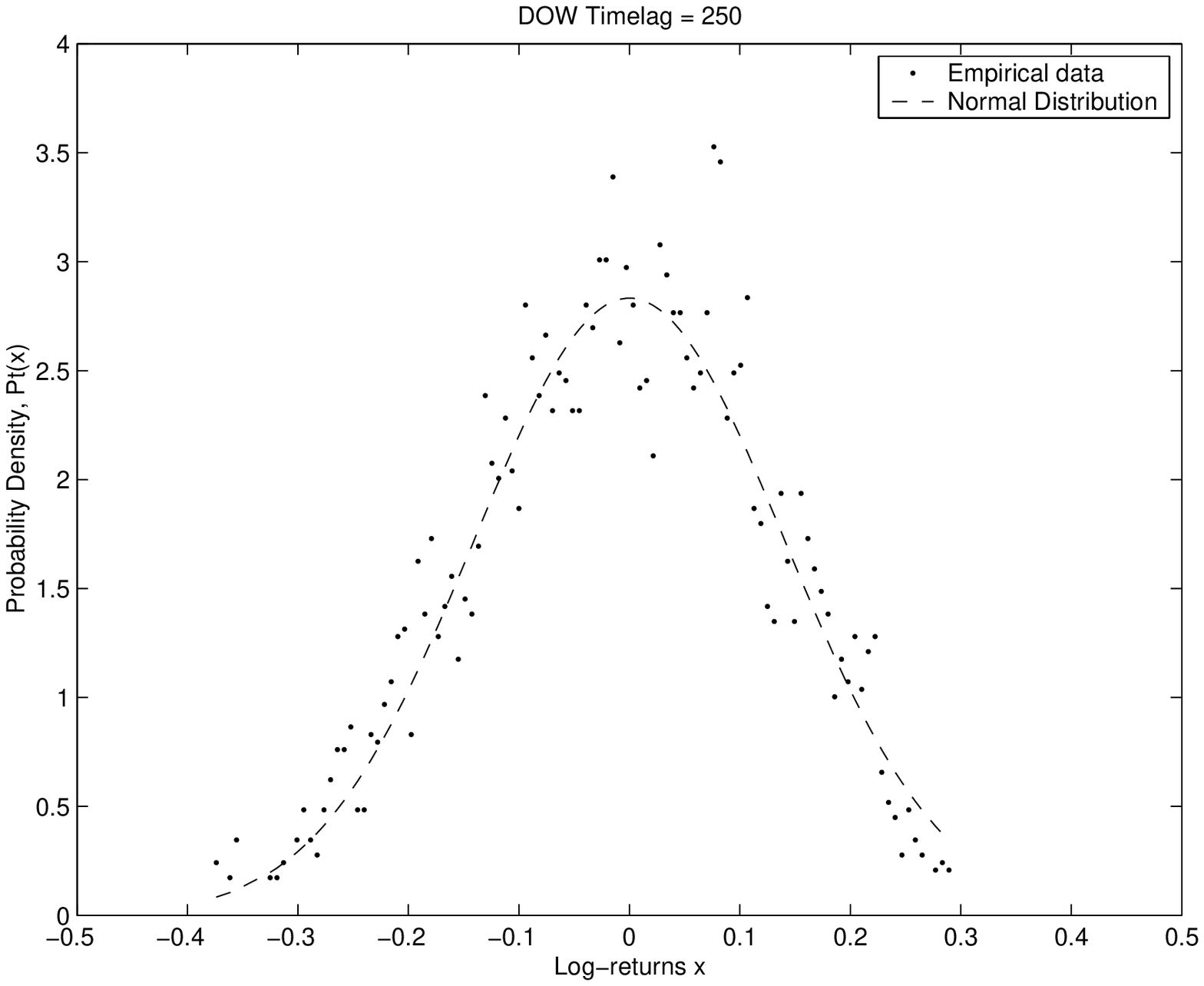} \\
\end{tabular}
\end{center}
\caption{First look at the fat tails}
\label{fig:fig-fatTails}  
\end{figure}

The issue is whether the Normal distribution fits empirical data sufficiently well, or whether the model should be rejected.

\subsection{Measure of kurtosis: Jarque-Bera Test}

If log returns really follow a Normal distribution, then we are expecting a null value for the Fisher kurtosis.\footnote{See Appendix B, Section ``Kurtosis"} Kurtosis is a measure of whether the data are peaked or flat relative to a normal distribution. That is, data sets with high kurtosis tend to have a distinct peak near the mean, decline rather rapidly, and have heavy tails.  We compute the kurtosis for each dataset, and for different time lags. As explained above, we can give an estimate of the standard deviation (indicated into parenthesis) of the kurtosis for time lags superior to 1, because we compute the kurtosis on many different datasets, or ``paths''. Are results are presented in Tables \ref{tab:kurtosis1}, \ref{tab:kurtosis2}, and \ref{tab:kurtosis3}.

\begin{table}[htbp]
\begin{center}
\begin{tabular}{ccc}
time lag & DJIA1982 & DJIA1988 \\
\begin{tabular}{r}
	 1 \\
	 5 \\
	 20 \\
	 40 \\
	 80 \\
	 100 \\
	 200 \\
	 250 \\
	\end{tabular} &
\begin{tabular}{r@{}l}
	69&.27  \\
	16&.87 $\pm$16.2 \\
	 7&.75 $\pm$4.77 \\
	 5&.69 $\pm$2.40 \\
	 2&.75 $\pm$1.21 \\
	 1&.68 $\pm$0.99 \\
	-0&.06 $\pm$0.57 \\
	-0&.52 $\pm$0.37 \\
	\end{tabular} &
\begin{tabular}{r@{}l}
	 5&.963  \\
	 3&.07 $\pm$0.91 \\
	 1&.43 $\pm$0.93 \\
	 1&.18 $\pm$0.74 \\
	 0&.86 $\pm$1.45 \\
	 0&.25 $\pm$0.88 \\
	-0&.59 $\pm$0.37 \\
	-0&.55 $\pm$0.58 \\
	\end{tabular} \\
\end{tabular}
\end{center}
\caption{Measure of kurtosis for DJIA1982 \& DJIA1988}
\label{tab:kurtosis1}
\end{table}

\begin{table}[htbp]
\begin{center}
\begin{tabular}{ccc}
time lag & DJIA1896 & DJIA 1930 \\
\begin{tabular}{r}
	 1 \\
	 5 \\
	 20 \\
	 40 \\
	 80 \\
	 100 \\
	 200 \\
	 250 \\
	\end{tabular} &
\begin{tabular}{r@{}l}
	26&.81            \\
	12&.55 $\pm$2.36  \\
	 8&.26 $\pm$1.60  \\
	 5&.85 $\pm$1.29  \\
	 3&.51 $\pm$1.30  \\
	 2&.65 $\pm$0.90  \\
	 3&.16 $\pm$2.49  \\
	 2&.65 $\pm$2.14  \\
	\end{tabular} &
\begin{tabular}{r@{}l}
	27&.38		 \\
	 9&.60 $\pm$3.82 \\
	 8&.40 $\pm$1.83 \\
	 7&.44 $\pm$2.05 \\ 
	 5&.43 $\pm$2.25 \\
	 4&.29 $\pm$1.68 \\
	 4&.99 $\pm$3.09 \\
	 4&.39 $\pm$2.77 \\
	\end{tabular} \\
\end{tabular}
\end{center}
\caption{Measure of kurtosis for DJIA1896 \& DJIA 1930}
\label{tab:kurtosis2}
\end{table}

\begin{table}[htbp]
\begin{center}
\begin{tabular}{ccc}
time lag & SP1965 & FTSE1984 \\ 
\begin{tabular}{r}
	 1 \\
	 5 \\
	 20 \\
	 40 \\
	 80 \\
	 100 \\
	 200 \\
	 250 \\
	\end{tabular} &
\begin{tabular}{r@{}l}
	42&.08           \\
	 9&.26 $\pm$8.46 \\
	 3&.90 $\pm$2.28 \\
	 2&.95 $\pm$1.26 \\
	 1&.18 $\pm$0.85 \\
	 0&.67 $\pm$0.62 \\
	 0&.07 $\pm$0.52 \\
	 0&.01 $\pm$0.70 \\
	\end{tabular} &
\begin{tabular}{r@{}l}
	12&.72 	  	 \\
	12&.34 $\pm$5.46 \\
	10&.67 $\pm$5.99 \\
	 5&.48 $\pm$2.44 \\
	 2&.70 $\pm$2.08 \\
	 2&.10 $\pm$1.70 \\
	 0&.14 $\pm$0.86 \\
	-0&.04 $\pm$0.74 \\
	\end{tabular} \\
\end{tabular}
\end{center}
\caption{Measure of kurtosis for SP1965 \& FTSE1984}
\label{tab:kurtosis3}
\end{table}

We can clearly see that for every dataset, empirical log returns exhibit high kurtosis for high frequencies (time lag = 1 and 5 days) and medium frequencies (time lag = 20, 40 and 80 days), and small (for the largest datasets only, DJIA1896 and DJIA1930) or no kurtosis for low frequencies (time lag = 200 and 250 days). But the standard deviation is very high compared with the mean value, which means that some paths exhibit very high kurtosis whereas others exhibit very little. For instance, let us have a look at the five paths of the 5 days time lag returns of the DJIA1982. Results are shown in Table \ref{tab:kurt1982}. 

\begin{table}[htbp]
\begin{center}
\begin{tabular}{cc}
path & kurtosis \\
\begin{tabular}{r}
	1 \\
	2 \\
	3 \\
	4 \\
	5 \\
	\end{tabular} &
\begin{tabular}{r@{}l}
	38&.20 \\ 
	30&.49 \\ 
	 5&.91 \\
	 5&.77 \\
	 3&.99 \\
	\end{tabular} \\
\end{tabular}
\end{center}
\caption{Example of kurtosis for each path, on DJIA1982, with $\tau=5$}
\label{tab:kurt1982}
\end{table}

The kurtosis goes from 3.99 to 38.20, with a mean and standard deviation of 16.87 and 16.20 respectively. Only two paths out of five exhibit very high kurtosis, which is enough to have a high mean, but the large standard deviation must remind us of the important heterogeneity of the different paths. 

On average, the probability mass of empirical log returns is leptokurtic\footnote{See appendix B, Section ``Measures of kurtosis''} for high frequencies. This departure from normality should enable us to reject the normal hypothesis.

To verify, we perform a Jarque-Bera test, which tests the goodness-of-fit to a normal distribution, according to the skewness and kurtosis.\footnote{See Appendix B, Section ``Jarque-Bera Goodness-of-Fit Test"} It tests a composite hypothesis, which means that the parameters of the tested distribution, viz. the mean and variance of the normal distribution,  can be derived from the empirical data, and do not need to be known in advance. For each path, the output of the test is 0 if we do not reject the null hypothesis (viz the normal hypothesis) at a significance level $\alpha=0.05$, and 1 if we reject it. We give in Table \ref{tab:jarqueberaTest} the average of the tests. For instance, for a time lag of 80 days, we perform the Jarque-Bera test 80 times, on 80 different log returns datasets. An average value of 0.9 means that the test rejected the null hypothesis 90\% of the time, i.e. 72 out of 80 datasets.

\begin{table}[h]
\begin{center}
\begin{tabular}{rllllll}
time lag  & DJIA1988 & DJIA1982 & DJIA1930 & DJIA 1892 & SP1965 & FTSE1984 \\ 
1         & 1        & 1        & 1        & 1         & 1      & 1        \\
5         & 1        & 1        & 1        & 1         & 1      & 1        \\
20        & 0.65     & 1        & 1        & 1         & 1      & 1        \\
40        & 0.625    & 1        & 1        & 1         & 1      & 0.975    \\
80        & 0.2375   & 0.9      & 1        & 1         & 0.5375 & 0.8      \\
100       & 0.08     & 0.6      & 1        & 1         & 0.23   & 0.54     \\
200       & 0        & 0        & 1        & 0.97      & 0.01   & 0        \\
250       & 0        & 0        & 1        & 0.964     & 0.064  & 0        \\
\end{tabular}
\end{center}
\caption{Proportion of paths for which the Jarque-Bera Goodness-of-Fit Test rejects $H_0$}
\label{tab:jarqueberaTest}
\end{table}

For each dataset, for high frequencies, the normal hypothesis is systematically rejected. For low frequencies, except for the largest datasets (DJIA1896 and DJIA1930), which exhibit small kurtosis, the normal distribution cannot be rejected. The conclusion is not straightforward for middle frequencies. 

\subsection{Fat tails: Normal Probability Plot and Lilliefors Test}

\begin{figure}
\begin{center}
\begin{tabular}{cc}
\includegraphics[width=5cm]{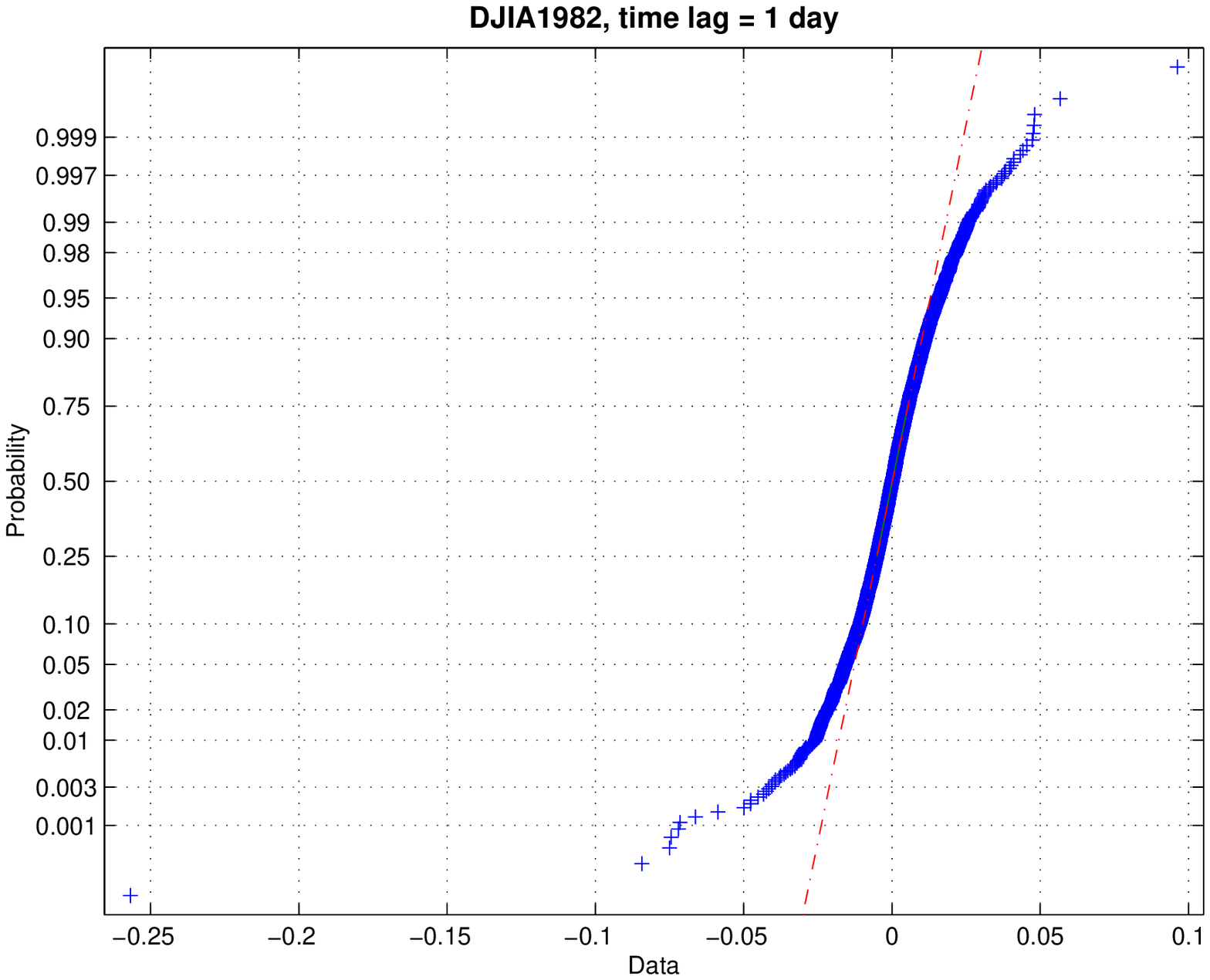} & \includegraphics[width=5cm]{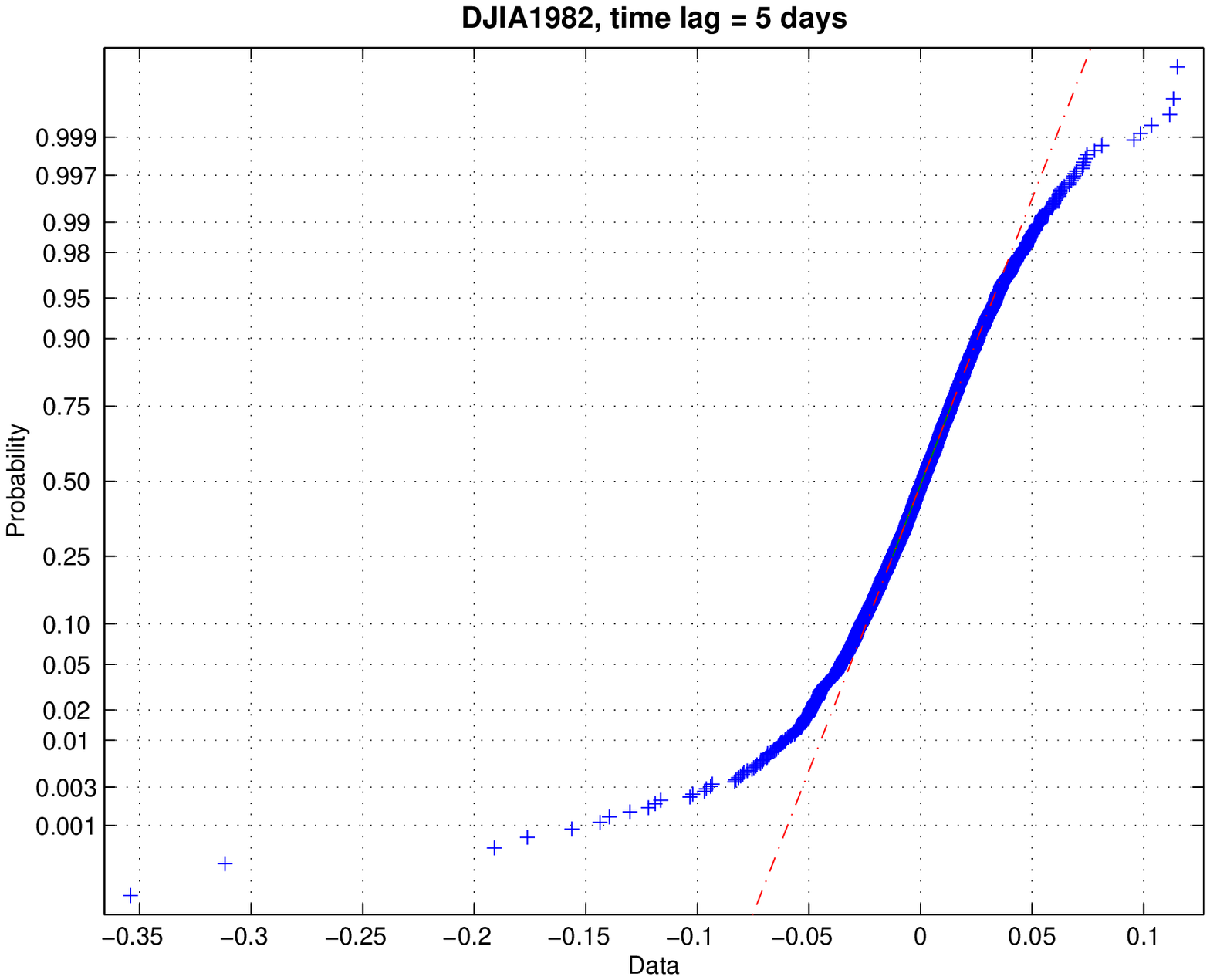} \\
\includegraphics[width=5cm]{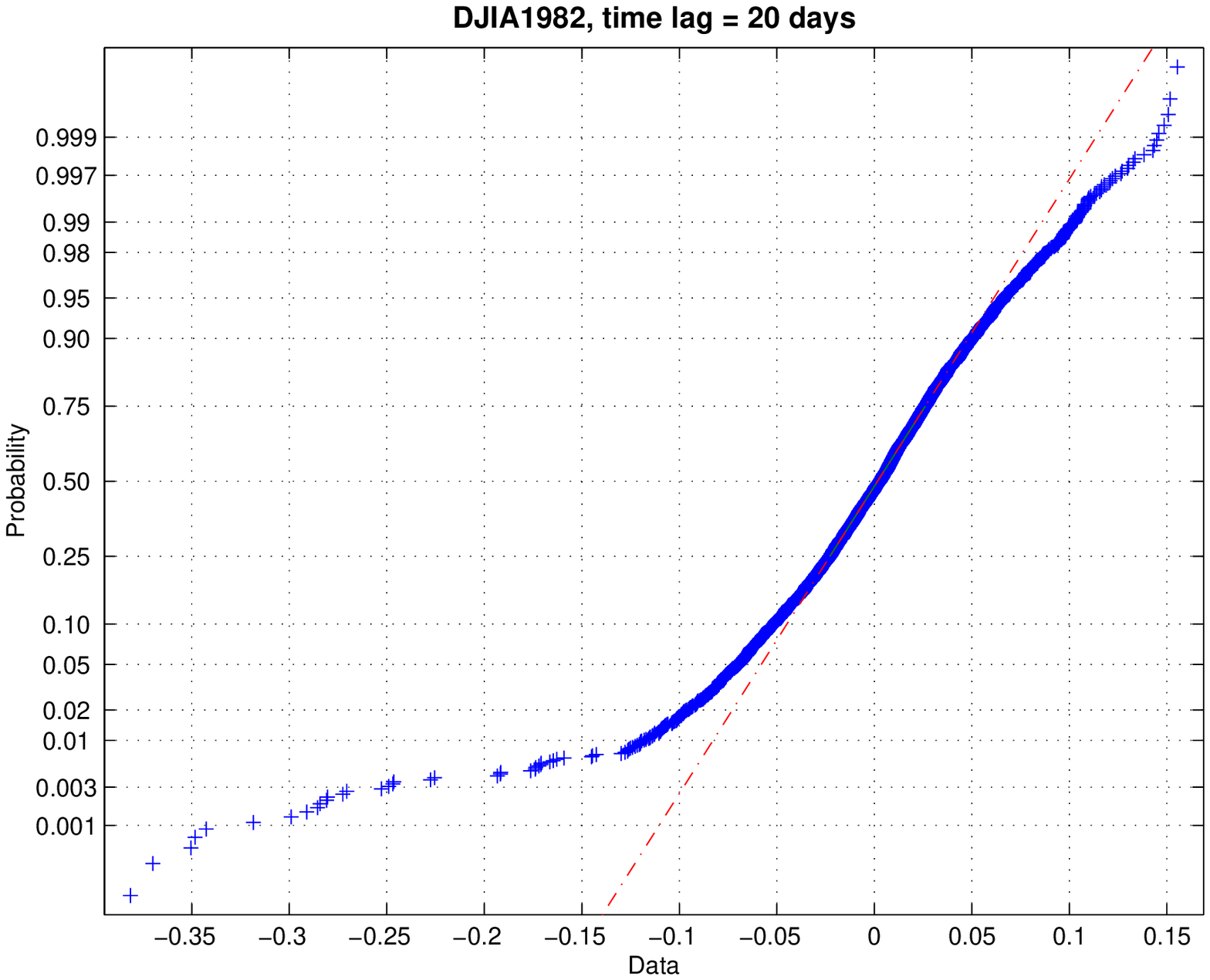} & \includegraphics[width=5cm]{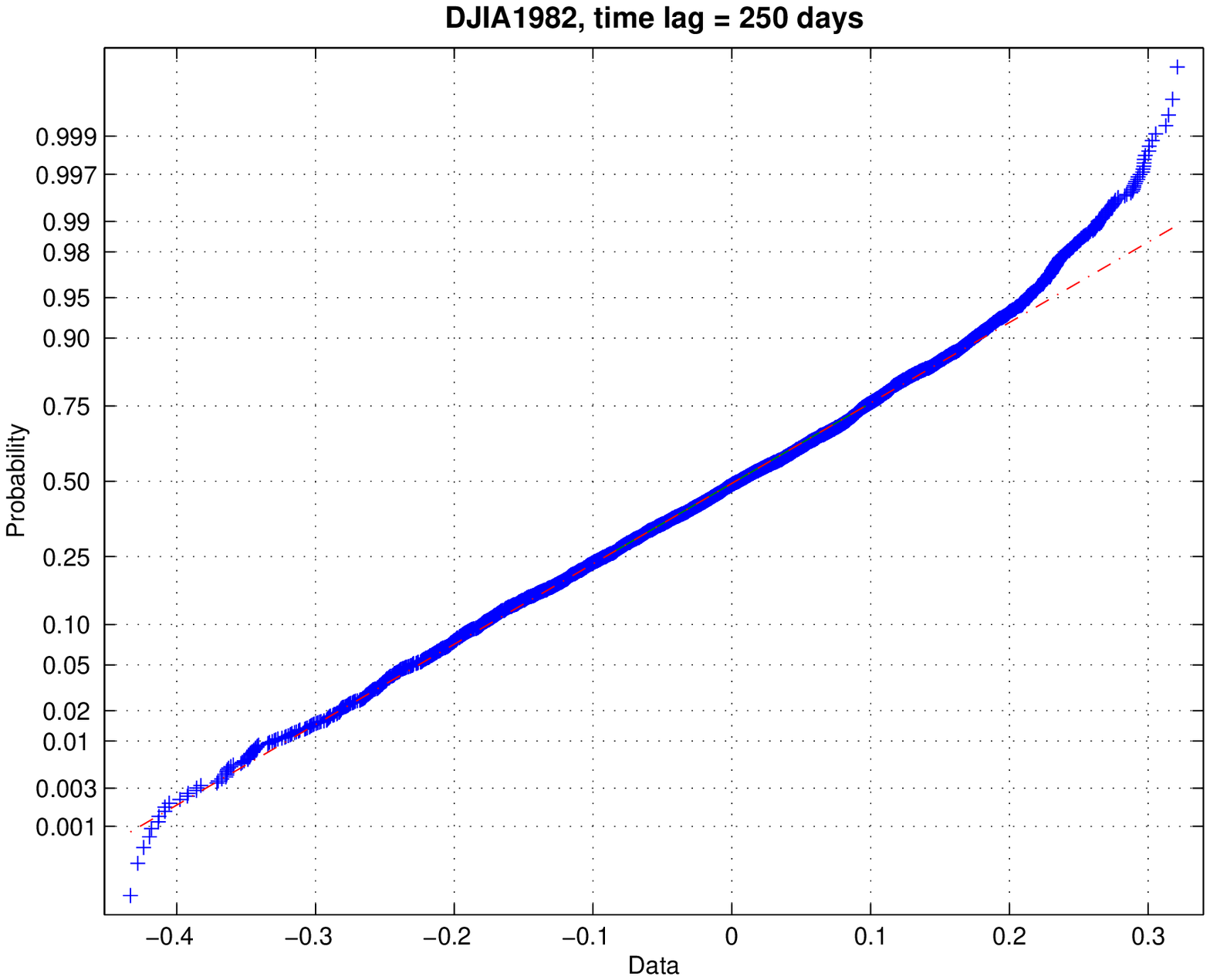} \\
\end{tabular}
\end{center}
\caption{Normal Probability Plot for $\tau=1,5,20 \  \mbox{and} \ 250$ days}
\label{fig:fig-normalProbPlot}  
\end{figure}

The other important departure from Normality consist of fat tails, that could be exhibited by performing a probability plot. This time, we do not perform a test on each dataset and each time lag, since a few examples should be enough. We select the first dataset, the Dow Jones Industrial Average, from January 04, 1982, to December 31, 2001 and draw the Normal Probability Plot on the log returns. The results are presented in Figure \ref{fig:fig-normalProbPlot}.

We can make the following conclusions from the above plot:
\begin{enumerate}
\item The normal probability plot shows a non-linear pattern;
\item The normal distribution is not a good model for these data.
\end{enumerate}

For data with short (less variance than expected in a normal distribution) or long (more variance than expected in a normal distribution) tails relative to the normal distribution, the non-linearity of the normal probability plot shows up in two ways. First, the middle of the data shows an S-like pattern. This is common for both short and long tails. Second, the first few and the last few points show a marked departure from the reference fitted line. For short tails, the first few points show increasing departure from the fitted line above the line and last few points show increasing departure from the fitted line below the line ($\tau=1,5 \  \mbox{and} \ 20$ days). For long tails, this pattern is reversed ($\tau=250$ days).

In this case, we can reasonably conclude that the normal distribution does not provide an adequate fit for this dataset, for high frequencies. To confirm this, we perform a Lilliefors Goodness-of-Fit Test.

Lilliefors tests the goodness of fit to a normal distribution. It is derivated from the Kolmogorov-Smirnof test, with the difference that it tests a composite hypothesis and not a simple hypothesis.\footnote{See Appendix B, Section ``Lilliefors Goodness-of-Fit Test"} The difference with the Jarque-Bera test is that this one is based on the maximum departure of the empirical distribution from the normal distribution, so this test will tend to reject the null hypothesis in the presence of kurtosis \emph{and} fat tails. We perform this test for each index and each dataset. For each path, the output of the test is 0 if we do not reject the null hypothesis (viz the normal hypothesis) at a significance level $\alpha=0.05$, and 1 if we reject it. We give in Table \ref{tab:lillieTest} the average of the tests. For instance, for a time lag of 80 days, we perform the Lilliefors test 80 times, on 80 different log returns datasets. An average value of 0.1875 means that we rejected the null hypothesis $0.1875*80=15$ times out of 80.   

\begin{table}[htbp]
\begin{center}
\begin{tabular}{rllllll}
time lag& DJIA1988& DJIA1982& DJIA1930 	& DJIA1896& SP1965 & FTSE1984   \\ 
1 	& 1 	  & 1 	    & 1 	& 1 	  & 1 	   & 1 		\\
5 	& 1 	  & 1 	    & 1 	& 1 	  & 1 	   & 1 		\\
20 	& 0.4 	  & 0.85    & 1 	& 1 	  & 0.8    & 0.65 	\\
40 	& 0.075   & 0.725   & 1 	& 1 	  & 0.675  & 0.425 	\\
80 	& 0.025   & 0.1875  & 1 	& 1 	  & 0.0625 & 0.1625 	\\
100 	& 0 	  & 0.06    & 1 	& 1 	  & 0.03   & 0.08 	\\
200 	& 0.035   & 0.04    & 0.61 	& 0.41 	  & 0.1    & 0.035 	\\
250 	& 0.02 	  & 0.024   & 0.532 	& 0.376   & 0.064  & 0.112 	\\
\end{tabular}
\end{center}
\caption{Proportion of paths for which the Lilliefors Goodness-of-Fit Test rejects $H_0$}
\label{tab:lillieTest}
\end{table}

The Lilliefors test rejects the normal hypothesis for high frequencies, but not for low frequencies. Again, for large datasets, the normal hypothesis is more often rejected, even for low frequencies. We believe rejection comes from the fact that kurtosis and fat tails are due to outliers, events that are expected to happen once in a century by the Bachelier-Osborne model, but that occur far more often. Even if these events happen more often, they are still rare enough to be absent from some too small datasets, specially for low frequencies where the number of points in each dataset is very low. This issue is not investigated in this thesis, and remains to be resolved.

\section{Conclusion}

We have described in this chapter the theory underlying most of statistical models of the stock market, the Random Walk Theory. The first model to use this theory was Bachelier-Osborne model (1959), that predicts a normal distribution for log returns. Even though this model remains widely used, specially by Black and Scholes in their famous model for option pricing, the empirical data show a clear departure from normality for high frequencies ($\tau=1$ and $5$ days): the observed distribution is leptokurtic and exhibits fat tails. For low frequencies ($\tau=200$ and $250$ days), the normal hypothesis cannot be rejected. The conclusion is not straightforward for medium frequencies ($\tau=20,\ 40,\ 80$ and $100$ days).

Since 1959, some attempts have been done to produce a better model for log returns (stable Pareto-Levy distributions \cite{Fama}, exponentially truncated power law, etc.), a model that would particularly fit the kurtosis and fat tails of the empirical distribution. But so far, all of them suffered from strong criticisms. We investigate in next chapter a recent model, proposed by Dragulescu et al. in 2002, based on a stochastic mean-reverting process for the volatility.

\chapter{Dragulescu's model}

\section{Introduction}

Mainly because of kurtosis and fat tails, we have to figure out better models for stock market returns than the Gaussian. Above all, the normal distribution fails to describe the most important phenomena: draw downs and bubbles, that occur far more often than expected, as shown by the fat tails.

Hence, we have to review some hypothesis that result in the classic Gaussian model, in order to improve upon it. Many models exist that try to explain or produce fat tails. The main assumptions of the Gaussian model concern (1) the independence of log-returns and (2) the finite constant volatility. If we assume that log-returns are really independent and identically distributed, then only a non constant or non finite volatility could explain the departure from normality observed in the empirical data.

In the Gaussian hypothesis, the instantaneous volatility takes the form of $\sigma S_t$, where $\sigma$ is a finite constant and $S_t$ is the security price at time \emph{t}. Some models, e.g. stable Paretian distributions (\cite{Fama}), consider the volatility as infinite and produce fat tails as in empirical data. Nevertheless, the assumption of infinite volatility does not appear to be relevant, since the volatility does not grow indefinitely with the sample size. Another innovation is to consider a finite stochastic volatility. This class of models has been introduced by Hull and White. We are studying here one of these models, proposed by A. Dragulescu and A. Yakovenko. We will reproduce their results following their methodology, make a few critical comments on the way they trim the data, and propose a methodology to test their model against the empirical distribution of log returns.   

\section{Mean-reverting stochastic volatility}

This model starts from a geometric Brownian motion stochastic differential equation for the price $S_t$
\begin{equation} \label{eqn:stochPrice}
dS_t = \mu S_t dt + \sigma_t S_t dW_t^{(1)}  
\end{equation}
where $W_t^{(1)}$ is a standard Wiener

Log returns $r_t=Log \frac{S_t}{S_0}$ and centred log returns $x_t = r_t - \mu t$ are introduced. From \ref{eqn:gbm}, we get
\begin{equation} \label{eqn:stochLogReturns1}
dr_t = (\mu-\frac{v_t}{2}) dt + \sqrt{v_t} dW_t^{(1)} \ \ \ \mbox{since} \ \sigma_t=\sqrt{v_t}
\end{equation}
and
\begin{equation} \label{eqn:stochLogReturns}
dx_t = -\frac{v_t}{2} dt + \sqrt{v_t} dW_t^{(1)}  
\end{equation}

Then instead of having a constant volatility $\sigma_t=\sigma$ as in the Bachelier-Osborne model, the authors assume the variance $v_t=\sigma_t^2$ obeys the following mean-reverting stochastic differential equation
\begin{equation} \label{eqn:stochVar}
dv_t = -\gamma (v_t - \theta) dt + k \sqrt{v_t} dW_t^{(2)} 
\end{equation}
\begin{tabular}{ll}
where & $v_t=\sigma_t^2$ \\
      & $\theta$ is the long time mean of $v_t$ \\
      & $\gamma$ is the rate of relaxation to this mean \\
      & $k$ is a constant parameter called the variance noise \\
      & $dW_t^{(2)}$ is another standard Wiener process, \\
      & not necessarily correlated with $dW_t^{(1)}$  \\
\end{tabular}

This model for the variance has been proposed first by Cox, Ingersoll and Ross \cite{Cox} in an attempt to price options, known as the CIR model.

\section{Forward Kolmogorov}

The authors solve the forward Kolmogorov (also called Fokker-Planck) equation that governs the time evolution of the joint probability $P_t(x,v|v_i)$ of having the log return $x$ and the variance $v$ for the time lag $t$, given the initial value $v_i$ of the variance

\begin{eqnarray}\label{eqn:forwardKolmogorov}
\frac{\delta}{\delta t} P & = & \gamma \frac{\delta}{\delta v} [(v-\theta)P] + \frac{1}{2} \frac{\delta}{\delta x} (vP) \nonumber \\
& = & + \rho k \frac{\delta^2}{\delta x \delta v} (vP) + \frac{1}{2} \frac{\delta^2}{\delta x^2} (vP) + \frac{k^2}{2} \frac{\delta^2}{\delta v^2} (vP)
\end{eqnarray}

They introduce a Fourier transform to solve analytically this equation, and obtain the following expression for the probability distribution of centred log-returns x for a time lag t:
\begin{equation} \label{eqn:draguPDF}
P_t(x) = \frac{1}{2 pi} \int_{-\infty}^{+\infty}{dp_x e^{ip_x+F_t(p_x)}}
\end{equation}
with

\begin{equation} \label{eqn:Ft}
F_t(p_x) = \frac{\gamma \Gamma \theta t}{k^2}
- \frac{2\gamma\theta}{k^2}ln[\frac{cosh\frac{\Omega t}{2} + \frac{\gamma}{\Omega}sinh\frac{\Omega t}{2}}{cosh\frac{\Omega t}{2} + \frac{\Gamma}{\Omega}sinh\frac{\Omega t}{2}}] 
- \frac{2 \gamma \theta}{k^2}ln[cosh\frac{\Omega t}{2} + \frac{\Omega ^2 - \Gamma ^2 + 2\gamma \Gamma}{2 \gamma \Omega}sinh\frac{\Omega t}{2}] 
\end{equation}

\begin{tabular}{ll}
where & $\Gamma = \gamma + i \rho k p_x$ \\
      & $\rho$ is the correlation coefficient between the two Wieners $W_t^{(1)}$ and $W_t^{(2)}$ \\
      & $\Omega = \sqrt{\Gamma^2 + k^2 (p_x^2 - ip_x)}$ \\
      & $\gamma$, $\theta$, $k$ and $\mu$ are the parameters of the model \\
\end{tabular}

Eqn. \ref{eqn:draguPDF} is the central result of their model. It gives, for a given time lag $t$, the expected probability density of centred log returns x. An asymptotic analysis\footnote{See \cite{Dragu} Part VI} of $P_t(x)$ shows that it predicts a Gaussian distribution for small values of $x$, and exponential, time dependent tails for large values of $|x|$.

To confront their model with observed log returns, they train the four parameters of the model, $\gamma$, $\theta$, $k$ and $\mu$, to fit the empirical index (DOW JONES from January 04, 1982 to December 31, 2002) by minimising the following square-mean deviation error 
$$
E = \sum_{x,t} |log P_t^*(x) - log P_t(x)|
$$
for all available values of log returns x, and time lag $t = 1, 5, 20, 40$ and $250$ days, \\
where $P_t^*(x)$ is the empirical probability mass\\

In their model, the authors set the correlation coefficient $\rho$ to zero, since (i) their trained parameter $\rho^trained$ is almost null ($\rho^trained \simeq 0$) and (ii) they do not observe any difference, in the fitting of empirical data, between taking $\rho^trained$ or $\rho = 0$. Hence, they reduce the complexity of their model.

Minimising the deviation of the log instead of the absolute difference $|P_t^*(x) - P_t(x)|$ forces the parameters to fit the fat tails instead of the middle of the distribution, where the probability mass is very high.

\begin{figure}
\begin{center}
\begin{tabular}{cc}
\includegraphics[width=6cm]{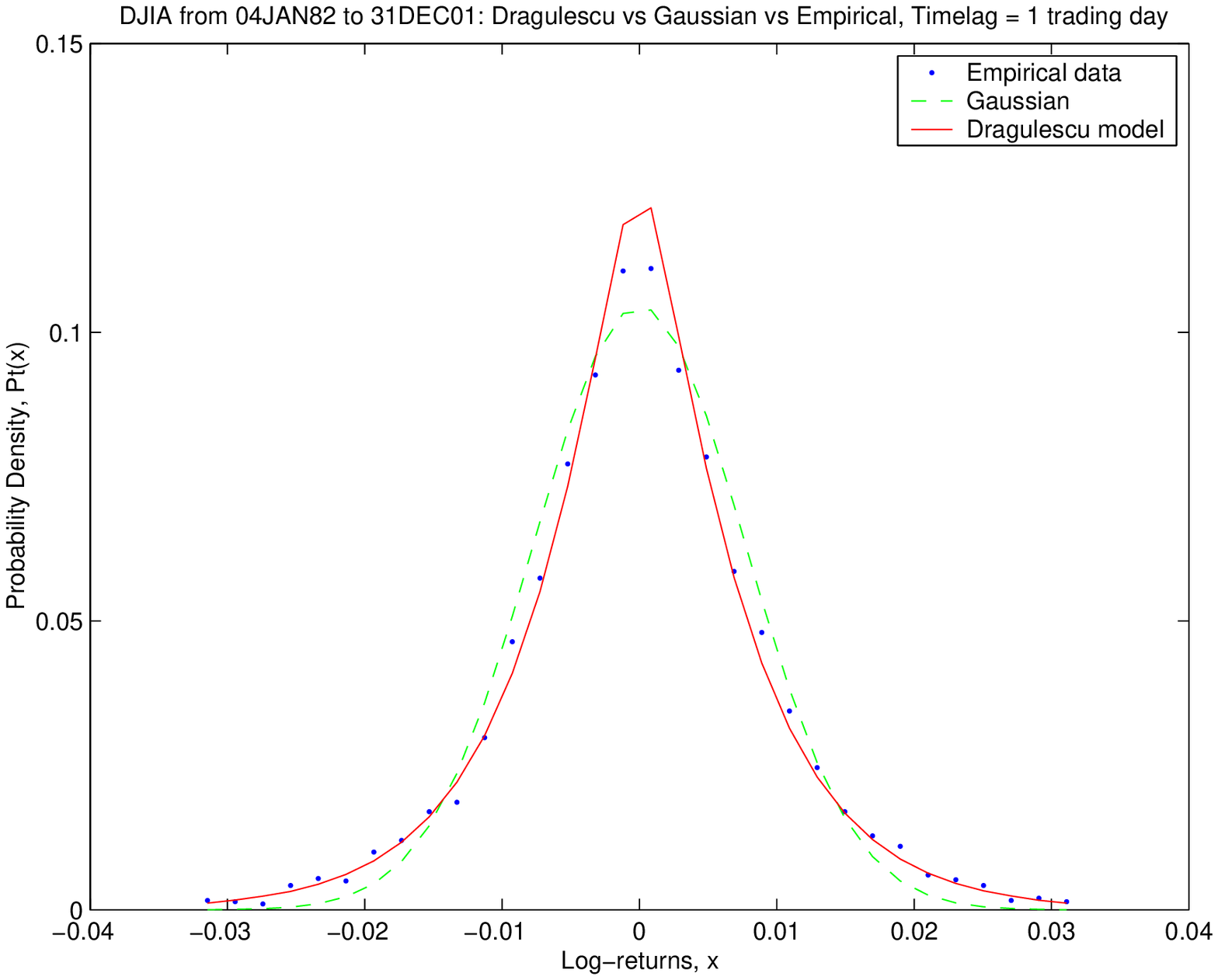} & \includegraphics[width=6cm]{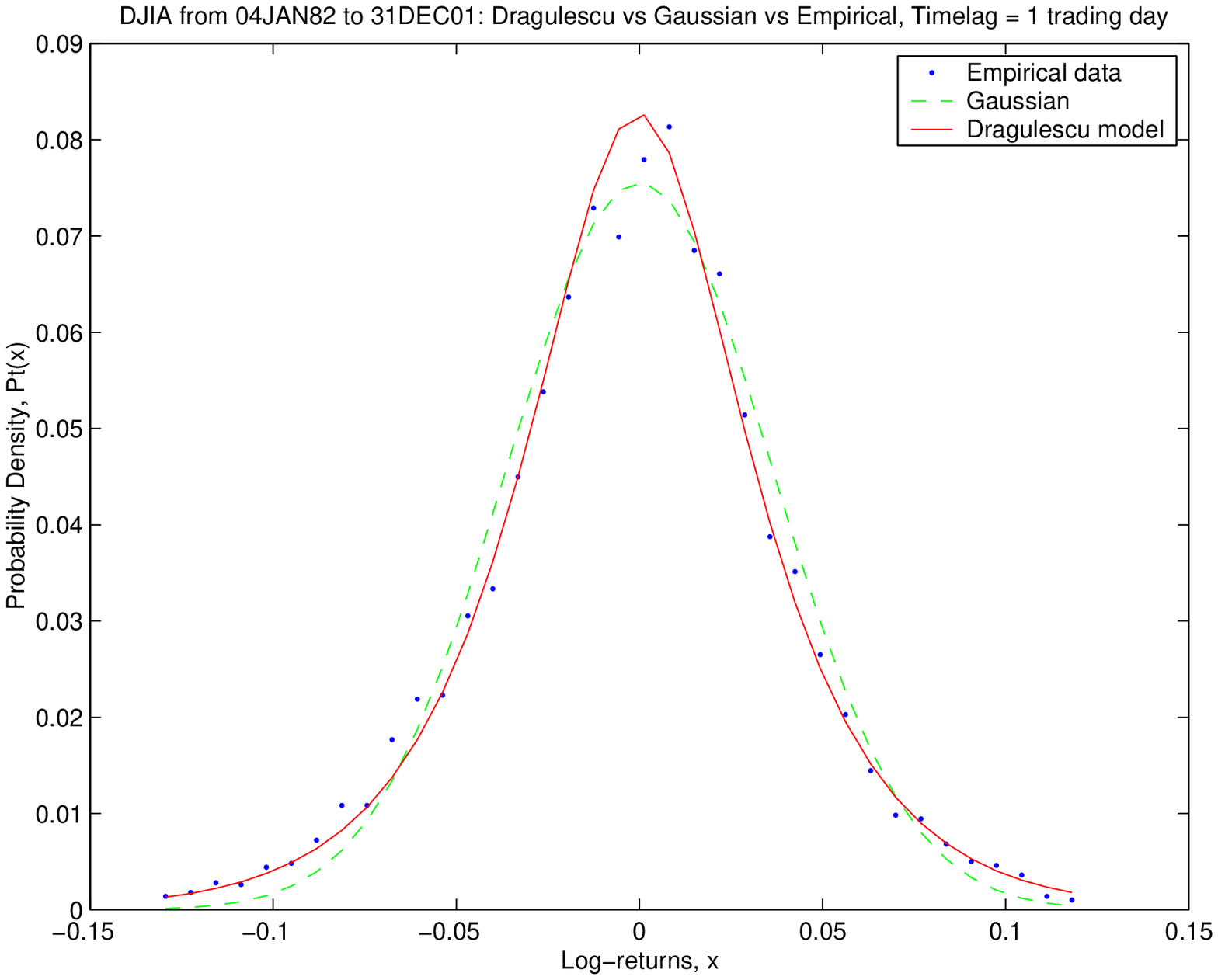} \\
\includegraphics[width=6cm]{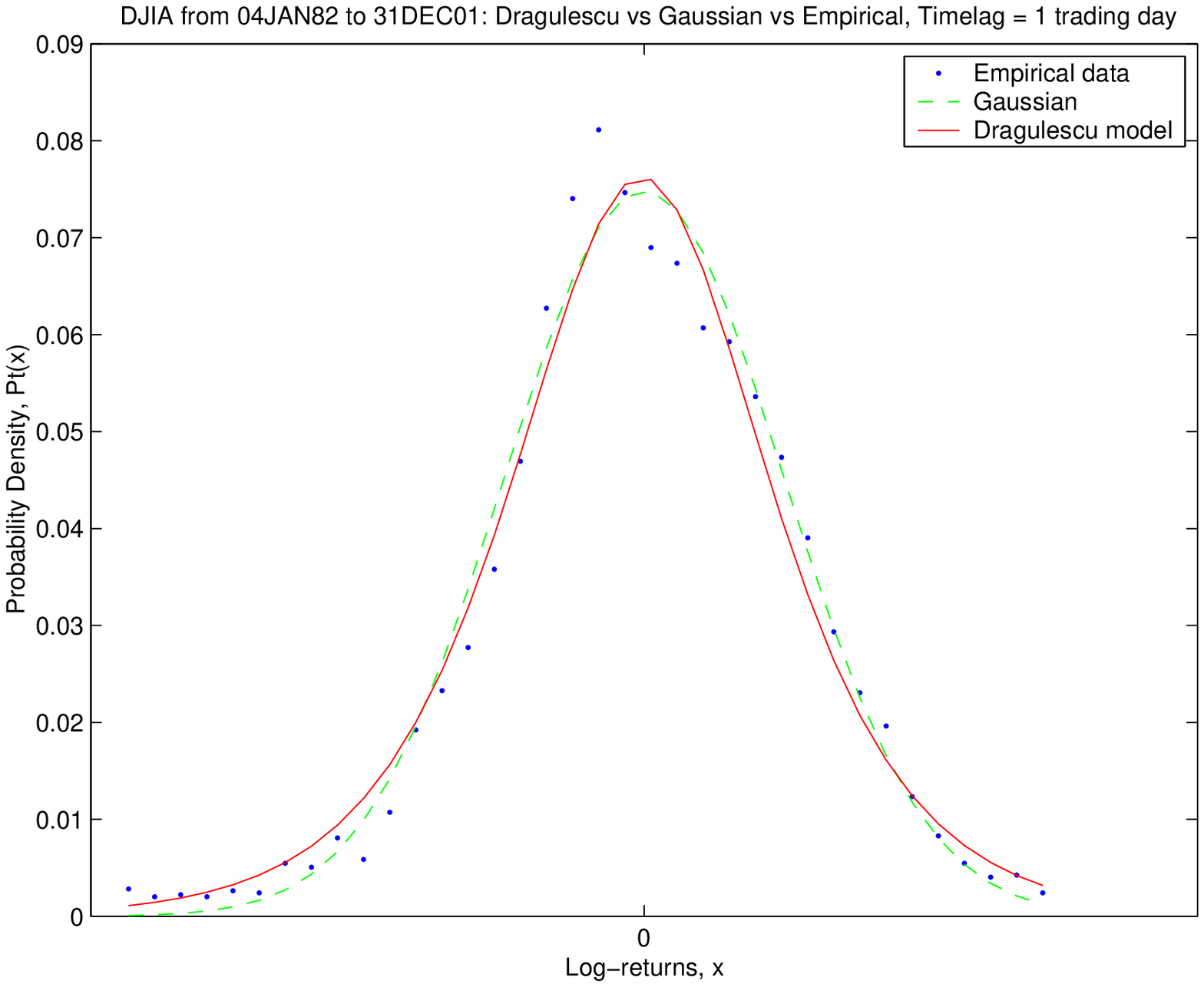} & \includegraphics[width=6cm]{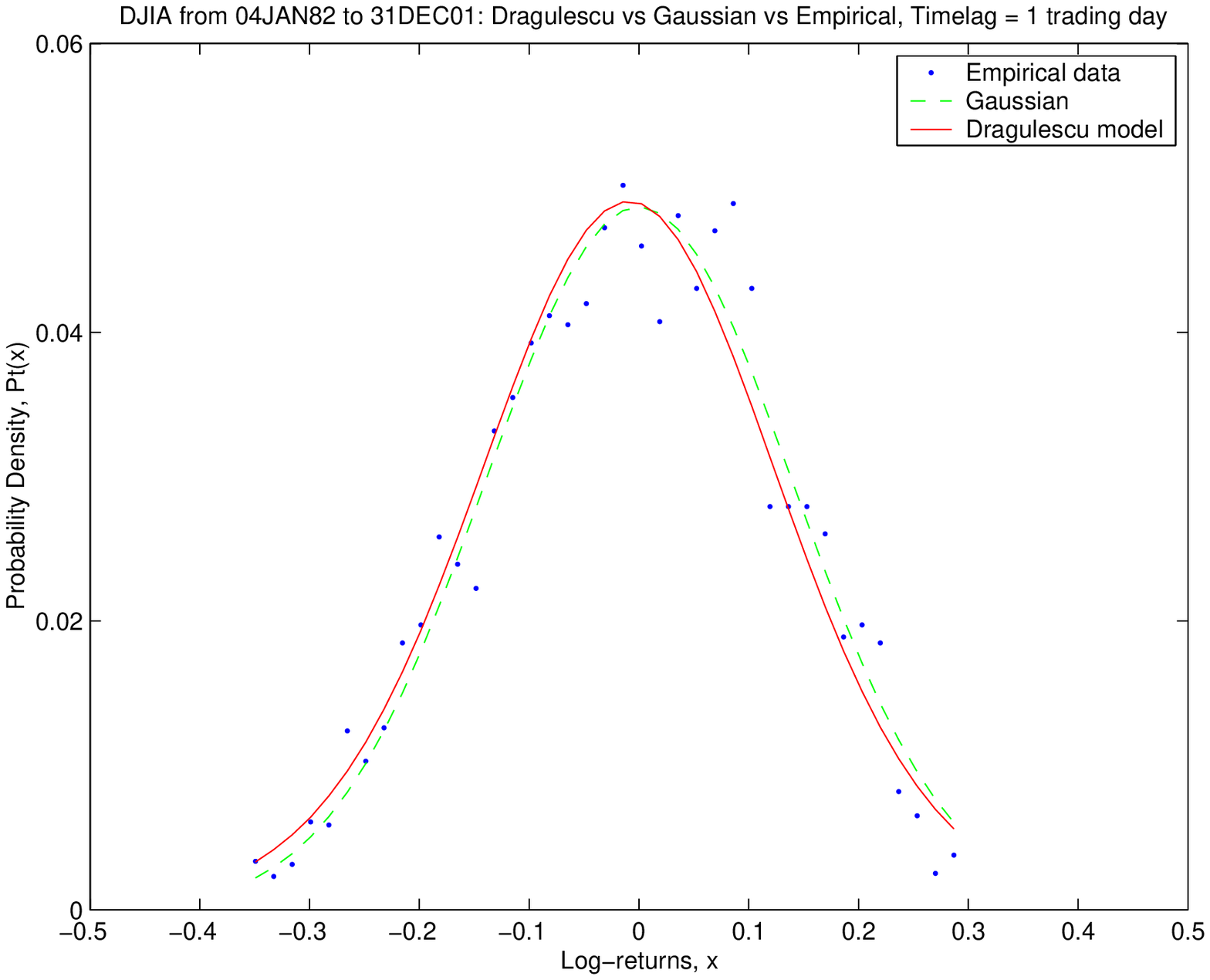} \\
\end{tabular}
\end{center}
\caption{Dragulescu vs Normal vs Empirical data}
\label{fig:fig-draguPDF}  
\end{figure}

The results are shown in figure \ref{fig:fig-draguPDF}.

Apparently, their model (plain line) fits the empirical data (dots) far better than the Gaussian (dash line), specially if we look at the fat tails.

\section{Conclusion}

In his attempt to improve the classical Bachelier-Osborne model, which does not handle the kurtosis and the fat tails of the empirical probability mass, Dragulescu and Yakovenko started from a geometrical Brownian motion for the stock price $S_t$, assumed a mean-reverting stochastic process for the variance $v_t$, and solved analytically the forward Kolmogorov equation that governs the joint probability of this two-dimensional stochastic process. Then, by integrating over the variance, they derived the probability density $P_t(x)$ of log returns $x$ for a given time lag $t$.

Once the four parameters of the model are trained, the resulting distribution seems to fit the empirical data far better than the Normal. We explain in Chapter 4 the methodology we used to obtain these results. Then we replicate these results on different datasets (different indexes) and perform some statistical tests to measure the goodness-of-fit of this model.

\chapter{Experiments}
\section{Introduction}

Our aim is to replicate Dragulescu and Yakovenko results and to see if they are reproducible on other datasets, with different time periods and/or different indexes. The model is supposed to fit any stock index, provided we set the value of the four parameters correctly. We will test their model itself and some assumptions such as the ergodicity of the dataset. First, we use exactly the same methodology as described in the paper, and see that strange points appear on our results. By clarifying the origin of these points directly with Dragulescu and Yakovenko, we make a few critical comments concerning the way they reuse and trim the data, and propose our own methodology based on the conservation of all the data (specially outliers, that occur during crashes and form fat tails). As a benchmark, we test this model against the classical Gaussian model and against a simple Neural Network.

\section{Datasets}

Our datasets will be, first, the Dow Jones Industrial Average (DJIA) for different time periods: the period used in Dragulescu and Yakovenko paper (from January 04, 1982 to December 31, 2001), the period after the 1987 crash (from January 04, 1988 to December 31, 2001), after the 1929 crash (from January 02, 1930 to December 31, 2001) and finally since 1896 (from May 26, 1896 to December 31, 2001) to get the largest dataset possible. Indeed, we need a very large dataset to compute distributions for important time lags such as 250 days (we take one point out of 250). Thanks to these datasets, we will test the robustness of our potential patterns according to different periods. In particular, we will focus on the impact the presence of a crash in the dataset can have on the model.

Moreover, we will use other indexes to test these patterns against other markets: the Standard and Poor's 500 from January 04, 1965 to December 31, 2001, and the FTSE100 from January 02, 1984, to December 31, 2001.

We download the data from YAHOO (\cite{Yahoo}) and ECONOMY.COM (\cite{Economy}) for the Dow Jones and Standard and Poor's 500 and from Datastream for the FTSE100. 

By ``day", we mean trading day, since all of our datasets are composed of trading days only: week-ends and bank holidays have been removed. By ``time lag", we mean the number of trading days between two points used to compute a log return. If our initial dataset is composed of 1000 close prices, then for a time lag of five days, we will take one point every five to compute the log returns. As a consequence, our final dataset will will be composed of $[\frac{1000}{5}]=200$ log returns only. Nevertheless, we can begin by the first, the second, the third, the fourth or the fifth close price, so that, finally, we use five different datasets of 200 log returns. This allows us to give, for any computation, the average value and an estimate of the variance of the result, which will give more robustness to our statistical tests.

\section{Methodology}

We first describe and follow strictly the methodology proposed in their paper by Dragulescu et al., in order to reproduce their results. This methodology suffers from imperfections, specially because (i) they re-use the data and (ii) they trim the data during the pre-processing step. This leads us to propose our own methodology.

\subsection{Reusing the data}

\subsubsection{Introduction}

For a given index \emph{I} at a given period, let us say the Dow Jones Industrial Average from January 04, 1982, to December 31, 2001, and a given time lag $\tau$, let us say $\tau = 5$ days, the raw close price dataset \emph{closePrice} is composed of \emph{n} close prices, here $n=5050$. When Dragulescu and Yakovenko compute the log returns dataset \emph{logReturns} starting from \emph{closePrice}, they obtain the following time series:
$$
logReturns = \{r_t | t \in [1, \ n-\tau] \}
$$  
where $r_t = log \frac{P_{t+\tau}}{P_t}, \ \forall t \in [1, \ n-\tau]$

In our example, we would have
\begin{eqnarray*}
logReturns & = & \{ r_1, r_2, ..., r_{n-\tau} \} \\
& = & \{ log\frac{P_{1+\tau}}{P_1}, log\frac{P_{2+\tau}}{P_2}, ..., log\frac{P_n}{P_{n-\tau}}  \} \\
& = & \{ log\frac{P_6}{P_1}, log\frac{P_7}{P_2}, ..., log\frac{P_{5050}}{P_{5045}}  \} \\
\end{eqnarray*}

Thus, they obtain a single dataset of $n-\tau$ log returns. We believe this way of computing the log returns time series is unfair, because it ``re-uses" the data. Indeed, let us assume that a crash occurs at time $t^*$. Then they will take into account this specific event $\tau$ times exactly in their dataset in log returns $\{ r_{t^*-\tau}, r_{t^*-\tau+1}, ...,r_{t^*-1 } \}$.

This way to derive, from a raw close price time series of $n$ points, a single log returns time series of $n-\tau$ points, is strictly equivalent, in terms of shape parameters of the final distribution (sample mean $\mu$ and sample standard deviation $\sigma$), to deriving $m=[\frac{n}{\tau}]$ \footnote{where $[A]$ denotes the nearest integer less than or equal to A} log returns time series composed of $m$ log returns, and averaging them into a single time series $logReturn'$.
$$
\forall k \in [1 \ m], \ logReturns'(k) = \frac{1}{m} \sum_{i=1}^{m}logReturns_j(k)
$$ 
with
$$
\forall j \in [1 \ m], \ logReturns_j = \{r_t^j | t \in [1 \ m] \}
$$  
where $r_t^j = log \frac{P_{t+(j-1)+\tau}}{P_{t+(j-1)}}, \ \forall t \in [1 \ m]$

To put it in a nutshell, we derive \emph{m} log returns time series $logReturns_j$ of cardinality $m=[\frac{n}{\tau}]$, called ``paths", instead of a single log return time series $logReturn$ of cardinality $n-\tau$. Then we average these $m$ paths to obtain a final log return time series $logReturns'$, of cardinality $m$.
Obviously, we have:
\begin{itemize}
\item $E[logReturns] = E[logReturns']$
\item $Var[logReturns] = Var[logReturns']$
\end{itemize}

The fact that Dragulescu and Yakovenko re-use the data is then justifiable only if all of the paths are equivalent, viz. only if we assume that the system is ergodic.\footnote{``A collection of systems forms an ergodic ensemble if the modes of behaviour found in any one system from time to time resemble its behaviour at other temporal periods and if the behaviour of any other system when chosen at random also is like the one system. We do not require identical performance, only quite similar time averages and number averages. (If you cannot tell one youth from another or one adult from another, they belong to an ergodic ensemble.) In an ergodic population, any single individual is representative of the entire population. The salient characteristics of this individual are essentially identical with any other member of the group", \cite{MathWorld} }

\subsubsection{First test of ergodicity}

There is a simple way to test the ergodicity of the dataset: for each time lag, we compute the sample mean $\mu^j$ and the sample standard deviation $\sigma^j$ of each path $j$. If the system is really ergodic, then we expect these shape parameters to be almost constant from one path to the other. In other words, their variance should be almost null. To compare things that are comparable, instead of giving the variance alone, we give the standard deviation (square root of the variance) of the parameter divided by its mean, which gives us a ``variation rate''. Results are presented in Tables \ref{tab:ergodic1} and \ref{tab:ergodic2} for DJIA1982 and DJIA1896 respectively.

\begin{table}[htbp]
\begin{center}
\begin{tabular}{ccc}
time lag  & \% variation on $\mu$ & \% variation on $\sigma$ \\ 
\begin{tabular}{r}
	 5 \\
	 20 \\
	 40 \\
	 80 \\
	 100 \\
	 200 \\
	 250 \\
	\end{tabular} &
\begin{tabular}{l}
	0.58 \\
	0.76 \\
	1.73 \\
	2.22 \\
	3.43 \\
	4.29 \\
	5.44 \\
	\end{tabular} &
\begin{tabular}{l}
	5.22 \\
	4.31 \\
	7.80 \\
	6.48 \\
	9.30 \\
	9.38 \\
	11.5 \\
	\end{tabular} \\
\end{tabular}
\end{center}
\caption{Variation of shape parameters $\mu$ and $\sigma$ over different paths, DJIA1982}
\label{tab:ergodic1}
\end{table} 

\begin{table}[htbp]
\begin{center}
\begin{tabular}{ccc}
time lag  & \% variation on $\mu$ & \% variation on $\sigma$ \\
\begin{tabular}{r}
	 5 \\
	 20 \\
	 40 \\
	 80 \\
	 100 \\
	 200 \\
	 250 \\
	\end{tabular} &
\begin{tabular}{l}
	0.12 \\
	0.53 \\
	1.84 \\
	1.52 \\
	2.43 \\
	2.23 \\
	2.18 \\
	\end{tabular} &
\begin{tabular}{l}
	1.46 \\
	2.26 \\
	2.46 \\
	5.15 \\
	2.82 \\
	2.33 \\
	4.84 \\
	\end{tabular} \\
\end{tabular}
\end{center}
\caption{Variation of shape parameters $\mu$ and $\sigma$ over different paths, DJIA1896}
\label{tab:ergodic2}
\end{table} 

There is no variance, and then no variation, for a time lag of one day, since we have only one single log returns time series. 

The variation rate is low for very large datasets like the Dow Jones Industrial Average from 1896 to 2001; it is always inferior to 5.5 \%. However, for relatively small datasets (DJIA 1982, DJIA1988), the variation rate can reach 11.5 \%, and even 17 \% for DJIA1988. This comes from the fact the distribution of an average tends to be normal (CLT\footnote{See Appendix B, Section ``Central Limit Theorem''}) when the sample size increases, with variance decreasing proportionally to $\sqrt{n}$. The more points we have, the less the variation rate, whatever the initial distribution. 

\subsubsection{Second test of ergodicity}

Given that this test is not conclusive, we perform a Kruskal-Wallis Test, which is a nonparametric version of the One-Way Analysis of Variance (``ANOVA'').\footnote{Se Appendix B, Section ``ANOVA Test''} The purpose of a one-way analysis of variance is to find out whether data from several datasets have a common mean. The assumption behind this test is that the measurements come from a continuous distribution, but not necessarily a normal distribution.\footnote{Se Appendix B, Section ``Kruskal-Wallis Test''} If the p-value is near zero, this casts doubt on the null hypothesis and suggests that at least one sample mean is significantly different than the other sample means.

\begin{table}[htbp]
\begin{center}
\begin{tabular}{cccc}
time lag  & Chi-Square & df & p-value \\ 
\begin{tabular}{r}
	   5 \\
	  20 \\
	  40 \\
	  80 \\
	 100 \\
	 200 \\
	 250 \\
	\end{tabular} &
\begin{tabular}{r}
	0.33  \\ 
	0.37  \\ 
	1.35  \\ 
	2.39  \\ 
	4.49  \\ 
	12.53 \\ 
	9.61  \\ 
	\end{tabular} &
\begin{tabular}{r}
	  4  \\ 
	 19  \\ 
	 39  \\ 
	 79  \\ 
	 99  \\ 
	199  \\ 
	249  \\ 
	\end{tabular} &
\begin{tabular}{l}
	0.9875 \\ 
	1 \\ 
	1 \\ 
	1 \\ 
	1 \\ 
	1 \\ 
	1 \\ 
	\end{tabular} \\
\end{tabular}
\end{center}
\caption{Kruskal-Wallis Test on means $\mu$, DJIA1982}
\label{tab:kruskalWallis1}
\end{table} 

\begin{table}[htbp]
\begin{center}
\begin{tabular}{cccc}
time lag  & Chi-Square & df & p-value \\ 
\begin{tabular}{r}
	   5 \\
	  20 \\
	  40 \\
	  80 \\
	 100 \\
	 200 \\
	 250 \\
	\end{tabular} &
\begin{tabular}{r}
	 0.07 \\ 
	 0.80 \\ 
	 0.80 \\ 
	 1.30 \\ 
	 3.37 \\ 
	 2.42 \\ 
	 11.79 \\ 
	\end{tabular} &
\begin{tabular}{r}
	  4  \\ 
	 19  \\ 
	 39  \\ 
	 79  \\ 
	 99  \\ 
	199  \\ 
	249  \\ 
	\end{tabular} &
\begin{tabular}{l}
	0.9995 \\ 
	1 \\ 
	1 \\ 
	1 \\ 
	1 \\ 
	1 \\ 
	1 \\ 
	\end{tabular} \\
\end{tabular}
\end{center}
\caption{Kruskal-Wallis Test on means $\mu$, DJIA1896}
\label{tab:kruskalWallis2}
\end{table}     

\begin{figure}
\begin{center}
\begin{tabular}{cc}
\includegraphics[width=5cm]{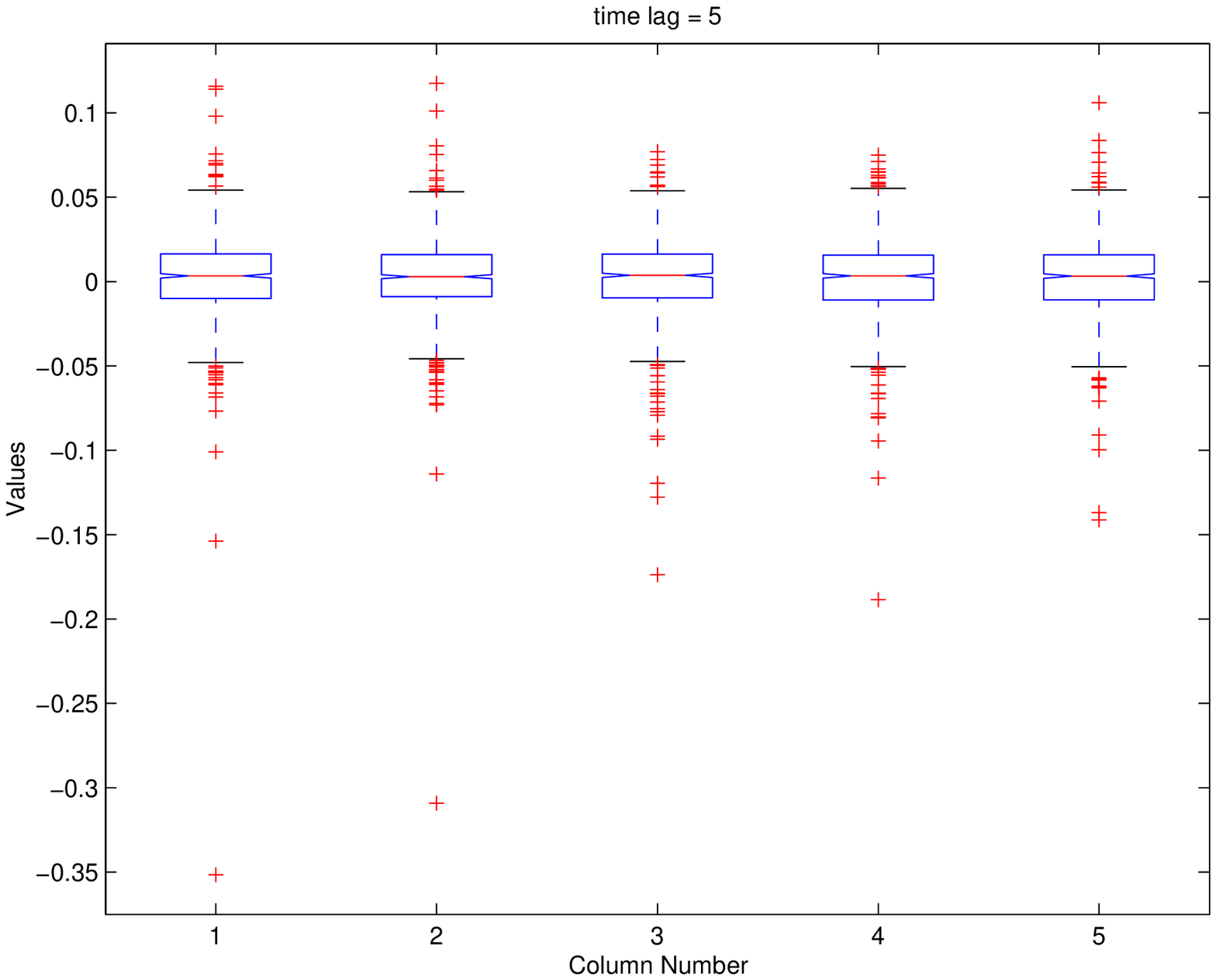} & \includegraphics[width=5cm]{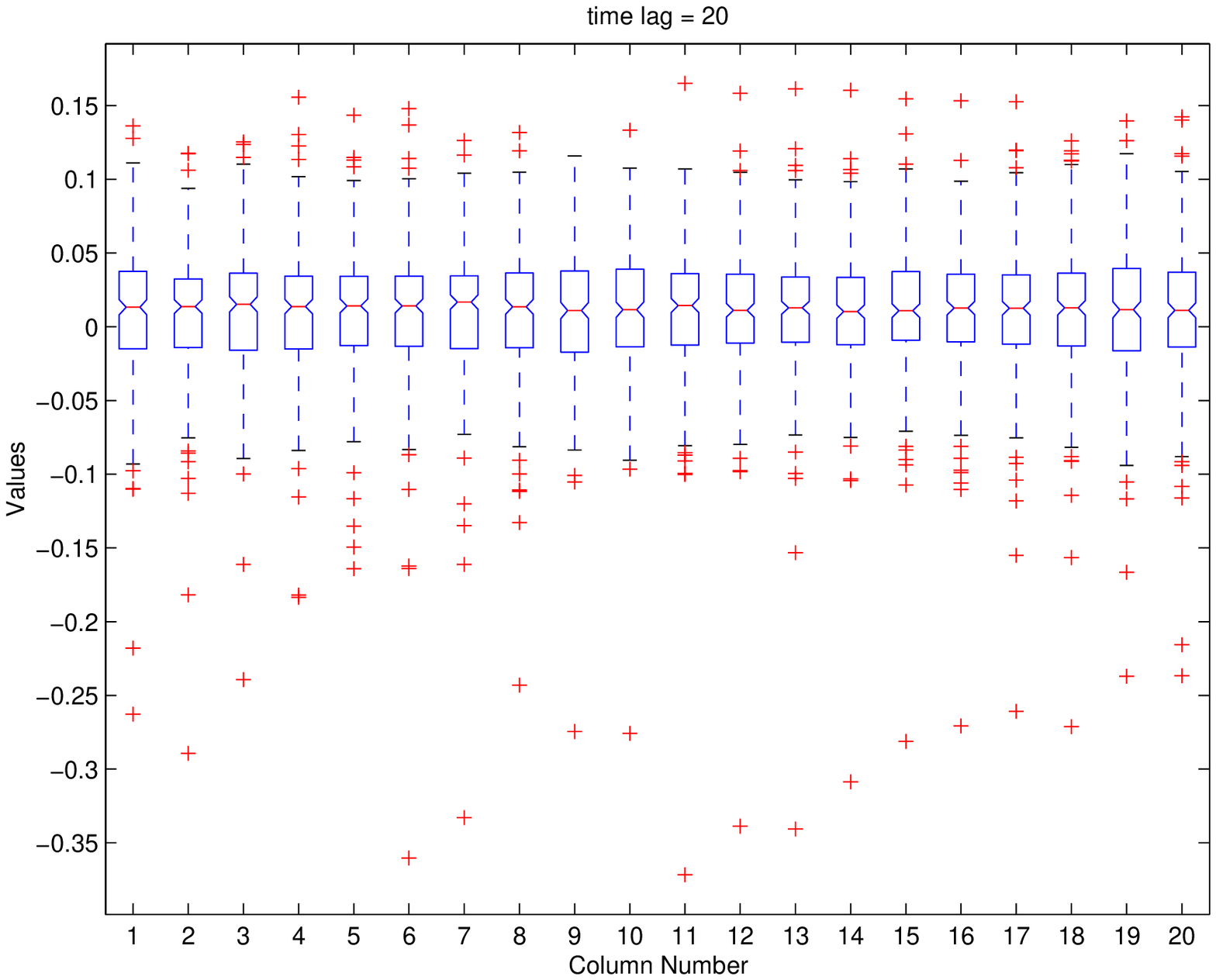} \\
\includegraphics[width=5cm]{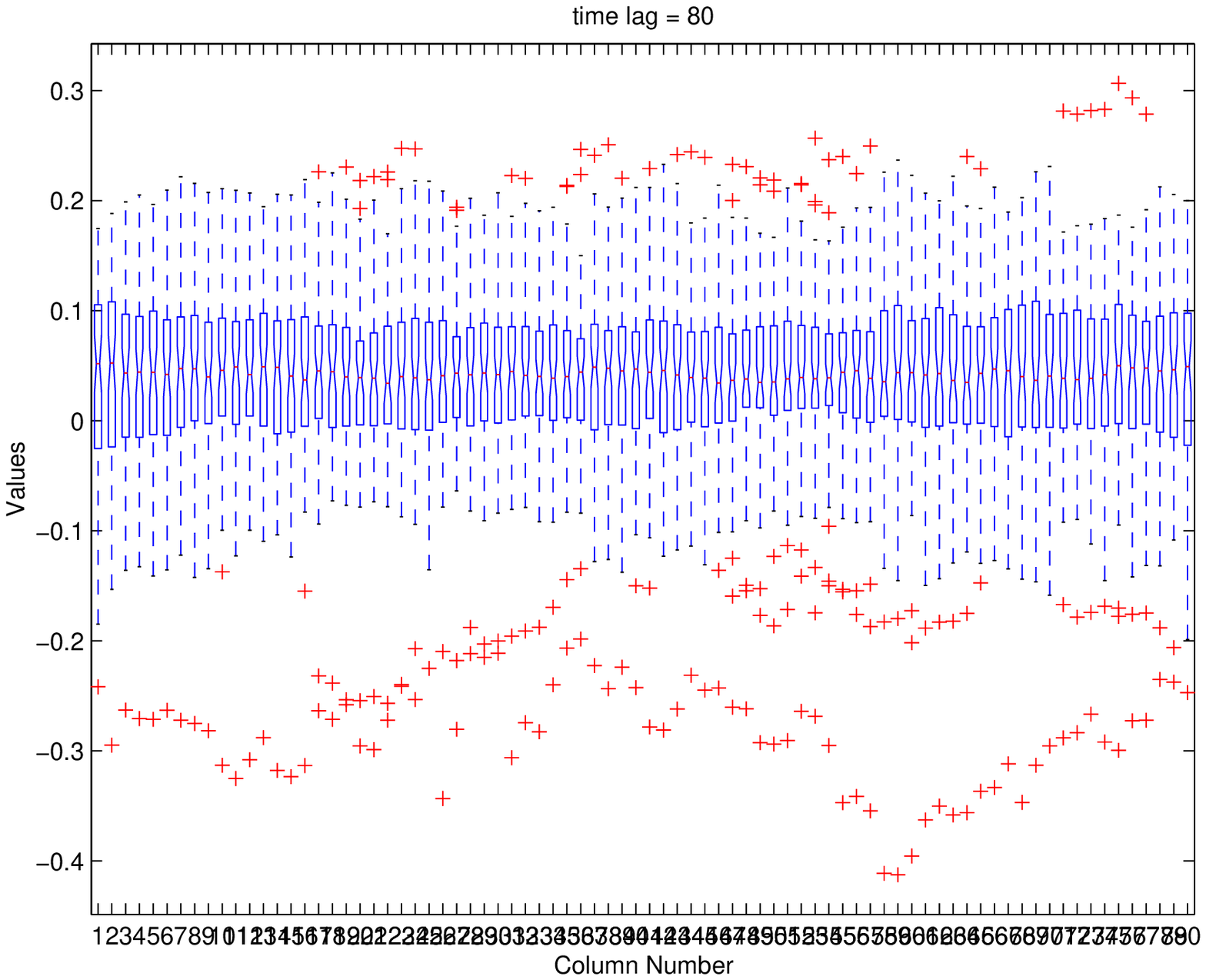} & \includegraphics[width=5cm]{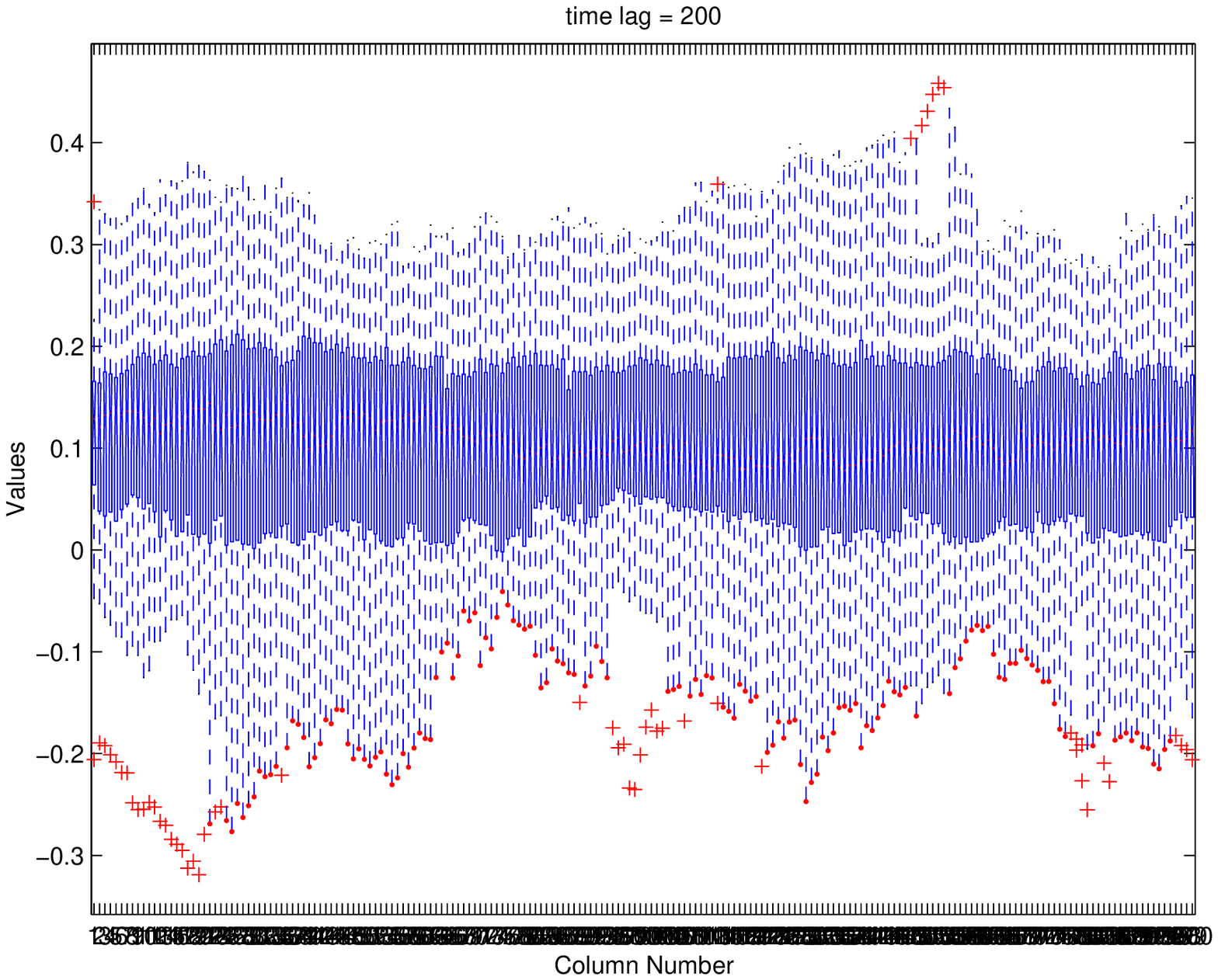} \\
\end{tabular}
\end{center}
\caption{Kruska-Wallis Test on standard deviation $\sigma$, DJIA1982}
\label{fig:fig-kruskaWallis}  
\end{figure}

The results, presented in Tables \ref{tab:kruskalWallis1} and \ref{tab:kruskalWallis2}, clearly demonstrate that there is no significant difference between the means of all the different paths, whatever the frequency. Indeed, the p-value is always very high. This tends to support the hypothesis that all the paths are equivalent. But if we have a look at the variance now (which is the core of Dragulescu and Yakovenko model), we see that it changes dramatically from a path to another. To show this, we plot in Figure \ref{fig:fig-kruskaWallis} the Box Plot of the different paths.\footnote{Box Plots are composed of different elements:\begin{itemize} \item The lower and upper lines of the ``box'' are the 25th and 75th percentiles of the sample \item The distance between the
top and bottom of the box is the interquartile range \item The line in the middle of the box is the sample median. If the median is not centred in the box, that is an indication
of skewness \item The ``whiskers'' are lines extending above and below the box. They show the extent of the rest of the sample (unless there are outliers). Assuming no outliers, the maximum of the sample is the top of the upper whisker. The minimum of the sample is the bottom of the lower whisker. By default, an outlier is a value that is more than 1.5 times the
interquartile range away from the top or bottom of the box \item The plus sign at the top of the plot is an indication of an outlier in the data. This point may be the result of a data
entry error, a poor measurement or a change in the system that generated the data \end{itemize}} 

We can clearly see that the different paths are not equivalent, even if they almost have the same mean. The number of outliers for instance vary dramatically from one path to another, as indicated by the number of red crosses outside the ``whiskers''. We have to indicate that this variance in the volatility of the log returns has no specific relation with the well documented ``seasonal effect'', since our paths are not based on calendar days but on trading days. For instance, the five paths obtained for a 5-days time lag are not composed of log returns from Mondays to Mondays, Tuesdays to Tuesdays, etc., but on consecutive trading days. Nevertheless, in the financial literature, analysts usually use realisations (called ``paths'' in this thesis) that do not reuse the data, because they are interested in investment strategies on a daily, weekly, monthly, etc., basis. They do not average the paths to get a final very large dataset, as Dragulescu and Yakovenko did.

\subsubsection{Conclusion}

For technical reasons (paths are different from each other, the system is \emph{not} ergodic) \emph{and} practical reasons (financial analysts do not do that), we think Dragulescu and Yakovenko should not reuse the data, which is equivalent to averaging the mean and the variance of each data. As a consequence, in our subsequent tests, we will present our results without reusing the data.

\subsection{Pre-processing the data}

We perform all of our tests for different time lags: 1, 5, 20, 40, 80, 100, 200, and 250 days. For a given time lag $\tau$ and a given dataset \emph{D}, we compute all of the log-return series $r_t = Log ( \frac{S_t}{S_{t-\tau}} )$. If the price dataset contains \emph{n} points (each point is the daily close price of the index considered), then we obtain $n - \tau$ log-returns. Dragulescu and Yakovenko trimmed the log returns, rejecting any value out of the boundaries presented in Table \ref{tab:boundTrim}.\footnote{This step is not mentioned in Dragulescu's paper. Before applying this trimming method, strange points used to appear in our results. Then we contacted the authors who informed us of trimming of the log returns, using those boundaries for the Dow Jones Industrial Average from January 04, 1982, to December 31, 2001 (``DJIA1982'').}

\begin{table}[h]
\begin{center}
\begin{tabular}{cc}
time lag  & trimming boundaries \\ 
\begin{tabular}{r}
	   1 \\
	   5 \\
	  20 \\
	  40 \\
	  80 \\
	 100 \\
	 200 \\
	 250 \\
	\end{tabular} &
\begin{tabular}{c}
	$[-0.04 \ 0.04]$ \\
	$[-0.08 \ 0.08]$ \\
	$[-0.13 \ 0.15]$ \\
	$[-0.17 \ 0.20]$ \\
	$[-0.18 \ 0.25]$ \\
	$[-0.20 \ 0.28]$ \\
	$[-0.22 \ 0.38]$ \\
	$[-0.22 \ 0.44]$ \\
	\end{tabular} \\
\end{tabular}
\end{center}
\caption{Boundaries used by Dragulescu et al. to trim empirical log returns}
\label{tab:boundTrim}
\end{table}

\begin{figure}[h]
\begin{center}
\includegraphics[width=10cm]{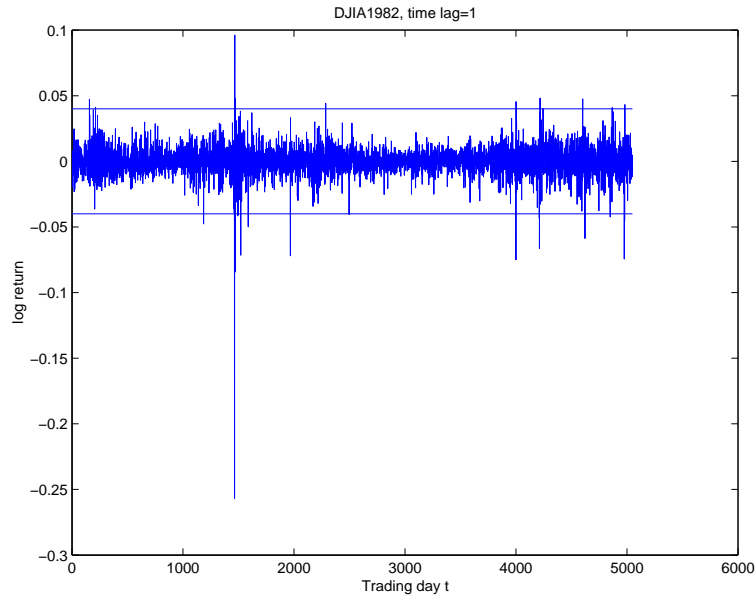}
\end{center}
\caption{Trimmed log returns, DJIA1982, $\tau=1$ day}
\label{fig:trimmedLogReturn}  
\end{figure}

We visualise in Figure \ref{fig:trimmedLogReturn} the effect of trimming the data: all of the log returns outside the boundaries, represented here by the two horizontal lines, were trimmed by the authors.

We believe this way of trimming the data is unfair, because it removes information from the dataset. Given that the model is supposed to outperform the Bachelier-Osborne model, and specially to fit the kurtosis and the fat tails, removing extreme values (that belong to the fat tails, and produce kurtosis) is counter-productive. Even the normal distribution could fit the data quite well in these conditions. To prove this, we perform the Jarque-Bera and Lillifors tests against the normal distribution, but using the trimmed data.

\begin{table}[p]
\begin{center}
\begin{tabular}{rrrrrrr}
\tt time lag  & DJIA1982 & DJIA1988 & DJIA1896 & DJIA 1930 & SP1965 & FTSE1984 \\ 
\tt 1   & 1 & 1 & 1 & 1 & 1 & 1 \\
\tt 5   & 1 & 1 & 1 & 1 & 1 & 0.6 \\
\tt 20  & 0 & 0.1 & 1 & 1 & 0.5 & 0.1 \\
\tt 40  & 0.175 & 0.375 & 0.225 & 0.575 & 0.175 & 0 \\
\tt 80  & 0 & 0.0125 & 0 & 0.025 & 0 & 0 \\
\tt 100 & 0 & 0.05 & 0 & 0 & 0 & 0 \\
\tt 200 & 0 & 0 & 0 & 0 & 0 & 0 \\
\tt 250 & 0 & 0 & 0 & 0 & 0 & 0 \\
\end{tabular}
\end{center}
\caption{Jarque-Bera Goodness-of-Fit Test to the Normal distribution, after trimming, DJIA1982}
\label{tab:jarqueberaTestTrim}
\end{table}

\begin{table}[p]
\begin{center}
\begin{tabular}{rrrrrrr}
\tt time lag  & DJIA1982 & DJIA1988 & DJIA1896 & DJIA 1930 & SP1965 & FTSE1984 \\ 
\tt 1   & 1 & 1 & 1 & 1 & 1 & 1 \\
\tt 5   & 1 & 1 & 1 & 1 & 1 & 0.8 \\
\tt 20  & 0.1& 0.35 & 1 & 0.9 & 0.35 & 0.1 \\
\tt 40  & 0.05 & 0.025 & 0.3 & 0.525 & 0.125 & 0.05 \\
\tt 80  & 0.0375 & 0.025 & 0.075 & 0.1375 & 0 & 0 \\
\tt 100 & 0 & 0 & 0.09 & 0.04 & 0.03 & 0.01 \\
\tt 200 & 0.025 & 0.035 & 0.01 & 0 & 0.065 & 0.04 \\
\tt 250 & 0.032 & 0.02 & 0.016 & 0 & 0.02 & 0.108 \\
\end{tabular}
\end{center}
\caption{Lilliefors Goodness-of-Fit Test to the Normal distribution, after trimming, DJIA1982}
\label{tab:lillieTestTrim}
\end{table}

Results in Tables \ref{tab:jarqueberaTestTrim} and \ref{tab:lillieTestTrim}, compared with the same test on untrimmed data (Tables \ref{tab:jarqueberaTest} and \ref{tab:lillieTest}) clearly show that trimming the data, as Dragulescu and Yakovenko do, rejects the normal hypothesis only for higher frequencies. This time the tests do not reject the normal hypothesis for medium frequencies as they did without trimming. 

For this reason, we decided to perform our statistical tests without trimming the data.

\subsection{Distributions}

To obtain the empirical distribution, we partition the log-returns into equal sized bins of length $\Delta_r$( $\Delta_r = \frac{max(r_t) - min(r_t)}{noBins + 1}$). Then we count the number of log-returns per bin, called \emph{occupation number} and remove the bins for which \emph{occupation number} is lower than a critical value of five. We initially choose the number of bins so that we globally remove less than one percent of the log-returns. This filtering technique is supposed to get rid of the outliers, viz. infrequent events. Thus, we obtain the frequency repartition of log-returns, also called in this paper the empirical probability density function, \emph{empPDF}.

In order to exhibit fat tails and kurtosis, we fit a Gaussian to the \emph{empPDF}, by estimating the sample mean and sample standard deviation of the set of log-returns $r_t$. We obtain the sample mean $\mu$ and the sample standard deviation $\sigma$ and plot the Gaussian \emph{normPDF}.

After having observed the departure from normality, we build the new model. We train the four parameters of Dragulescu and Yakovenko model by minimising the mean-square deviation $E = \sum_{x,t} |log P_t^*(x) - log P_t(x)|$, and compute the PDF, \emph{draguPDF} for different time lags.

Finally, we build and train a Neural Network to fit \emph{empPDF} as precisely as possible. A very simple structure is enough for this first approach. We will use this NN, \emph{nnPDF}, later in our tests as a benchmark.

We can now perform goodness-of-fit tests, measures of kurtosis and measures of fat tails on different models (Gaussian, Neural Network and Dragulescu).

\subsubsection{Empirical Distribution}
\begin{figure}[h]
\begin{center}
\begin{tabular}{cc}
\includegraphics[width=5cm]{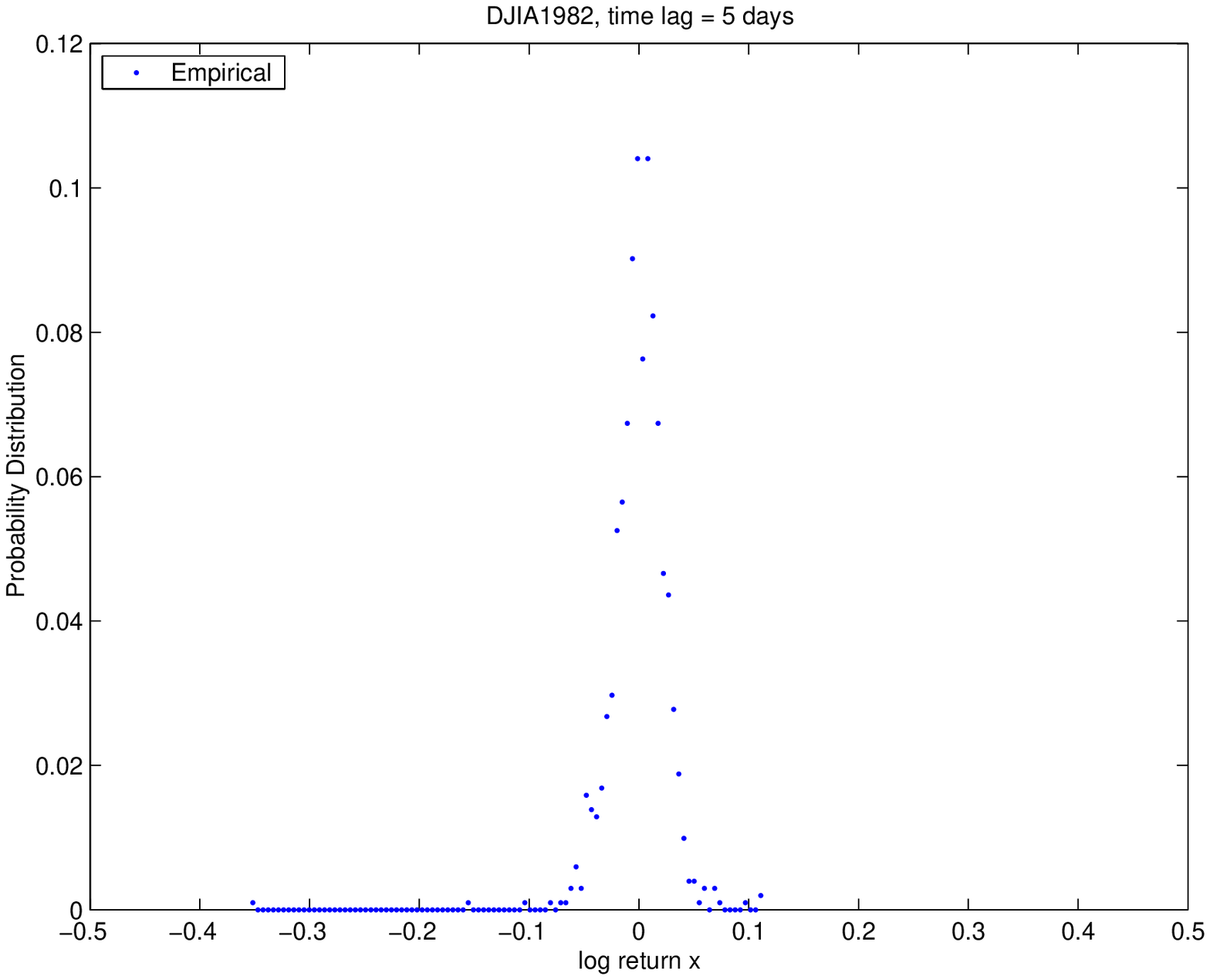} & \includegraphics[width=5cm]{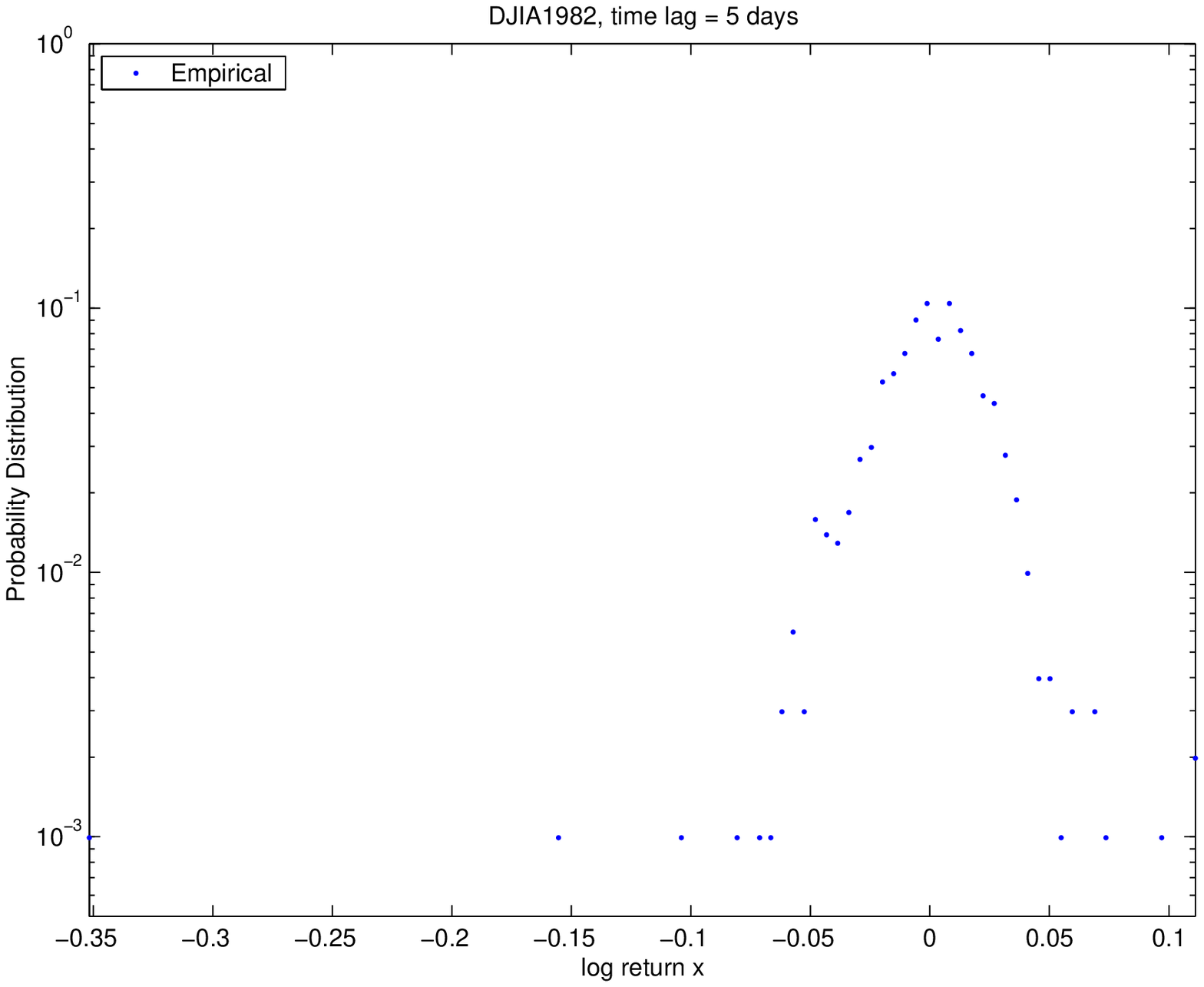}
\end{tabular}
\end{center}
\caption{Empirical distribution \emph{empPDF} in normal and logarithmic scale}
\label{fig:empPDF}  
\end{figure}

All we have to do to obtain the empirical pdf \emph{empPDF} is to divide each occupation number by the bin size $\Delta r$ and the total number of observations, once bins with less than five log-returns have been removed. Then we centre the result (we subtract the sample mean to the x axis) to obtain the final probability density function \emph{empPDF}. We plot in Figure \ref{fig:empPDF} the probability mass obtained. Since the log returns are almost normal, at least according to the classical Bachelier-Osborne model, we prefer plotting the PDFs on a semi-logarithmic graph. Thus, the slopes of the probability mass should look like straight lines. This has also the advantage of focusing on the tails, since on a normal plot any discrepancy in the tails looks negligible compared to the discrepancies in the middle of the distribution, that form the kurtosis.

On the semi-log plot, a close look at the tails, specially on the left side, exhibits a series of points aligned horizontally. These events happened once, and constitute a kind of long tail. Indeed, the probability mass (or empirical probability density) is bounded, and the inferior limit is simply $\frac{1}{number \ of \ events}$ where an event is a specific log return in the time series. We will see that our models are all unbounded: they cannot predict those extremely isolated events. We will have to pay attention to these long tails in our tests, and not confuse them with the fat tails.

\subsubsection{Gaussian Distribution}
In order to exhibit fat tails and kurtosis, we compute the sample mean and standard deviation of the log-returns data, and generate a normal distribution \emph{normPDF} with these parameters. We stress here that as we do not have any prior knowledge of the mean and variance of the Gaussian, we have to derive them from the empirical log returns. This will have an impact later in our statistical tests, since we will have to test a composite hypothesis instead of a simple hypothesis, usually easier to deal with. \\

\begin{figure}[h]
\begin{center}
\includegraphics[width=5cm]{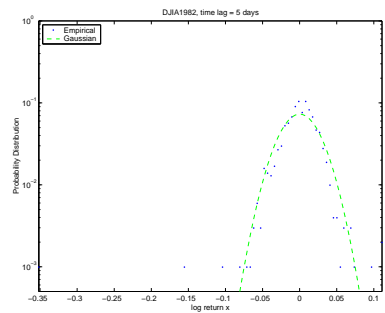}
\end{center}
\caption{Empirical distribution \emph{empPDF} vs Gaussian \emph{normPDF}}
\label{fig:normPDF}  
\end{figure}

The Gaussian seems to fit the empirical distribution quite well, but a simple look at the graph, even if it is often useful, cannot be used as a strong evidence. We need statistical tests to measure the goodness-of-fit of the model, as we will see later in this Chapter.

\subsubsection{Dragulescu's Distribution}
We can now compute the distribution expected by Dragulescu's model, \emph{draguPDF}. First, we have to set the value of the four parameters of the model. To do so, we minimise the mean-square deviation between the model and the empirical data. Once we have these values, we can generate \emph{draguPDF} by integrating between finite bounds the expression given in the model. Using finite instead of infinite bounds does not seem to modify the results, provided the bound is large enough.

\begin{figure}[h]
\begin{center}
\includegraphics[width=5cm]{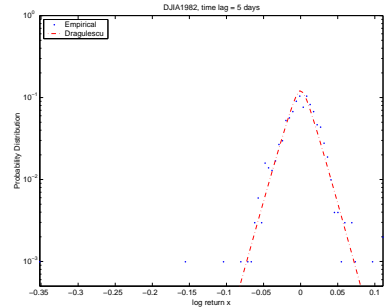}
\end{center}
\caption{Empirical distribution \emph{empPDF} vs Dragulescu \emph{draguPDF}}
\label{fig:draguPDF}  
\end{figure}

\subsubsection{Neural Network Distribution}
Even if \emph{draguPDF} is supposed to fit \emph{empPDF} better than \emph{normPDF}, we want to compare it with the best fit possible, the one obtained with a Neural Network. This Neural Network must be as simple as possible, but should fit the main characteristics of the empirical time series, fat tail and kurtosis. Th structure chosen was the following: it is a feed-forward back-propagation network, with a five node hidden layer and a single node output layer. The transfer functions are respectively $tansig$ and $purelin$, where $tansig(n) = \frac{2}{1+e^{-2*n}}-1$ and $purelin(n)=n$. The back-propagation function used is $trainscg$, a network training function that updates the weight and bias values according to Levenberg-Marquardt optimisation. It minimises a combination of squared errors and weights and then determines the correct combination so as to produce a network which generalises well. The process is called Bayesian regularisation. This structure appears to be a good trade-off between the complexity and the goodness of fit. 

We prefer not to complicate the structure, in order to have meaningful statistical tests: indeed, a model with many parameters will obviously manage to fit the data, but the goodness-of-fit will be very poor.\footnote{See Chapter 4, Section ``$\chi^2$ Statistic''}

\begin{figure}[h]
\begin{center}
\includegraphics[width=5cm]{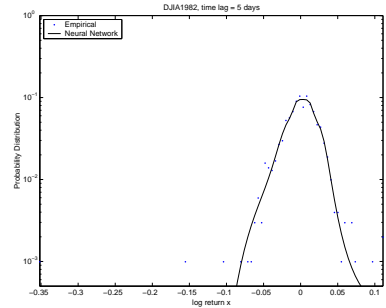}
\end{center}
\caption{Empirical distribution \emph{empPDF} vs Neural Network \emph{nnPDF}}
\label{fig:nnPDF}  
\end{figure}

\section{Comparison of the models}

Now that we have obtained different models, we can compare them. Our aim is to verify if the Dragulescu and Yakovenko model fits the empirical data better than the classical Gaussian. The Neural Network is used as a benchmark. We perform the following tests without re-using or trimming the data. \\
For each dataset D and for each time lag $t$, we obtain a set of distributions: the empirical distribution \emph{empPDF} computed from the empirical log-returns, the normal distribution \emph{normPDF} fitted on the empirical log-returns, Dragulescu and Yakovenko distribution fitted on \emph{empPDF} and finally the neural network distribution \emph{nnPDF} fitted on \emph{empPDF}.

\begin{figure}[h]
\begin{center}
\includegraphics[width=10cm]{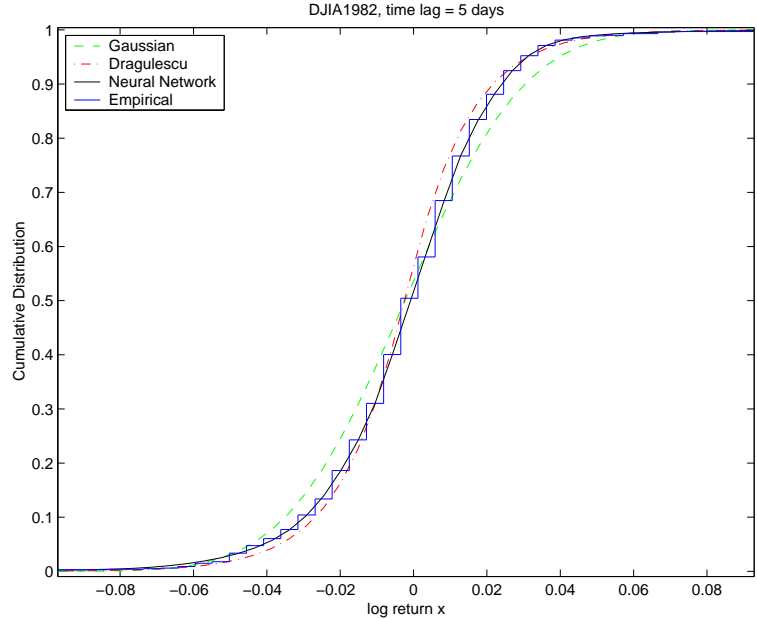} 
\end{center}
\caption{Cumulative Density function of log returns, together with the theoretical Gaussian (--), Dragulescu (-.) distributions and the Neural Network distribution (-), on DJIA1982 dataset, for $\tau=5$ days}
\label{fig:fig-cdf}  
\end{figure}

\begin{figure}[h]
\begin{center}
\includegraphics[width=10cm]{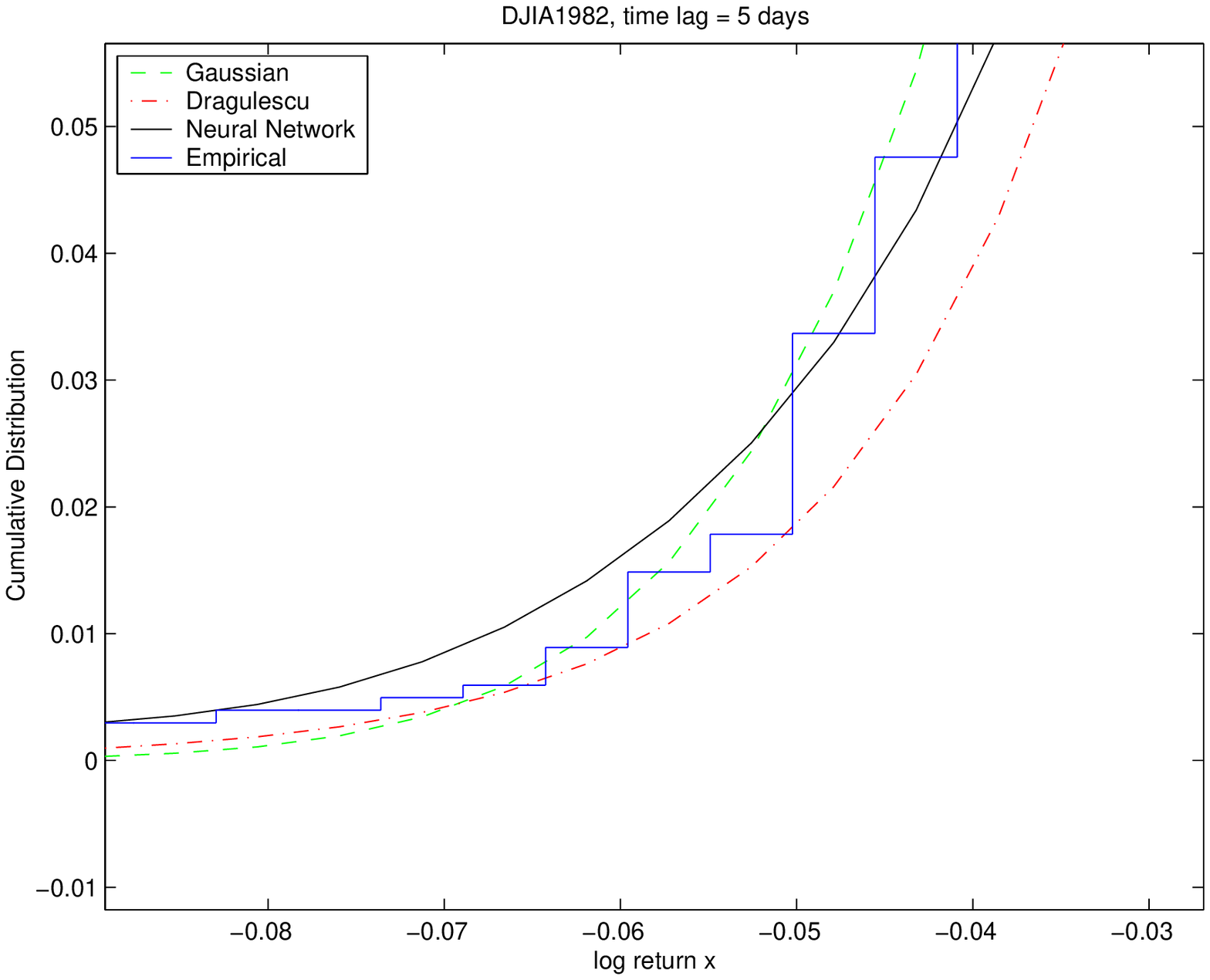} 
\end{center}
\caption{Zoom around the low tail}
\label{fig:fig-cdfTails}  
\end{figure}

The first thing to do when comparing different models is to have a close look at the empirical and expected cumulative distributions. Even if it can be misleading sometimes, this usually gives a good overview of the possible discrepancies between the theoretical and observed data. We plot in Figure \ref{fig:fig-cdf} the expected and observed cumulative distributions for the index DJIA1982 and a 5 days time lag. It seems that Dragulescu and Yakovenko curve fits the empirical distribution a bit better than the Gaussian, specially if we look at the tails in Figure \ref{fig:fig-cdfTails}. Even if it has a very simple structure, the Neural Network seems to be the best model, except in the law tail.

To test the goodness-of-fit of our models, we will first use the Kolmogorov-Smirnov Statistic. Mainly because this test poorly handles the fat tails and because it can only test a simple hypothesis, we will then perform  a $\chi^2$ goodness-of-fit test on equal expected frequency bins. Finally, we will use generated random data to focus on the kurtosis of the different models and then study the outliers that compose the fat and long tails. 

\subsection{Kolmogorov-Smirnov Statistic}

\subsubsection{Introduction}

Dragulescu and Yakovenko claim that their model fits the empirical data of DJIA1982 better than the Gaussian for any time lag. To check that, we use the Kolmogorov-Smirnov Statistic,\footnote{See Appendix B, Section ``Kolmogorov-Smirnov Goodness-of-Fit Test''} based on the maximal discrepancy between the expected and the observed cumulative distributions, for any log return x. We perform this test on the DJIA1982 index, for different time lags. This statistic is suitable for testing only a simple hypothesis, for instance a Gaussian with known $\mu$ and $\sigma$, but not a composite hypothesis (a class of Gaussians, or a Gaussian with $\mu$ and $\sigma$ derivated from the tested sample dataset itself). Unfortunately, whatever the model, we always derive the parameters ($\mu$ and $\sigma$ for \emph{normPDF}, $\gamma$, $\theta$, $k$ and $\mu$ for \emph{draguPDF}, the weights and biases for \emph{nnPDF}) from the initial dataset. 

By performing this test with parameters derived from the dataset, we expect the statistic to be large enough to reject the simple hypothesis, and \emph{a fortiori} the composite hypothesis (\cite{Bree}). But if the value of $Z$ is small enough to accept the simple hypothesis, it does not mean that we can accept the composite hypothesis. \\

\subsubsection{Methodology}

For each time lag, we compute the log returns dataset, and we divide it into paths. For each path and each model, we build the empirical cumulative density function \emph{empCDF} and the expected CDF \emph{modelCDF} (\emph{normCDF}, \emph{draguCDF} or \emph{nnCDF}), and we compute the KS-statistic $Z$. We present in Tables \ref{tab:normKolmo}, \ref{tab:draguKolmo} and \ref{tab:nnKolmo} the mean $\tilde{Z}$ and standard deviation $\sigma_Z$ of $Z$ over the different paths, and the associated p-value\footnote{The p-value is the probability of observing the given sample result under the assumption that the null hypothesis (the tested model) is true. If the p-value is less than the level of significance $\alpha$, then you reject the null hypothesis. For example, if $\alpha = 0.05$ and the p-value is 0.03, then you reject the null hypothesis. The converse is not true. If the p-value is greater than $\alpha$, you have insufficient evidence to reject the null hypothesis.} interval $p(\tilde{Z}+\sigma_Z) \leq p(\tilde{Z}) \leq p(\tilde{Z}-\sigma_Z)$. \\

\subsubsection{Results}

\begin{table}[p]
\begin{center}
\begin{tabular}{|c|c|c|}
\hline
time lag & $\tilde{Z}$ & p-value \\ 
\hline
\begin{tabular}{r}
	 1 \\
	 5 \\
	 20 \\
	 40 \\
	 80 \\
	 100 \\
	 200 \\
	 250 \\
	\end{tabular} &
\begin{tabular}{l}
       0.131         \\
       0.081 $\pm$0.020 \\
       0.089 $\pm$0.013 \\
       0.104 $\pm$0.020 \\
       0.113 $\pm$0.021 \\
       0.113 $\pm$0.021 \\
       0.148 $\pm$0.038 \\
       0.170 $\pm$0.047 \\ 
	\end{tabular} &
\begin{tabular}{r@{}lrr@{}lrr@{}l}
	&&& 2&.93e-75 &&& \\
	1&.75e-09 & $\leq$ & 2&.98e-06 & $\leq$ & 9&.88e-04\\
	9&.51e-03 & $\leq$ & 0&.036    & $\leq$ & 0&.112\\
	0&.038    & $\leq$ & 0&.122    & $\leq$ & 0&.321\\
	0&.199    & $\leq$ & 0&.385    & $\leq$ & 0&.649\\
	0&.322    & $\leq$ & 0&.533    & $\leq$ & 0&.778\\
	0&.339    & $\leq$ & 0&.630    & $\leq$ & 0&.917\\
	0&.291    & $\leq$ & 0&.598    & $\leq$ & 0&.919\\
	\end{tabular} \\
\hline
\end{tabular}
\end{center}
\caption{KS-Test on the Gaussian, DJIA1982}
\label{tab:normKolmo}
\end{table}

\begin{table}[p]
\begin{center}
\begin{tabular}{|c|c|c|}
\hline
time lag & $\tilde{Z}$ & p-value \\ 
\hline
\begin{tabular}{r}
	 1 \\
	 5 \\
	 20 \\
	 40 \\
	 80 \\
	 100 \\
	 200 \\
	 250 \\
	\end{tabular} &
\begin{tabular}{l}
	0.109 \\        
	0.087 $\pm$0.019  \\
	0.089 $\pm$0.014 \\
	0.094 $\pm$0.010 \\
	0.116 $\pm$0.018 \\
	0.128 $\pm$0.019 \\
	0.163 $\pm$0.048 \\
	0.186 $\pm$0.046 \\
	\end{tabular} &
\begin{tabular}{r@{}lrr@{}lrr@{}l}
	&&& 1&.2e-53  &&& \\
	2&.08e-10 & $\leq$ & 3&.64e-07 & $\leq$ & 1&.48e-04\\
	8&.75e-03 & $\leq$ & 0&.033    & $\leq$ & 0&.104\\
	0&.125    & $\leq$ & 0&.211    & $\leq$ & 0&.337\\
	0&.197    & $\leq$ & 0&.355    & $\leq$ & 0&.576\\
	0&.215    & $\leq$ & 0&.372    & $\leq$ & 0&.585\\
	0&.209    & $\leq$ & 0&.512    & $\leq$ & 0&.893\\
	0&.224    & $\leq$ & 0&.481    & $\leq$ & 0&.816\\
	\end{tabular} \\
\hline
\end{tabular}
\end{center}
\caption{KS-Test on Dragulescu model, DJIA1982}
\label{tab:draguKolmo}
\end{table}

\begin{table}[p]
\begin{center}
\begin{tabular}{|c|c|c|}
\hline
time lag & $\tilde{Z}$ & p-value \\ 
\hline
\begin{tabular}{r}
	 1 \\
	 5 \\
	 20 \\
	 40 \\
	 80 \\
	 100 \\
	 200 \\
	 250 \\
	\end{tabular} &
\begin{tabular}{l}
	0.106  \\
	0.048 $\pm$0.014 \\
	0.047 $\pm$0.009 \\
	0.071 $\pm$0.059 \\
	0.075 $\pm$0.061 \\
	0.076 $\pm$0.039 \\
	0.116 $\pm$0.034 \\
	0.137 $\pm$0.041 \\
	\end{tabular} &
\begin{tabular}{r@{}lrr@{}lrr@{}l}
	&&& 3&.27e-42 &&& \\
	1&.64e-03 & $\leq$ & 0&.026 & $\leq$ & 0&.204\\
	0&.430 & $\leq$ & 0&.615 & $\leq$ & 0&.778\\
	0&.153 & $\leq$ & 0&.746 & $\leq$ & 1&\\
	0&.729 & $\leq$ & 0&.919 & $\leq$ & 0&.995\\
	0&.532 & $\leq$ & 0&.871 & $\leq$ & 0&.999\\
	0&.126 & $\leq$ & 0&.453 & $\leq$ & 0&.932\\
	0&.190 & $\leq$ & 0&.502 & $\leq$ & 0&.904\\
	\end{tabular} \\
\hline
\end{tabular}
\end{center}
\caption{KS-Test on the Neural Network, DJIA1982}
\label{tab:nnKolmo}
\end{table}

First, we observe an important variance, over the different paths, in the statistic $Z$: the standard deviation $\sigma_Z$ is not negligible in comparison with the mean $\tilde{Z}$. It is another evidence that the paths are not equivalent. We had observed a similar phenomenon on empirical data\footnote{See Chapter 2, ``Measure of kurtosis: Jarque-Bera Test''} when computing their kurtosis. It comes from the high heterogeneity of the dataset, which makes our tests less robust. But any test performed on this heterogeneous dataset would suffer from the same issue. Even though this apparent lack of consistency prevents us from drawing any strong and global conclusion, the knowledge of the mean and standard deviation $\tilde{Z} \pm \sigma_Z$ provides us with a fair overview of the statistic $Z$. 

On plots, Dragulescu and Yakovenko model seems to fit the empirical cumulative distribution better than the Gaussian. But in fact, on average, both models are rejected for high  frequencies (for $\tau=1 \ \mbox{and}\  5$ days) at the 0.01 level of significance. Even the Neural Network is rejected for a one day time lag. This rejection of the three models may come from the fact that this test is based on the maximum discrepancy between the empirical and the theoretical cumulative distributions, for \emph{any} x. To pass this test, a model must fit the observed data sufficiently well everywhere, i.e. in the tails (problem of fat tails) and in the middle (problem of high kurtosis for high frequencies) of the distribution. We will test the kurtosis and the tails of the models separately later in this Chapter. 

We point out that even if Dragulescu and Yakovenko model is rejected for a one day time lag, the statistic $Z$ is smaller than the Gaussian one (0.109 vs 0.131), which is an indication that the model fits the data a bit better. For other time lags, the p-value are equivalent: both models are systematically rejected for 5 days ($p \ll 0.01$), sometimes rejected for 20 days ($p(\tilde{Z}+\sigma_Z) \leq 0.01$, but $p(\tilde{Z}-\sigma_Z) \geq 0.05$) and never rejected for higher frequencies ($p \gg 0.05$). For medium and low frequencies, the fact that the simple hypothesis is not rejected does not guarantee that the composite hypothesis can be accepted. 

\subsubsection{Conclusion}

The Kolmogorov-Smirnov Goodness-of-Fit Test rejects both the Gaussian and Dragulescu and Yakovenko model for high frequencies ($\tau=1 \ \mbox{and}\  5$ days). For medium and low frequencies, we cannot conclude because of the theoretical limits of this test. To go on investigating, we need a more powerful statistical test that can be used even if the parameters of the model are derived from the tested dataset itself. The $\chi^2$ statistic is suitable in those conditions.

\subsection{$\chi^2$ Statistic}

\subsubsection{Introduction}

The $\chi^2$ Goodness-of-Git Test, based on binned data, is a powerful statistical tool to test if an empirical distribution comes from a given distribution.\footnote{See Appendix B, Section ``Chi-Square Goodness-of-Fit Test''} Contrary to the Kolmogorov-Smirnov test, it is designed to evaluate a composite hypothesis, i.e. the parameters of the model can be derivated from the empirical dataset tested. This test is a good trade-off between the goodness-of-fit of a model (the better fit, the smaller the $\chi^2$ statistic) and its complexity (the more complex, the larger number of parameters $m$). Indeed, even if a model fits the empirical data very well, a too large complexity may penalise its p-value, so that it still can be rejected.

Finally, to be meaningful, this test must be performed using relatively large bins, and a critical value of 5 expected observations per bin is regarded as a minimum.

\subsubsection{Methodology}

If we perform this test with equal size bins, then the fats tails will be trimmed (there are less than 5 expected log returns per bin in tails) and will not participate in the value of the statistic, making the the test inaccurate. Instead, we split the log return axis into \emph{equal expected frequency bins}, so that all of the log returns participate in the value of the statistic. We use an expected frequency of 5 log returns per bin. 

Unfortunately, this test cannot be performed for large time lags, because of the lack of data. Indeed, in the DJIA1982 index for instance, we have initially around 5000 close prices, which means that for a time lag of 250 days, each path will have only about 20 log returns. In those conditions, because of the critical value of 5 log returns per bin, we will have finally only 4 bins, which is too small to perform a relevant test.

\subsubsection{Results}

We present our results of the $\chi^2$ Goodness-of-Fit Test in Tables \ref{tab:normChi}, \ref{tab:draguChi} and \ref{tab:nnChi}. The degree of freedom is given by $df=noBins-1-m$, where $m$ is the number of parameters of the model ($m=2$ for the Gaussian, $m=4$ for Dragulescu and Yakovenko model, and $m=11$ for the Neural Networks if we count the weights and the biases). For large time lags, $df$ becomes smaller and smaller because $noBins$ decreases, as explained above. \\

\begin{table}[p]
\begin{center}
\begin{tabular}{|c|c|c|c|}
\hline
time lag & $\tilde{\chi^2}$ & df & p-value \\ 
\hline
\begin{tabular}{r}
	1 \\
	5 \\
	20 \\
	40 \\
	80 \\
	\end{tabular} &
\begin{tabular}{r@{}ll} % norm
	1790&& \\
	255&& $\pm$30\\
	61&& $\pm$12\\
	29&.1 &$\pm$7.0\\
	10&.4 &$\pm$4.6\\
	\end{tabular} & 
\begin{tabular}{r}
	1010 \\
	198 \\
	47 \\
	22 \\
	9 \\
	\end{tabular} & 
\begin{tabular}{r@{}lrr@{}lrr@{}l}
	&&& 6&.29e-11 &&&\\
	5&.38e-05 & $\leq$ & 4&.07e-03 & $\leq$ & 0&.0931\\
	7&.99e-03 & $\leq$ & 0&.0819 & $\leq$ & 0&.409\\
	0&.0295 & $\leq$ & 0&.141 & $\leq$ & 0&.451\\
	0&.0915 & $\leq$ & 0&.32 & $\leq$ & 0&.76\\
	\end{tabular} \\
\hline
\end{tabular}
\end{center}
\caption{$\chi^2$ Test on the Gaussian, DJIA1982}
\label{tab:normChi}
\end{table}

\begin{table}[p]
\begin{center}
\begin{tabular}{|c|c|c|c|}
\hline
time lag & $\tilde{\chi^2}$ & df & p-value \\ 
\hline
\begin{tabular}{r}
	1 \\
	5 \\
	20 \\
	40 \\
	80 \\
	\end{tabular} &
\begin{tabular}{r@{}ll} % dragu
	\tt 1420 \\
	\tt 244& & $\pm$26 \\
	\tt 48&.5 & $\pm$11.5 \\
	\tt 27&.3 & $\pm$6.1 \\
	\tt 9&.7 & $\pm$4.4 \\
	\end{tabular} & 
\begin{tabular}{r}
	\tt 1000 \\
	\tt 196  \\
	\tt 45 \\
	\tt 20  \\
	\tt 7 \\
	\end{tabular} & 
\begin{tabular}{r@{}lrr@{}lrr@{}l}
	&&& 1&.16e-04 &&& \\
	3&32e-04 & $\leq$ & 0&.0108 & $\leq$ & 0&.133\\
	0&.0663 & $\leq$ & 0&.333 & $\leq$ & 0&.796\\
	0&.0301 & $\leq$ & 0&.126 & $\leq$ & 0&.385\\
	0&.049 & $\leq$ & 0&.206 & $\leq$ & 0&.624\\
	\end{tabular} \\
\hline
\end{tabular}
\end{center}
\caption{$\chi^2$ Test on Dragulescu model, DJIA1982}
\label{tab:draguChi}
\end{table}

\begin{table}[p]
\begin{center}
\begin{tabular}{|c|c|c|c|}
\hline
time lag & $\tilde{\chi^2}$ & df & p-value \\ 
\hline
\begin{tabular}{r}
	1 \\
	5 \\
	20 \\
	40 \\
	80 \\
	\end{tabular} &
\begin{tabular}{r@{}ll} % nn
	\tt 2230 \\
	\tt 232& & $\pm$38 \\
	\tt 45&.9 & $\pm$11.1 \\
	\tt 21&.5 & $\pm$6.3 \\
	\tt 7&.6 & $\pm$6.3 \\
	\end{tabular} & 
\begin{tabular}{r}
	\tt 997 \\
	\tt 189 \\
	\tt 38 \\
	\tt 13 \\
	\tt 0 \\
	\end{tabular} & 
\begin{tabular}{r@{}lrr@{}lrr@{}l}
	&&& 0&.0839 &&& \\
	0&.0559 & $\leq$ & 0&.346 & $\leq$ & 0&.817\\
	0&.0473 & $\leq$ & 0&.25 & $\leq$ & 0&.688\\
	0&.0057 & $\leq$ & 0&.0552 & $\leq$ & 0&.333\\
	&NaN & $\leq$ & &NaN & $\leq$ & &NaN\\
	\end{tabular} \\
\hline
\end{tabular}
\end{center}
\caption{$\chi^2$ Test on the Neural Network, DJIA1982}
\label{tab:nnChi}
\end{table}

Concerning the Neural Network, we cannot perform this test for time lags higher than 40 days, or else the degree of freedom decreases to zero. This is due to the relatively high number of parameters ($m=11$). With a structure even more complicated, we could not have performed the test at all, except for high frequencies. \\

First we notice that the Neural Network's $\chi^2$ statistic is slightly smaller than Dragulescu's one, itself smaller than the Gaussian's one, for all paths with a time lag from $\tau=1$ to $\tau=80$ days. It means that the Neural Networks fits empirical data better than Dragulescu model, which itself is better than the Gaussian. But there is a price to pay, in terms of complexity: due to too many parameters (and then a lower degree of freedom), the p-value of the Neural Networks and Dragulescu model are not systematically higher than the p-value of the Gaussian. And it is precisely the p-value that is used to accept or reject a model, not directly the $\chi^2$ statistic. \\

If we look at the p-value in detail, we observe that
\begin{itemize}
\item For $\tau=1$, the three models are rejected at a 0.05 level of significance
\item For $\tau=5$, only the Neural Network is systematically accepted. The Gaussian and Dragulescu model are only accepted in the best situation ($p(\tilde{\chi^2}) < 0.05 < p(\tilde{\chi^2}-\sigma_{\chi^2})$)
\item For $\tau=20$, the three models are accepted and Dragulescu model is better than the Neural Network and the Gaussian
\item For $\tau=40$ and $\tau=80$, Dragulescu model is still accepted, but its p-value is smaller than the one of the Gaussian 
\end{itemize}

\subsubsection{Conclusion}

Thanks to the $\chi^2$ Goodness-of-Fit Test, we can assert that Dragulescu and Yakovenko model fits empirical data slightly better than the Gaussian, for high and medium frequencies.
Nevertheless, both models are rejected for high frequencies ($\tau=1 \ \mbox{and}\  5$ days), at a 0.05 level of significance. In this sense, these results are consistent with the Kolmogorov-Smirnov Goodness-of-Fit Test.

We also observe a clear shift in the goodness-of-fit of the models around $\tau=40$ days: the probability of accepting the Gaussian becomes larger than the probability to accept Dragulescu model (and even the Neural Network) due to the lower complexity of the Normal model (two parameters instead of four and eleven respectively). \\

To put it in a nutshell, using a complex model, such as the Dragulescu model or a Neural Network, is only worth for $\tau=1, 5 \ \mbox{and} \ 20$ days. For lower frequencies ($\tau \geq 40$ days), the Gaussian is preferable because it is simpler. Given that for these frequencies, we had observed neither fat tails nor kurtosis in the empirical datasets, the Gaussian represents the best trade-off between goodness-of-fit and complexity.  

\subsection{Measure of kurtosis}

\subsubsection{Introduction}

As attested by Figures \ref{fig:fig-cdf} and \ref{fig:fig-cdfTails} and by the results of the $\chi^2$ Goodness-of-Fit Test, the Dragulescu and Yakovenko model fits their empirical data slightly better than the Gaussian, for any time lag. Nevertheless, both models are rejected for high frequencies, characterised by prominent fat tails and high kurtosis, as exposed in Section 2. Hence, we should try to find out if the models are rejected mainly because of fat tails, kurtosis, or both. First, let us have a look at the kurtosis, as it is easy to test. We will concentrate on the tails in the next Section. 

\subsubsection{Methodology}

We perform our tests on the DJIA1982 index, without reusing or trimming the data. For each time lag, we compute the log returns dataset, and we divide it into paths. For each path $i$, we start by computing the observed kurtosis, exactly as we did in Chapter 2. Then, for each model, we build the PDF (\emph{empPDF}, \emph{normPDF}, \emph{draguPDF} or \emph{nnPDF}, where \emph{empPDF} is the empirical PDF), and use it to generate random data, i.e. plausible log returns time series. More precisely, we generate $N=100$ random datasets of $noLogReturns$ elements, where $noLogReturns$ is the number of log returns in the initial paths from which we derivated the model distribution. Finally, we compute the kurtosis of these time series and obtain, for each path $i$, a mean value $\tilde{k_i}$ and a standard deviation $\sigma_{k_i}$ over the $N$ simulations. \\

We already know that the empirical data exhibit kurtosis mainly for $\tau=1 \mbox{and} \ 5$ days, and that even for those time lags, they do not exhibit kurtosis consistently for each path\footnote{See Chapter 2, Section ``Measure of kurtosis: Jarque-Bera Test''}. As a consequence, we expect a good model to produce kurtosis only when the empirical data does. For each path, we give in Tables \ref{tab:simulKurtosis1} and \ref{tab:simulKurtosis2} the average kurtosis $\tilde{k_i}$, and its standard deviation $\sigma_{k_i}$, produced by the $N$ simulations. 

We expect almost the same results as in Table \ref{tab:kurtosis1} for \emph{empPDF}, no kurtosis for \emph{normPDF}, and similar kurtosis as in Table \ref{tab:kurtosis1} for \emph{draguPDF} and \emph{nnPDF}, since these models are supposed to fit the data sufficiently well.

\subsubsection{Results}

\begin{table}[htp]
\begin{center}
\begin{tabular}{ccccc}
observed & from \emph{empPDF} & from \emph{normPDF} &from \emph{draguPDF} &from \emph{nnPDF} \\ 
69.27 & 22.21 $\pm$ 20.48 &-0.01 $\pm$0.07 &107.69 $\pm$ 16.66 & 30.55 $\pm$ 24.9964 \\
\end{tabular}
\end{center}
\label{tab:simulKurtosis1}
\caption{Kurtosis of generated data, DJIA1982, $\tau=1$ day}
\end{table}

\begin{table}[htp]
\begin{center}
\begin{tabular}{ccccc}
observed & from \emph{empPDF} & from \emph{normPDF} & from \emph{draguPDF} & from \emph{nnPDF} \\ 
\begin{tabular}{r@{}l}
    38&.20 \\
    30&.49 \\
    5&.91  \\
    5&.77  \\
    3&.99  \\
	\end{tabular} &
\begin{tabular}{r@{}l}
    14&.87 $\pm$ 13.35 \\
    10&.06 $\pm$ 10.17 \\
     4&.12 $\pm$ \ 1.76 \\
     3&.13 $\pm$ \ 1.97 \\
     3&.08 $\pm$ \ 1.30  \\
	\end{tabular} &
\begin{tabular}{r@{}l}
    0&.07 $\pm$ 0.20 \\
    0&.11 $\pm$ 0.24 \\
   -0&.03 $\pm$ 0.14 \\
   -0&.05 $\pm$ 0.14 \\
    0&.03 $\pm$ 0.17 \\
	\end{tabular} &
\begin{tabular}{r@{}l}
    2&.35 $\pm$ 1.99 \\
    3&.38 $\pm$ 4.40 \\
    1&.70 $\pm$ 0.68 \\
    1&.80 $\pm$ 1.03 \\
    2&.04 $\pm$ 0.59 \\
	\end{tabular} &
\begin{tabular}{r@{}l}
   28&.43 $\pm$ 13.12 \\
   14&.74 $\pm$ 10.72 \\
    5&.10 $\pm$ \ 2.31 \\
    1&.63 $\pm$ \ 0.44 \\
    3&.08 $\pm$ \ 1.09 \\
	\end{tabular} \\
\end{tabular}
\end{center}
\label{tab:simulKurtosis2}
\caption{Kurtosis of generated data, DJIA1982, $\tau=5$ day}
\end{table}

The variance in the kurtosis is extremely important, for observed data and for generated random data. Because the aim of this thesis is to compare a model with observed data, the analysis of the origin itself of this variance is out of our scope. Actually, we are only interested here in the capacity of different models to reproduce or not a high kurtosis, when it is exhibited by observed data. 

As expected, the Gaussian never exhibits kurtosis, since by definition the kurtosis is a departure from normality. Moreover, the simulated time series generated from the empirical PDF \emph{empPDF} and the Neural Network \emph{nnPDF} exhibit high kurtosis in accordance with observed data, even if their kurtosis is in general smaller ($22.21 \pm 20.48 < 69.27$ for $\tau=1$, $14.87 \pm 13.35 < 38.20$ and $10.06 \pm 10.17 < 30.49$ for $\tau=5$). 

\begin{figure}[h]
\begin{center}
\includegraphics[width=10cm]{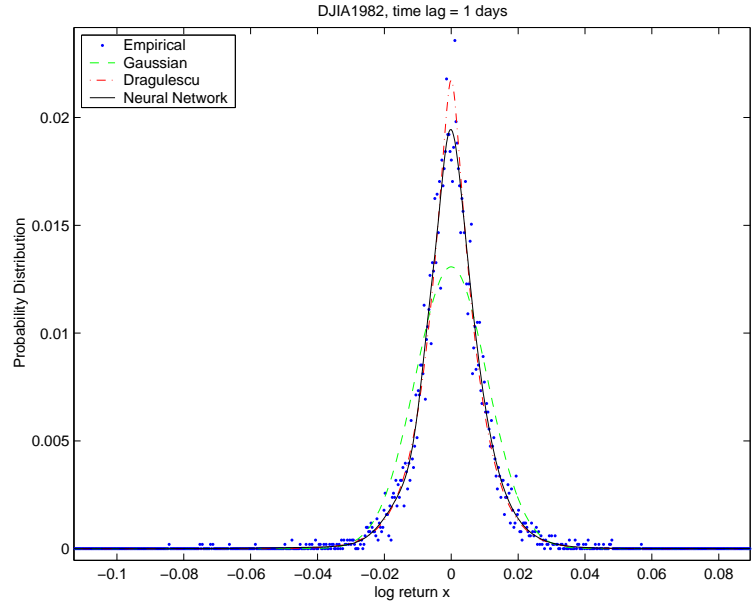}
\end{center}
\caption{Empirical and expected PDFs, $\tau=1$ day}
\label{fig:draguKurt}  
\end{figure}

To the opposite, random data generated from \emph{draguPDF} exhibit a very high kurtosis for a one day time lag, even higher than expected ($107.69 \pm 16.66 > 69.27$), as attested by the high peakedness of the distribution in Figure \ref{fig:draguKurt}. Moreover, for a five day time lag, this model clearly fails to produce kurtosis, whatever the path, as shown by column (4) of Table \ref{tab:simulKurtosis2}. 

This could explain why Dragulescu and Yakovenko model is rejected by both the Kolmogorov-Smirnov and the $\chi^2$ goodness-of-fit tests for high frequencies, whereas the Neural Network, for instance, is not rejected for a five day time lag. 

\subsubsection{Conclusion}

Our conclusion is that the Dragulescu and Yakovenko model does not handle correctly the kurtosis: it exhibits too high kurtosis for a one day time lag, and not enough for a five day time lag. In this sens, although it is better than the Gaussian because at least, it can produce some kurtosis, this model is still rejected for high frequencies. In terms of plot, it is generated by a too large probability mass in the centre and in the tails of the distribution, which is translated by an important peakedness and by fat tails. We will focus on the fat tails in next Section.   

\subsection{Measure of the fat tails}

\subsubsection{Introduction}

Fat tails are very difficult to handle, because they correspond to exceptional events, statistically not significant, but terribly important for stock markets. Indeed, they are caused by crashes and bubbles that happen far more often than predicted by the Gaussian. By trimming the data, Dragulescu and Yakovenko removed some of them, which explains why their plots look so smooth. Using equal expected frequency bins in the $\chi^2$ goodness-of-fit test was the only way to keep them. We will investigate in this Section these extreme events and try to capture them as precisely as possible. 

\subsubsection{Methodology}

In 1962, E. Fama defended Mandelbrot's stable-Paretians against the Gaussian specially because stable-Paretians could produce fat tails. He was the first to exhibit clearly those fat tails, and used a simple test (among others) to do so. The idea is to count the number of outliers, viz. the number of log returns outside $\mu \pm 2\sigma$, $\mu \pm 3\sigma$ and $\mu \pm 4\sigma$. If the log returns really followed a normal distribution, then the number of outliers should be respectively $noLogReturns*0.0455, noLogReturns*0.0027$ and $noLogReturns*0.0000063$, where $noLogReturns$ is the total number of log returns in the given path. If we compare this expected value with the real number of outliers for \emph{normPDF}, then we should capture the fat tails. In Tables \ref{tab:tails1}, \ref{tab:tails2} and \ref{tab:tails3}, we indicate, for a given level of deviation ($\mu \pm 2, 3, \ \mbox{and} \ 4 \sigma$) , the expected number of outliers \emph{if the log returns were normally distributed} in Column (2) and the observed (regarding \emph{normPDF}) number of outliers in Column (3). For instance, for a one day time lag, if the log returns followed a normal distribution, we would expect around 13.63 out of 5049 log returns outside the boundaries $\mu \pm 3\sigma$. We observed, however, 50 outliers (i.e. observations outside three standard deviations of \emph{normPDF}). It indicates that the Gaussian dramatically underestimates the number of outliers.

\subsubsection{Results}

\begin{table}[p]
\begin{center}
\begin{tabular}{ccc}
time lag & expected & observed in \emph{normPDF} \\ 
\begin{tabular}{r}
	1 \\
	5 \\
	20 \\
	40 \\
	80 \\
	100 \\
	200 \\
	250 \\
	\end{tabular} &
\begin{tabular}{r@{}l}
229&.72 \\ 
45&.90  \\
11&.42  \\
5&.68   \\ 
2&.82   \\ 
2&.23   \\ 
1&.09   \\ 
0&.86   \\ 
	\end{tabular} &
\begin{tabular}{r}
205 \\ 
30  \\
12  \\
5   \\ 
2   \\ 
2   \\ 
2   \\ 
1   \\ 
	\end{tabular} \\
\end{tabular}
\end{center}
\caption{Expected and observed number of outliers in fat tails, out of $\mu \pm 2\sigma$}
\label{tab:tails1}
\end{table}

\begin{table}[p]
\begin{center}
\begin{tabular}{ccc}
time lag & expected & observed in \emph{normPDF} \\ 
\begin{tabular}{r}
	1 \\
	5 \\
	20 \\
	40 \\
	80 \\
	100 \\
	200 \\
	250 \\
	\end{tabular} &
\begin{tabular}{r@{}l}
13&.63 \\ 
2&.72  \\
0&.67  \\
0&.33  \\
0&.16  \\
0&.13  \\
0&.06  \\
0&.05  \\
	\end{tabular} &
\begin{tabular}{r}
50 \\ 
7  \\
2  \\
1  \\
1  \\
1  \\
0  \\
0  \\
	\end{tabular} \\
\end{tabular}
\end{center}
\caption{Expected and observed number of outliers in fat tails, out of $\mu \pm 3\sigma$}
\label{tab:tails2}
\end{table}

\begin{table}[p]
\begin{center}
\begin{tabular}{ccc}
time lag & expected & observed in \emph{normPDF} \\ 
\begin{tabular}{r}
	1 \\
	5 \\
	20 \\
	40 \\
	80 \\
	100 \\
	200 \\
	250 \\
	\end{tabular} &
\begin{tabular}{r@{}l}
0&.3181  \\ 
0&.0636  \\
0&.0158  \\
0&.0079  \\
0&.0039  \\
0&.0031  \\
0&.0015  \\
0&.0012  \\
	\end{tabular} &
\begin{tabular}{r@{}l}
22 \\ 
5  \\
2  \\
1  \\
0  \\
1  \\
0  \\
0  \\
	\end{tabular} \\
\end{tabular}
\end{center}
\caption{Expected and observed number of outliers in fat tails, out of $\mu \pm 4\sigma$}
\label{tab:tails3}
\end{table}

In Table \ref{tab:tails1}, the real number of outliers is systematically inferior to the expected number: it means that we don't have fat tails at a level of $2\sigma$. In fact the tails are not fatter than expected by a Gaussian. But for high frequencies ($\tau=1 \ \mbox{and} \ 5$ days), fat tails appear at a level of $3\sigma$ ($50 \gg 13.63, 7>2.72$) and $4\sigma$ ($22 \gg 0.32, 5 \gg 0.06$). For medium frequencies, the expected number of outliers is inferior to one, and the real is around one or two: these log returns correspond to extremely rare events, that appear very far from the mean (more than 4 standard deviations!); we classify them as long tails. 

To summarise, for high frequencies, the Gaussian exhibits fat tails outside $\mu \pm 3 \sigma$ and long tails after $\mu \pm 4 \sigma$. Fat tails correspond to crashes (bubbles) and occur far more often than predicted, whereas long tails correspond to exceptional huge crashes (resp. huge bubbles). For medium frequencies, the Gaussian exhibits long tails at $\mu \pm 3 \sigma$, but no fat tails. Finally, for low frequencies, the Gaussian exhibits neither fat nor long tails.

\begin{figure}[p]
\begin{center}
\includegraphics[width=10cm]{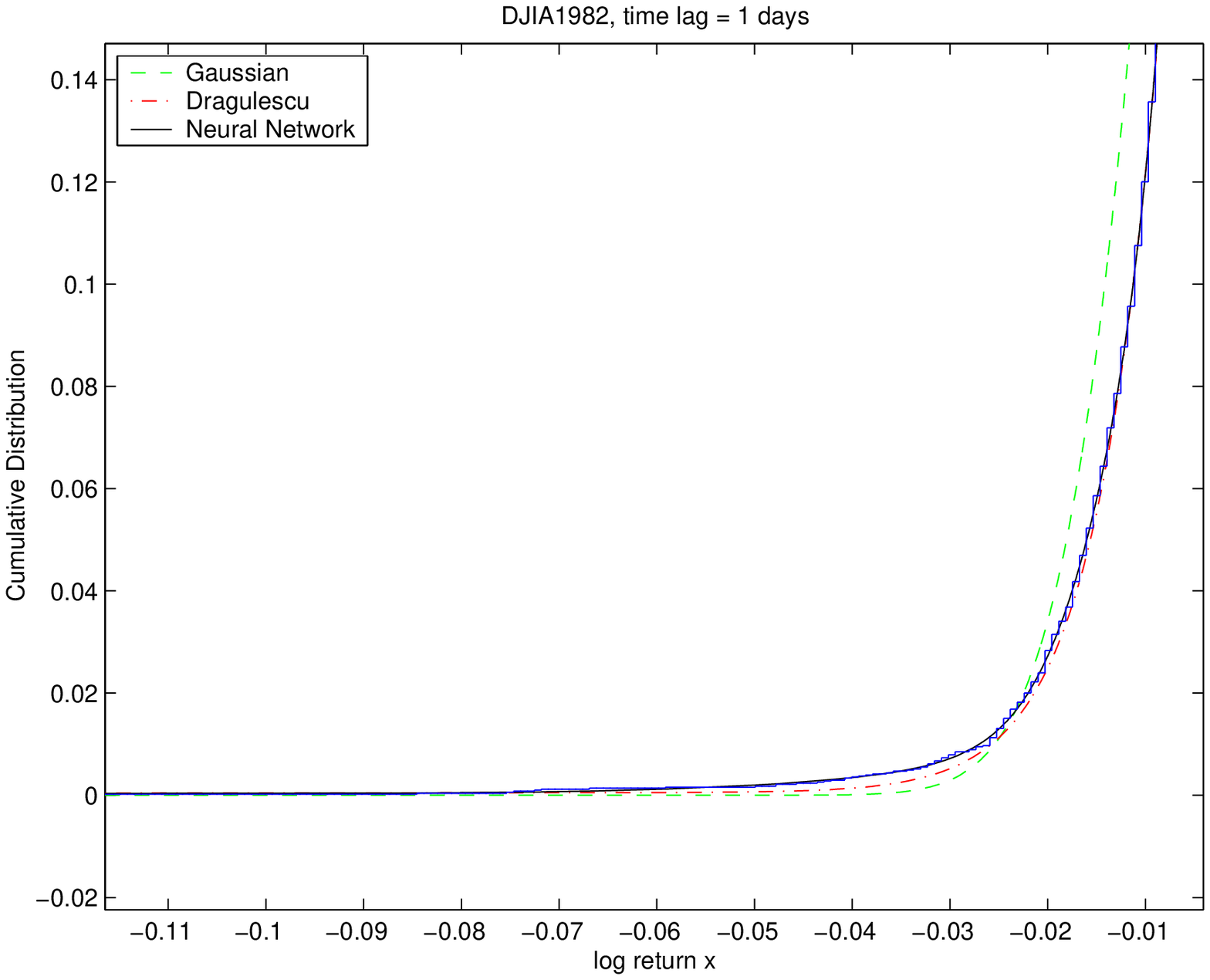}
\end{center}
\caption{Empirical and expected CDFs, $\tau=1$ day}
\label{fig:draguTails1}  
\end{figure}

Unfortunately, the tails are not as easy to isolate statistically for the other models, Dragulescu and the Neural Network. Nevertheless, given that both these models outperform the Gaussian, we expect them to fit the tails a bit better. We can verify that by a mere observation of the plots, for instance the left tails (that corresponds to crashes) of the CDFs of the different models, in Figure \ref{fig:draguTails1}. 

\begin{figure}[p]
\begin{center}
\includegraphics[width=10cm]{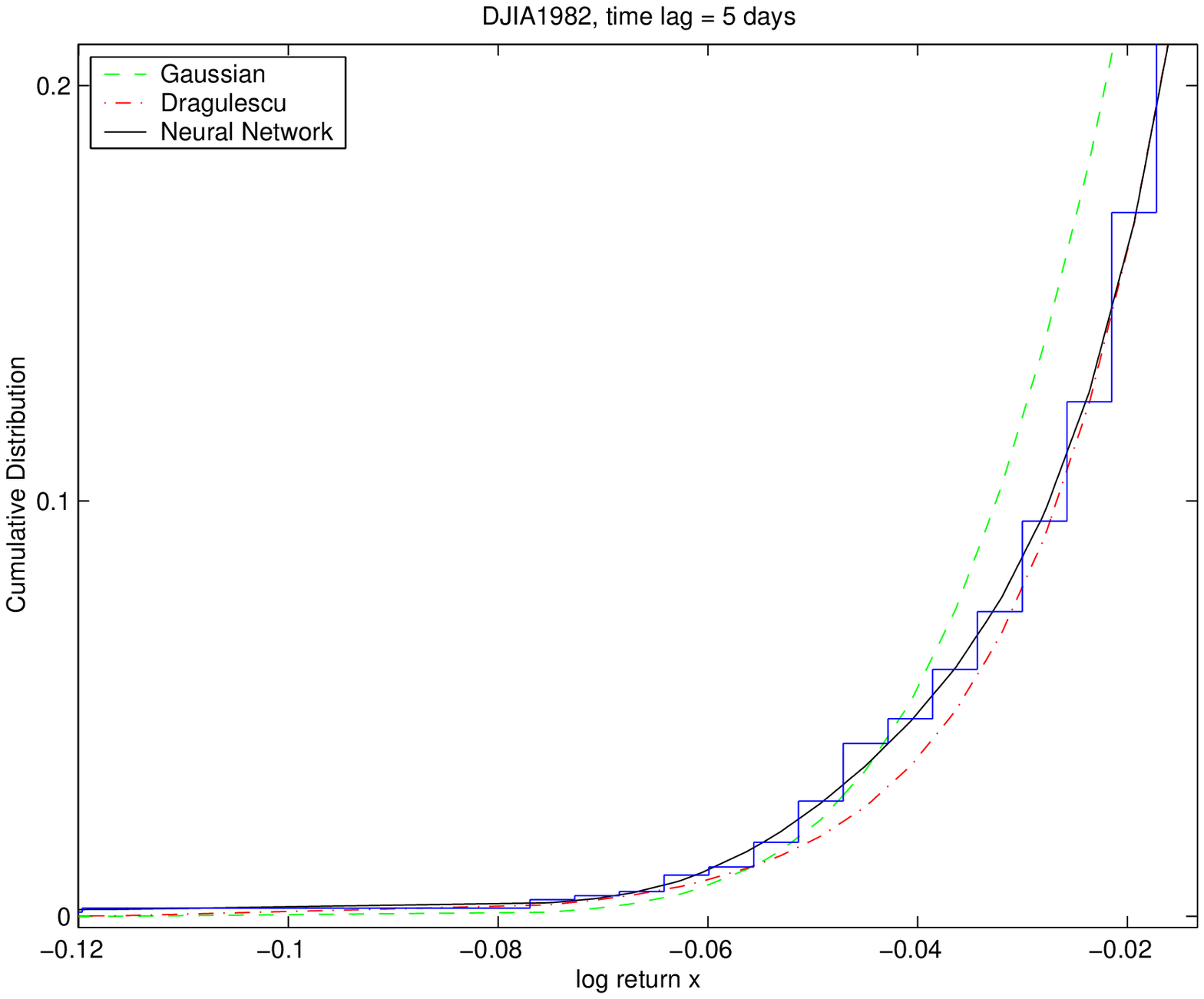}
\end{center}
\caption{Empirical and expected CDFs, $\tau=5$ day}
\label{fig:draguTails2}  
\end{figure}

\begin{figure}[h]
\begin{center}
\includegraphics[width=10cm]{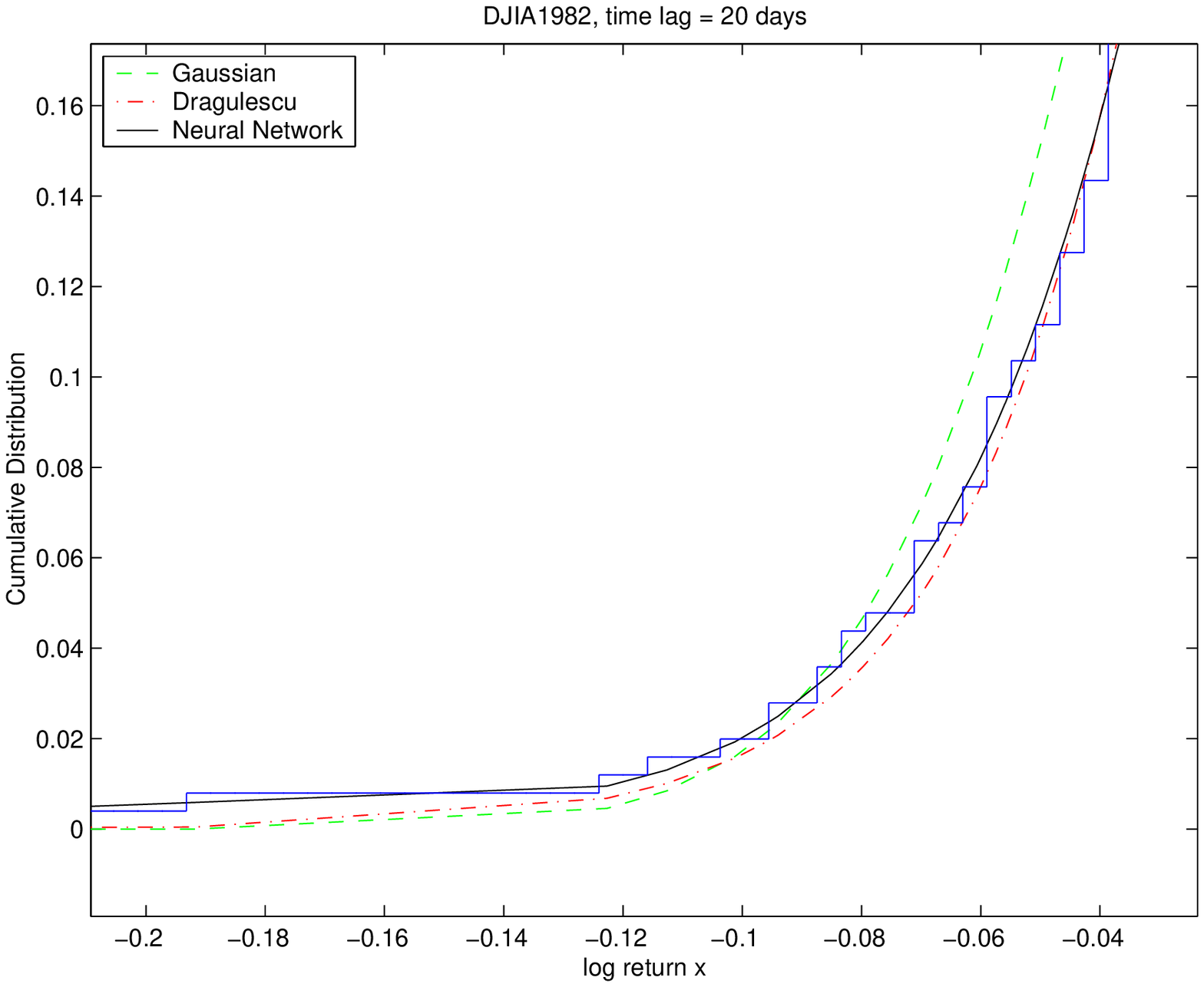}
\end{center}
\caption{Empirical and expected CDFs, $\tau=20$ day}
\label{fig:draguTails3}  
\end{figure}

We clearly see on this plot the poor fit of the Gaussian, the slightly better fit of the Dragulescu and Yakovenko model, and the very good fit of the Neural Network. But we have to keep in mind that the complexity of those models is greater as well, which explains why the Gaussian is a still preferable for medium and low frequencies. For five days, this phenomenon still subsists, but is less prominent (\ref{fig:draguTails2}). Finally, by a way of comparison, we plot the same figure for medium frequencies ($\tau=20$ days), where all of the models are accepted(\ref{fig:draguTails3}). We can see that gradually, the difference of fit between the Dragulescu and Yakovenko distribution and the Gaussian becomes smaller and smaller.

\section{Conclusion}

We performed in this Chapter many tests on the Dragulescu and Yakovenko model, claimed to outperform the Gaussian distribution (Bachelier-Osborne model) for any time lag. We used a simple Neural Network as a benchmark. Our first conclusion was that the data should be neither re-used nor trimmed, as the authors do, because it makes them so smooth that any model could fit them sufficiently well to be accepted, even the Gaussian itself. Then, thanks to two goodness-of-fit tests, based on the Kolmogorov-Smirnov and the Chi-Square Statistics, we could reject Dragulescu and Yakovenko model for high frequencies ($\tau=1 \ \mbox{and} \ 5$ days), mainly because it is unable to produce the correct kurtosis, even if the fit of fats tails is improved. 

For medium and low frequencies, the fit of the model is better than the Gaussian, but the price to pay in terms of complexity is too high (we introduced four parameters instead of two), so that finally the Normal distribution appears to be the best trade-off between goodness-of-fit and complexity.

\chapter{Conclusions and further work}

\section{Conclusions}

The goal of this research was to evaluate quantitatively a new model of stock market returns, based on a stochastic volatility. \\

To accomplish such task, we introduced, during early stages, the notion of a Random Walk, followed by an overview of the Geometrical Brownian Motion, the mathematical theory underlying the statistical models of stock markets, and in particular the classical Bachelier-Osborne model, according to which the log returns follow a Normal distribution (a Gaussian). \\

At this point, two principal departures from this model were identified and analysed: fat tails and kurtosis, exhibited by empirical data mainly for high frequencies (time lag $\tau=1 \ \mbox{and} \ 5$ days). \\

This evaluation was followed by a description of the new model, published in March 2002 by Dragulescu and Yakovenko \cite{Dragu}, and claimed to outperform the Gaussian for any time lag.
Statistical tests, such as the Jarque-Bera test and Lilliefors test, and goodness-of-fit tests, based on the Kolmogorov-Smirnov statistic and the $\chi^2$ statistic, were then performed. \\

First, we demonstrated that the way Dragulescu and Yakovenko trim and reuse their dataset is unfair, and we proposed our own methodology, based on the conservation of all the data. \\

Then, we found out that their model effectively fits the empirical data slightly better than the Gaussian for any time lag. Nevertheless, the Gaussian is preferable to any more complicated model (Dragulescu model \emph{or} a Neural Network) for medium and low frequencies (time lag $\tau \geq 40$ days), essentially because empirical data  exhibit neither fat tails nor kurtosis for these frequencies. Hence, the Gaussian represents the best trade-off between goodness-of-fit and complexity. \\

Finally, we tried to investigate why both models were rejected for high frequencies (time lag $\tau=1$ and $5$ days) at a 0.05 level of significance, and concentrated first on the kurtosis and then on the fat tails.

\section{Further work}

All along this research, we realised that none of the statistical model could handle the extreme events, huge crashes or bubbles, that generate long tails, beyond the fat tails. We reach here one of the limits of a statistical approach, where the expected probability density is unbounded, contrary to the observed one. If used in risk management or option pricing, those models, to be complete, should absolutely be coupled with \emph{ad hoc} rules for extreme events. Indeed, people tend to overweight outcomes that are considered certain, which lead Dragulescu and Yakovenko to trim the long tails, considered as too rare to deserve attention. This is known as the \emph{certainty effect}, described by D. Kahneman and A. Tversky in their Prospect Theory  about decision under risk\cite{ProspectTheory}. And just like the classical Expected Utility Theory cannot handle this \emph{certainty effect}, classical statistical models cannot handle the phenomenon of long tails. \\

This leads us to a second possible improvement: with the recent amazing expansion of Artificial Intelligence techniques, a new class of models, called ``agent models'' appeared. Those models are based on the representation of clusters of traders who can communicate and interact with each others, following very simple rules. Here, the traditional i.i.d. hypothesis for stock returns is rejected, and replaced by basic assumptions about the inter-dependence of traders. Those models, \cite{HerdBehavior, Edmonds}, have a strong explanatory power, contrary to the statistical ones, and can handle very extreme events that constitute the long tails. We might investigate these models very soon, as a PhD student.

\bibliography{dissertation}             % this causes the references to be
                                % listed

\bibliographystyle{alpha}       % this determines the style in which
                                % the references are printed, other
                                % possible values are plain and abbrv
% Appendices start here
\appendix
\chapter{Elements of stochastic calculus}

\section*{Markov processes}

The stochastic processes we consider here are random vectors that depend on time t and take their value in $\Re^n$: $$ \underline{X}(t) \in \Re^n $$ where $\underline{X}$ represents a $n*1$ vector. We suppose that $\underline{X}(t)$ represents the state of the world at time t, i.e. contains all of the characteristics of the market (price, mean, volatility, etc.).

Let us consider a process $\underline{X}(t)$, a series of successive instants $t_0$, $t_1$, ..., $t_{m-1}$ in [0, T], real vectors $\underline{x}_0$, $\underline{x}_1$, ... $\underline{x}_{m-1}$ and $\underline{x}$, and the probability that $\underline{X}(t_m)$ is inferior or equal to $\underline{x}$ given that $\underline{X}(t_0)=\underline{x}_0$, $\underline{X}(t_1)=\underline{x}_1$, ...,$\underline{X}(t_{m-1})=\underline{x}_{m-1}$; this conditional probability is written: 

$$
P(\underline{X}(t_m)\leq\underline{x} \ | \ \underline{X}(t_0)=\underline{x}_0, \ \underline{X}(t_1)=\underline{x}_1, ..., \underline{X}(t_{m-1})=\underline{x}_{m-1})
$$

In a Markov process, by definition, the conditional probabilities of $\underline{X}(t_m)$ depend only on ($t_{m-1}$, $\underline{x}_{m-1}$), whatever the series ($t_0$, $t_1$, ..., $t_m$) and $\underline{x}$ are. To sum up, given the present, the future is conditionally independent of the past.

The conditional probability above is then simplified:

$$
P(\underline{X}(t_m)\leq\underline{x} \ | \  \underline{X}(t_{m-1})=\underline{x}_{m-1})
$$ 

This definition is valid in two distinct situations: the state of the system can be defined for discrete values of the time $t$ , or it can be defined for any time in a given interval ($\forall t \in [0,T]$). Processes of the first type are called "discrete", whereas processes of the second type are called "processes in continuous time".

Processes in continuous time can be split in two categories:``continuous processes'' and ``jump processes''.

In continuous processes, also called ``diffusion processes'' or ``It\^o processes'', only infinitesimal variations of $\underline{X}$ are possible during the time interval dt, whereas jump processes are characterised by some discontinuities.

Many models have been developed to study the time series of the price, mean, volatility, etc. of securities. Continuous and jump processes have been used frequently.   

\section*{Brownian motion}

The theory of diffusion processes has been developed by mathematicians and physicists at the end of $XVIII^{th}$ century. A diffusion process can be seen as the limit of a discrete Markov process. This vision enables us to highlight the different properties of such processes and to justify there use in Financial Market models. First, let us have a look at a specific diffusion process: the Brownian motion.

The description "Brownian motion" comes from the fact that the same process describes the physical motion of a particle subject to random shocks, a phenomenon first noted by the British physicist Brown in 1828, observing irregular movement of pollen suspended in water.

Consider a unidimensional discrete Markov process observed at different times, with values $X(0), X(1), ..., X(t), ...,$ characterised by independent and identically distributed (i.i.d.) increments of mean $\mu$ and standard deviation $\sigma$. This particular process, called "Brownian motion", is used in Physics to describe the motion of a particle in suspension, and in Financial Markets, the evolution of security yields.

$$
N(t) \equiv \frac{X(t+1) - X(t) - \mu}{\sigma}
$$

Then $\forall t \leq T$, $E[N(t)]=0$, $Var[N(t)]=1$ and $N(t)$ are i.i.d. $X$ follows the equation

$$
X(t+1) = X(t) + \mu + \sigma N(t)
$$

Let us consider now the partition of the interval $[t, t+1]$ constituted of $n$ sub-intervals of length $l=1/n$ $[t+(i-1)l, t+il] \forall i=1, ...,n$.
Let us assume that the variation $X(t+1)-X(t)$ results in the sum of the $n$ elementary variations $\Delta_iX$ of each sub-interval:

$$
X(t+1) - X(t) = \sum_{i=1}^{n}X(t+il)-X(t+(i-1)l) = \sum_{i=1}^{n}\Delta_iX
$$

To conclude, let us assume that the elementary variations $X(t+il)-X(t+(i-1)l) \equiv \Delta_iX$ are i.i.d. themselves, with mean $\lambda$ and standard-deviation $\upsilon$:

$$
\Delta_iX = \lambda + \upsilon U(i)
$$

\begin{tabular}{ll}
where & $E[(U(i)]=0$ \\
&$Var[U(i)]=1$ \\
&$U(i)$ are i.i.d. \\
\end{tabular}

Then:

\begin{itemize}
\item $E[X(t+1)-X(t)] = \sum_{i=1}^{n}E(\Delta_iX) = n \lambda = \mu$ so $\lambda = \frac{\mu}{n}= \mu l$
\item $Var[X(t+1)-X(t)] = \sum_{i=1}^{n}Var(\Delta_iX) = n\upsilon^2 = \sigma^2$ so $\upsilon=\frac{\sigma}{\sqrt{n}}=\sigma \sqrt{l}$ since $\upsilon \geq 0$ and $\sigma \geq 0$ 
\end{itemize}

As a consequence:

\begin{equation} \label{eqn:Brownian}
\Delta_iX \equiv X(t+il) - X(t+(i-1)l) = \mu l + \sigma \sqrt{l} U(i)
\end{equation}

This means that:
\begin{itemize}
\item the increments of $X$ have a mean and variance constant by unit of time equal to $\mu$ and $\sigma^2$ respectively. Mean and variance of $X(t)-X(s)$ are then proportional to $t-s, \forall t<s$
\item to the limit, if $X$ is a continuous Markov process, $X(t)$ is normally distributed. Indeed, when the number $n$ of sub-intervals tends to infinity:
$$
X(t+1)-X(t) = lim_{n \rightarrow +\infty} \sum_{i=1}^{n}\Delta_iX
$$
Given that the $\Delta_iX$ are i.i.d., the central limit theorem (CLT) states that the increments $X(t+1)-X(t)$ are normally distributed. Furthermore, if we consider that the elementary variations $\Delta_iX$ are themselves made of an infinite number of i.i.d. variations on even smaller intervals, then the $\Delta_iX$ are normally distributed themselves.
\end{itemize}

\section*{Wiener processes}

The first mathematically rigorous construction of Brownian motion was carried out by Wiener in 1923.

In the limit, \ref{eqn:Brownian} takes the form of the following differential equation:

\begin{equation} \label{eqn:wiener}
dX = \mu dt + \sigma dW
\end{equation}

\begin{tabular}{ll}
where & $dW = U(t)\sqrt{t}$ \\
& $U(t)$ standard normal, independent of U(t') for t $\neq$ t' \\
\end{tabular}

$\mu$ and $\sigma^2$ are called respectively the instantaneous expectation (or "drift") and instantaneous variance of $X$. The process $W$, which increments are independent and normally distributed with a null expectation and an instantaneous variance of 1, is called a "Wiener process" or a "standardised Brownian motion". The trajectories of the Brownian motion $X$ are:

\begin{itemize} 
\item continuous
\item not derivable nearly surely 
\end{itemize}

Hence, trajectories of $X$ are continuous but characterised by a change of slope at each time. Moreover, the process is stationary. The properties of the Wiener processed are described in \cite{Poncet}

The Brownian motion is a process whose increments are i.i.d., following a Gaussian distribution with constant instantaneous expectation and variance. It can be used when the motion of a system results of a constant strength that imposes a drift ($\mu dt$) associated with a succession of random and independent shocks ($\sigma dW$) that impose erratic motions.

We can generalise \ref{eqn:wiener} easily to a multidimensional vector $\underline{X}(t)$:

\begin{equation} \label{eqn:wiener-gen}
d\underline{X} = \underline{\mu} dt + \underline{\underline{\sigma}} d\underline{W}(t)
\end{equation}

\begin{tabular}{ll}
where & $\underline{\mu}$ is a constant vector, $\underline{\mu} \in \Re^n$ \\
      & $\underline{W}$ is a Wiener process of m independent elements (m $\leq$ n) \\
      & $\underline{\underline{\sigma}}$ is a $nxm$ matrix of constant elements  \\
\end{tabular}

\section*{It\^o processes}

The integral with respect to Brownian motion was developed by It\^o in 1944.

The Brownian motion described above is very particular, specially because instantaneous expectation and variance ($\underline{\mu}$, $\underline{\sigma}$) are supposed to be constant.

Let's extend \ref{eqn:wiener-gen} to the case where $\underline{\mu}$ and $\underline{\sigma}$ are not constant but depend on the time t and the value of $\underline{X}$:

\begin{equation} \label{eqn:ito}
d\underline{X}(t) = \underline{\mu}(t,\underline{X}(t)) dt + \underline{\underline{\sigma}}(t,\underline{X}(t)) d\underline{W}(t)
\end{equation}

If a solution to \ref{eqn:ito} exists, then it should take the form:

\begin{equation} \label{eqn:ito-integral}
\underline{X}(t) = \underline{X}(s) + \int_s^t \underline{\mu}(r,\underline{X}(r)) dr + \int_s^t \underline{\underline{\sigma}}(r,\underline{X}(r)) d\underline{W}(r)
\end{equation}

where s$<$t is the current instant

If $\underline{\mu}(t,\underline{X}(t))$ and $\underline{\underline{\sigma}}(t,\underline{X}(t))$ respect two conditions, called It$\hat{o}$'s conditions, then \ref{eqn:ito-integral} admits a unique solution $\underline{X}(t)$, and this solution is a Markov process.

It\^o's conditions are:\\

$\forall t \in [0,T]$, $\underline{x} \in Q$, where Q is an open of $\Re^n$, $\exists$ two constants $C$ and $K$ such as:

\begin{itemize}
\item $\parallel \underline{\mu}(t,\underline{x}) \parallel \leq C(1+\parallel \underline{x} \parallel); \ \parallel \underline{\underline{\sigma}}(t,\underline{x}) \parallel \leq C(1+\parallel \underline{x} \parallel)$
\item $\parallel \underline{\mu}(t,\underline{x}) - \underline{\mu}(t,\underline{y}) \parallel \leq K(\parallel \underline{x} \parallel - \parallel \underline{y} \parallel); \ \parallel \underline{\underline{\sigma}}(t,\underline{x}) - \underline{\underline{\sigma}}(t,\underline{y}) \parallel \leq K(\parallel \underline{x} \parallel - \parallel \underline{y} \parallel)$
\end{itemize}

Stochastic processes that obey \ref{eqn:ito} and whose instantaneous expectation and variance respect It$\hat{o}$'s conditions are called "diffusion processes" or "It$\hat{o}$'s processes".

\section*{It\^o's lemma}

Let's consider a unidimensional diffusion process $X(t)$:

$$
dX = \mu(t,X)dt + \sigma(t,X) dW
$$
where $\mu(t,X)$ and $\sigma(t,X)$ respect It\^o's conditions and $W$ is a Wiener.

Let f be a function from $\Re^2$ to $\Re$, once continuously derivable with respect to $t$ and twice continuously derivable with respect to $X$:

$$
(t,X) \stackrel{f}{\longmapsto} f(t,X)
$$

It\^o's lemma is:

\begin{equation} \label{eqn:ito-lemma}
df = \frac{\delta f}{\delta t}dt + \frac{\delta f}{\delta x}dX + \frac{1}{2}\frac{\delta^2f}{\delta x^2}(dX)^2
\end{equation}

Equation \ref{eqn:ito-lemma} is very important since it is the basis of the differential calculus of stochastic functions. The main difference with usual differential calculus is the presence of the term $\frac{1}{2}\frac{\delta^2f}{\delta x^2}(dX)^2$.

\chapter{Elements of statistics}

\section*{Central Limit Theorem (CLT)}
The central limit theorem considers a series of random variables $X_1$, $X_2$, ..., $X_n$ independent and identically distributed ("i.i.d.") with finite mean $\mu$ and variance $\sigma$, and states that:
$$
Z_n \equiv \frac{\sqrt{n}}{\sigma} \left [ \frac{\sum_{i=1}^{n}Y_i}{n} - \mu \right ] \sim_{n \rightarrow +\infty} N(0,1)
$$
This capital theorem indicates that the sum of a large number of independent events is approximatively normal.
In other words, the distribution of an average will tend to be normal as the sample size $n$ increases, regardless of the distribution from which the average is taken (the parent distribution), except when the central moments of the parent distribution do not exist (viz. are not finite).

\section*{Kurtosis}

The degree of peakedness of a distribution, also called the "excess" or "excess coefficient." Kurtosis is a normalised form of the fourth central moment of a distribution. There are several types of kurtosis commonly encountered, including Fisher kurtosis (denoted $\gamma_2$ and also known as the kurtosis excess)
$$
\gamma_2 = \frac{\mu^4}{\mu_2^2} - 3
$$
and Pearson kurtosis (denoted or $\beta_2$)
$$
\beta_2 = \frac{\mu^4}{\mu_2^2}
$$
Here, $\mu_i$ denotes the ith central moment
$$
\mu_i = E[(X-\mu)^i]
$$
where $\mu$ is the mean of the distribution.

If not specifically qualified, then term "kurtosis" is generally taken to refer to Fisher kurtosis. A distribution with a high peak is called leptokurtic, a flat-topped curve is called platykurtic, and the normal distribution is called mesokurtic.

\section*{Normal Probability Plot}
The normal probability plot is a graphical technique for assessing whether or not a data set is approximately normally distributed.
The data are plotted against a theoretical normal distribution in such a way that the points should form an approximate straight line. Departures from this straight line indicate departures from normality.

The normal probability plot is formed by:
\begin{itemize}
\item Vertical axis: Ordered response values
\item Horizontal axis: Normal order statistic medians
\end{itemize}
The observations are plotted as a function of the corresponding normal order statistic medians which are defined as:
$$
N(i) = G(U(i))
$$
where $U(i)$ are the uniform order statistic medians (defined below) and $G$ is the percent point function of the normal distribution. The percent point function is the inverse of the cumulative distribution function (probability that $x$ is less than or equal to some value). That is, given a probability, we want the corresponding $x$ of the cumulative distribution function.

The uniform order statistic medians are defined as:
\begin{eqnarray*}
      m(1) &=& 1 - m(n) \\
      m(i) &=& (i - 0.3175)/(n + 0.365) \ \rm for \ i = 2, 3, ..., n-1 \\
      m(n) &=& 0.5(1/n)
\end{eqnarray*}
In addition, a straight line can be fitted to the points and added as a reference line. The further the points vary from this line, the greater the indication of departures from normality. The correlation coefficient of the points on the normal probability plot can be compared to a table of critical values to provide a formal test of the hypothesis that the data come from a normal distribution.

The underlying assumptions for a measurement process are that the data should behave like:
\begin{itemize}
\item random drawings
\item from a fixed distribution
\item with fixed location
\item with fixed scale
\end{itemize}

Probability plots are used to assess the assumption of a fixed distribution. In particular, most statistical models are of the form:

response = deterministic + random\\
\noindent where the deterministic part is the fit and the random part is error. This error component in most common statistical models is specifically assumed to be normally distributed with fixed location and scale. This is the most frequent application of normal probability plots. That is, a model is fit and a normal probability plot is generated for the residuals from the fitted model. If the residuals from the fitted model are not normally distributed, then one of the major assumptions of the model has been violated.

\section*{Statistical tests terminology}

Here are a few definitions of the terminology used in our statistical tests:

\begin{itemize}
\item The \emph{null hypothesis} H0 is the original assertion. For instance, H0 = The empirical points follow a normal distribution. The null hypothesis is always tested against an alternative hypothesis H1. For instance, H1 = The empirical points do \emph{not} follow a normal distribution
\item The \emph{significance level} is related to the degree of certainty you require in order to reject the null hypothesis in favour of the alternative. By taking a small sample you cannot be certain about your conclusion. So you decide in advance to reject the null hypothesis if the probability of observing your sampled result is less than the significance level. For a typical significance level of 5\%, the notation is $\alpha = 0.05$. For this significance level, the probability of incorrectly rejecting the null hypothesis when it is actually true is 5\%. If you need more protection from this error, then choose a lower value of $\alpha$.
\item The \emph{p-value} is the probability of observing the given sample result under the assumption that the null hypothesis is true. If the p-value is less than $\alpha$, then you reject the null hypothesis. For example, if $\alpha = 0.05$ and the p-value is 0.03, then you reject the null hypothesis.
The converse is not true. If the p-value is greater than $\alpha$, you have insufficient evidence to reject the null hypothesis.
\end{itemize}

\section*{Kolmogorov-Smirnov Goodness-of-Fit Test}

The Kolmogorov-Smirnov test (K-S, \cite{KS}), is used to decide whether a sample comes from a population with a specific distribution. The K-S test is based on the empirical cumulative distribution function (ECDF). 

The empirical distribution function is compared with the model cumulative distribution function. The K-S test is based on the maximum distance between 
these two curves. It tests a simple hypothesis, which means that the parameters of the expected distribution must \emph{not} be derived from the empirical data, but must be specified in advance.

The Kolmogorov-Smirnov test is defined by:
\begin{itemize}
\item H0:  The data follow a specified distribution  
\item Ha:  The data do not follow the specified distribution  
\item Test Statistic:  The Kolmogorov-Smirnov test statistic is defined as  
$$
D = \max_{1 \leq i \leq N}{\mid F(Y_i) - \frac{i}{N} \mid}
$$ 
where F is the theoretical cumulative distribution of the distribution being tested which must be a continuous distribution (i.e., no discrete distributions 
such as the binomial or Poisson), and it must be fully specified (i.e., the location, scale, and shape parameters cannot be estimated from the data). 
\item Significance Level:  $\alpha$  
\item Critical Values:  The hypothesis regarding the distributional form is rejected if the test statistic, $D$, is greater than the critical value obtained from a table.  
\end{itemize}

An attractive feature of this test is that the distribution of the K-S test statistic itself does not depend on the underlying cumulative distribution 
function being tested. Another advantage is that it is an exact test (the chi-square goodness-of-fit test depends on an adequate sample size for the 
approximations to be valid). Despite these advantages, the K-S test has several important limitations:
\begin{enumerate}
\item It only applies to continuous distributions.
\item It tends to be more sensitive near the center of the distribution than at the tails.
\item Perhaps the most serious limitation is that the distribution must be fully specified. That is, if location, scale, and shape parameters are estimated from the data, the critical region of the K-S test is no longer valid. It typically must be determined by simulation. 
\end{enumerate}

\section*{Chi-Square Goodness-of-Fit Test}
The chi-square test is used to test if a sample of data came from a population with a specific distribution.

An attractive feature of the chi-square goodness-of-fit test is that it can be applied to any univariate distribution for which you can calculate the cumulative distribution function. The chi-square goodness-of-fit test is applied to binned data (i.e., data put into classes).

The chi-square test is an alternative to the  Anderson-Darling and Kolmogorov-Smirnov goodness-of-fit tests. The chi-square goodness-of-fit test can be applied to discrete distributions such as the binomial and the Poisson. The Kolmogorov-Smirnov and Anderson-Darling tests are restricted to continuous distributions.

The test statistic follows, approximately, a chi-square distribution with $k - c$ degrees of freedom where k is the number of non-empty cells
and $c$ = the number of estimated parameters (including location and scale parameters  and shape parameters) for the distribution + 1. For example,
for a 3-parameter Weibull distribution, c = 4.

Therefore, the hypothesis that the data are from a population with the specified distribution is rejected if $\chi^2 > \chi^2(\alpha,k-c)$, where $\chi^2(\alpha,k-c)$ is the chi-square percent point function with $k - c$ degrees of freedom and a significance level of $\alpha$.

\section*{Jarque-Bera Goodness-of-Fit Test}

The Bera-Jarque test is a parametric hypothesis test of composite normality. It determines if the null hypothesis of composite normality is a reasonable assumption regarding the population distribution of the observed data X, at a given significance level $\alpha$.
  
The Bera-Jarque hypotheses are: 
\begin{itemize}
\item    Null Hypothesis:        $X$ is normal with unspecified mean and variance.
\item    Alternative Hypothesis: $X$ is not normally distributed.
\end{itemize} 

The Bera-Jarque test is a 2-sided test of composite normality with sample mean and sample variance used as estimates of the population mean and variance, respectively. The test statistic is based on estimates of the sample skewness and kurtosis of the normalised data (the standardised $Z$-scores computed from $X$ by subtracting the sample mean and normalising by the sample standard deviation). Under the null hypothesis, the standardised 3rd and 4th moments are asymptotically normal and independent, and the test statistic has a Chi-square distribution with two degrees of freedom. Note that the Bera-Jarque test is an asymptotic test, and care should be taken with small sample sizes.

\section*{Lilliefors Goodness-of-Fit Test}

The Lilliefors test for goodness of fit to a normal distribution. It evaluates the hypothesis that observed data X have a normal distribution with unspecified mean and variance, against the alternative that $X$ do not have a normal distribution. This test compares the empirical distribution of X with a normal distribution having the same mean and variance as X. It is similar to the Kolmogorov-Smirnov test, but it adjusts for the fact that the parameters of the normal distribution are estimated from X rather than specified in advance.  Thus, it determines if the null hypothesis of composite normality is a reasonable assumption regarding the population distribution of the observed data X.

Let S(x) be the empirical c.d.f. estimated from the sample vector $X, F(x)$ be the corresponding true (but unknown) population c.d.f., and CDF be a normal c.d.f. with sample mean and standard deviation taken from $X$. The Lilliefors hypotheses and test statistic are:
\begin{itemize}
\item Null Hypothesis: $F(x)$ is normal with unspecified mean and variance
\item Alternative Hypothesis: $F(x)$ is not normally distributed.
\item Test Statistic:         $T = max|S(x) - CDF|$
\end{itemize}
The decision to reject the null hypothesis occurs when the test statistic exceeds the critical value.

\section*{Maximum Likelihood Ratio Test}

Let $L1$ be the maximum value of the likelihood of the data without the additional assumption. In other words, $L1$ is the likelihood of the data with all the parameters unrestricted and maximum likelihood estimates substituted for these parameters.

Let $L0$ be the maximum value of the likelihood when the parameters are restricted (and reduced in number) based on the assumption. Assume $k$ parameters were lost (i.e., $L0$ has $k$ less parameters than $L1$).

Form the ratio $\lambda = L0/L1$. This ratio is always between 0 and 1 and the less likely the assumption is, the smaller $\lambda$ will be. This can be quantified at a given confidence level as follows:
\begin{enumerate}
\item Calculate $\chi^2= -2 \ln \lambda$. The smaller $\lambda$ is, the larger $\chi^2$ will be.
\item We can tell when $\chi^2$ is significantly large by comparing it to the upper $100 * (1-\alpha)$ percentile point of a Chi Square distribution with $k$ degrees of freedom.  $\chi^2$ has an approximate Chi-Square distribution with $k$ degrees of freedom and the approximation is usually good, even for small sample sizes.
\item The likelihood ratio test computes $\chi^2$ and rejects the assumption if $\chi^2$ is larger than a Chi-Square percentile with $k$ degrees of freedom, where the percentile corresponds to the confidence level chosen by the analyst.
\end{enumerate}

\section*{Kruskal-Wallis Test}

The Kruskal-Wallis test is a nonparametric version of one-way analysis of variance. The assumption behind this test is that the measurements come from a continuous distribution, but not necessarily a normal distribution. The test is based on an analysis of variance using the ranks of the data values, not the data values themselves. The function returns the p-value for the null hypothesis that all samples are drawn from the same population (or from different populations with the same mean). It is based on the Chi-Square distribution. \\

If the p-value is near zero, this casts doubt on the null hypothesis and suggests that at least one sample mean is significantly different than the other sample means.

\section*{Analysis of Variance}

The purpose of one-way analysis of variance (``ANOVA'') is to find out whether data from several groups have a common mean. That is, to
determine whether the groups are actually different in the measured characteristic.

One-way ANOVA is a simple special case of the linear model. The one-way ANOVA form of the model is
$$
y_{ij} = \alpha_{.j} + \epsilon_{ij}
$$
\begin{tabular}{rl}
where & $y_{ij}$ is a matrix of observations in which each column represents a different group \\
& $\alpha_{.j}$ is a matrix whose columns are the group means. (The "dot j" notation means \\
& that applies to all rows of the jth column. That is, the value ij is the same for all i.) \\
& $\epsilon_{ij}$ is a matrix of random disturbances
\end{tabular}

The model posits that the columns of $y$ are a constant plus a random disturbance. You want to know if the constants are all the same. The p-values returned by the ANOVA test depend on assumptions about the random disturbances $\epsilon_{ij}$ in the model equation. For the p-values to be correct these disturbances need to be independent, normally distributed, and have constant variance. Some nonparametric methods like the Kruskal-Wallis Test do not require a normal distribution.

\end{document}